\documentclass{aa}

\usepackage{graphicx}
\usepackage{txfonts}

\usepackage{silence}
\usepackage{subcaption}

\usepackage{hyperref}      
\hypersetup{colorlinks=true,linkcolor=blue,citecolor=blue,filecolor=blue,urlcolor=blue,}
\makeatletter
\renewcommand*\aa@pageof{, page \thepage{} of \pageref*{LastPage}}
\makeatother

\newcommand{\kms}{$\textrm{km\:s}^{-1}$}
\newcommand{\kpvsys}{$K_p-V_{sys}\ $}

\begin{document}

   \title{The \textbf{Mantis Network} \texttt{III}: Expanding the limits of chemical searches within ultra-hot Jupiters}

    \subtitle{New detections of Ca\,I, V\,I, Ti\,I, Cr\,I, Ni\,I, Sr\,II, Ba\,II, and Tb\,II in KELT-9\,b}

   \author{N. W. Borsato
          \inst{1}
          \and
          H. J. Hoeijmakers\inst{1}
          \and
          B. Prinoth\inst{1}
          \and
          B. Thorsbro\inst{1,2} 
          \and
          R. Forsberg \inst{1}
          \and
          D. Kitzmann\inst{3}
          \and 
          K. Jones \inst{3}
          \and
          K. Heng \inst{4}
          }

   \institute{Lund Observatory, Department of Astronomy and Theoretical Physics, Lund University, Box 43, 221 00 Lund, Sweden\\
              \email{n.borstato@astro.lu.se}
            \and
            Department of Astronomy, School of Science, The University of Tokyo, 7-3-1 Hongo, Bunkyo-ku, Tokyo 113-0033, Japan
            \and
            University of Bern, Center for Space and Habitability, Gesellschaftsstrasse 6, CH-3012, Bern, Switzerland
            \and
            Universitäts-Sternwarte, Fakultät für Physik, Ludwig-Maximilians-Universität München, Scheinerstr. 1, 81679 München, Germany}

   \date{Received October 3, 2022; accepted April 1, 2022}

\abstract{Cross-correlation spectroscopy is an invaluable tool in the study of exoplanets. However, aliasing between spectral lines makes it vulnerable to systematic biases. This work strives to constrain the aliases of the cross-correlation function to provide increased confidence in the detections of elements in the atmospheres of ultra-hot Jupiters (UHJs) observed with high-resolution spectrographs. We use a combination of archival transit observations of the UHJ KELT-9\,b obtained with the HARPS-N and CARMENES spectrographs and show that it is possible to leverage each instrument's strengths to produce robust detections at a substantially reduced signal-to-noise. Aliases that become present at low signal-to-noise regimes are constrained through a linear regression model. We confirm previous detections of H\,I, Na\,I, Mg\,I, Ca\,II, Sc\,II, Ti\,II, Cr\,II, Fe\,I, and Fe\,II, and detect eight new species, Ca\,I, Cr\,I, Ni\,I, Sr\,II, and Tb\,II, at the 5$\sigma$ level, and Ti\,I, V\,I, and Ba\,II above the 3$\sigma$ level. Ionised terbium (Tb\,II) has never before been seen in an exoplanet atmosphere. We further conclude that a $5\sigma$ threshold may not provide a reliable measure of confidence when used to claim detections, unless the systematics in the cross-correlation function caused by aliases are taken into account.}



   \maketitle
%
\section{Introduction}
Exoplanets and their natural diversity have revolutionised the field of astronomy. It is now a well-understood fact that most stars host exoplanets. In addition to detecting these worlds, it is now possible to observe them in great detail and uncover information about their atmospheric and orbital characteristics, formation histories, and occurrence rates. Observations of exoplanet atmospheres hold the interest of observers and theorists alike, as they are the next step in further characterising these objects. However, the coupling of signals between an exoplanet and its host star presents a challenge when observing its atmosphere. Since the first successful detection of the atmosphere of the ultra-hot Jupiter HD 209458 b \citep{2002ApJ...568..377C}, improvements in instrumentation have been made such that it is now possible to detect the absorption signatures of individual ionised, atomic, and molecular species.

KELT-9\,b, which was discovered by \cite{Gaudi_2017_KELT9b}, is an ultra-hot Jupiter (UHJ) with the highest recorded equilibrium temperature of 4000\,K. The planet has been observed with many instruments, including the HARPS-North \citep{Hoeijmakers_2019_Fe_Detections,Pino_2022_K9_Dayside}, CARMENES \citep{Yan_2018}, PEPSI \citep{Cauley_2019_K9_MgTriplet,Asnodkar_2022_Single_Lines}, FIES \citep{Bello-Arufe_2022_Iron_NOT}, and TRES \citep{Lowson_2023} spectrographs. These observations have led to the detection of the following species: H\,I, Na\,I, Mg\,I, Ca\,II, Sc\,II, Ti\,II, Cr\,II, Fe\,I, Fe\,II, and Y\,II. 

Detecting heavier elements, such as those produced by neutron-capture processes \citep{burbidge1957} in exoplanets, is desirable due to the information they can reveal about planet formation and the host stars they orbit \citep{ek2020}. Studies of meteorites in our Solar System have shown that the inner Solar System (e.g. the Earth) is relatively enriched in isotopes produced by the slower neutron-capture process (s-process) compared to the much more rapid neutron-capture process (r-process), which contradicts current established theories on planet formation \citep{burkhardt2011_mo,budde2016_mo,budde2019_mo,saji2020_nd_ppd_evolution}. The inhomogeneity of these heavy isotopes in Solar System meteorites may be driven from where the protoplanetary material, which makes up the meteorites, originates \citep{dauphas2010_sn_dust,lugaro2016,ek2020}. In conjunction with this observation, massive planets are commonly known to migrate inwards towards their host stars, and form on faster timescales than terrestrial planets \citep{Formation_Rate_of_Jupiter_Mass_planets,Formation_Rate_Of_Terrestrials}. Therefore, giant planets should sample the protoplanetary material used in the formation of terrestrial planets as they migrate, leading to the postulation that there could be a coupling of the inhomogeneity trend with the Jupiter-mass planet's atmosphere.

Furthermore, \citet{Melendez2009} noted that the Sun, as compared to solar twins, is depleted in several refractory elements – elements with a high condensation temperature – such as yttrium and barium. The possibility that planet formation imprints chemical signatures on the host star is an active area of research, where differences in observed stellar abundance trends with condensation temperature are likely associated with the histories of planet formation \citep[e.g.][]{Booth2020_giantplanets,Liu202_possible_planetsign,Liu2021_planetformation_binaries}.  Additionally, \citet{kruijer2017_jupiter_age} show that measurements of heavy elements have proven to be a useful tool in determining the age of Jupiter, which is directly linked to the age of the Sun. Age determination is an ongoing challenge in stellar astrophysics \citep{Age_Detemination_Of_Stars,Sahlholdt_2019}, and an independent approach through planet age determination could help to validate current age estimate approaches. Thus, the study of heavy trans-iron elements in exoplanets has the potential to help explain how planets form, and to infer more information about the host stars they inhabit. Detection studies and further analysis of heavier elements in UHJs would present the first steps in constructing a sample size to further test the ideas of planet formation in a context outside of the Solar System.

In order to study the chemical composition of exoplanets, it is necessary to remove auxiliary signals, such as telluric and stellar signals. The removal of auxiliaries results in a noise-dominated spectrum with the planet's absorption spectrum buried inside but all for the deepest transmission lines. These deep lines, however, are routinely analysed \citep[e.g.][]{EXM_Ehrenreich_2008,Wyttenbach_2017_Na_HEARTS,EXM_Salz_2018,Cauley_2019_K9_MgTriplet,EXM_Alonso-Floriano_2019,EXM_Zguy_2019,Seidel_2020_Nadetections_Hearts_2,EXM_Lamp_2021,Carmenes_Paper_2_Method,Seidel_Na_Detection_2022,Asnodkar_2022_Single_Lines}, and they can directly measure the mass-loss rate, exosphere, temperature profile, and atmospheric dynamics, and constrain where in the atmosphere a particular species is located. The cross-correlation technique \citep{Snellen_2010_CC} can be used to leverage all the known lines of a given species in a spectrum to produce an average line profile. This method increases the signal-to-noise ratio (S/N) enough so that the species becomes detectable, although it removes information about the individual lines and prevents studies of individual line profiles. Nevertheless, cross-correlation enables the acquisition of chemical inventories and the analysis of their atmospheric characteristics, such as temperature-pressure profiles, chemical abundances, and surface gravity \citep[e.g.][]{Brogi_Line_2019,Fisher_2020,Pelletier_2021,Gibson_2022}. 

The primary use of the cross-correlation technique is to detect chemical species in an exoplanet's atmosphere. With it, a diverse range of ionised, atomic, and molecular species have been identified, including metals such as iron and titanium \citep[e.g.][]{Hoeijmakers_2018_Fe_Detection,Hoeijmakers_2019_Fe_Detections,Casasayas-Barris_2019_Iron_And_Lines,Ehrenreich_2020,Pino_2020,Jens_Detectic_Atoms_on_Mascara_1B_2020,Jens_WASP_121b_Transit_Detections_of_Atoms_2020,Cabot_2021_Fe_Detections,Merritt_2021_Chemical_Inventory_WASP_121b,Bello-Arufe_2022_Iron_NOT}, and molecules such as water, carbon monoxide, hydroxide radicals \citep[e.g.][]{Brogi_Welch_ttest,Brogi_2013_Molecules,Brogi_2014,Birkby_2017,deKok_2017,Alonso_Floriano_2019,Landman_2020_OH_Radical,Giacobbe_2021,Nugroho_2021,vanSluijs_2022,Guilluy_New_Molecules_Detections_Gaicobbe_2022}, and titanium-oxide \citep{Nugroho_2017_DaysideTiO,Cont_2021,Prinoth_2022}.

A substantial proportion of chemical species detections have come from the same class of planets, UHJs, which are short-period giant planets with equilibrium temperatures comparable to the effective temperatures of M dwarfs. Their extreme equilibrium temperature creates an extended, cloud-free atmosphere that maximises the potential flux contribution that can come from the optically thin region of an exoplanet's atmosphere \citep{Heng_2016,Stevenson_2016,Kitzman_KELT9b_Introduction}. Thus, UHJs have effectively become laboratories for experimenting with exoplanet spectra, enabling the field to develop and refine techniques. Provided that an adequate line list exists, it is possible to search for any element, although not all are equally detectable \citep{Kesseli_2022}. For heavier elements, abundances generally decrease, entailing weaker absorption from the given species. Furthermore, absorption is dictated by the combined line opacity of a species \citep{HengAKitz_2017}, so species with fewer lines may also be difficult to detect. If the scientific goal is to detect a particular species, as much spectroscopic information as possible should be leveraged to obtain the highest chance of detection.

Cross-correlation results rely heavily on the chosen line list, with different lists containing slight variations in the calculated cross-sections of a given species \citep{GharibNezhad_2021_LineLists}. Incorrect cross-sections can lead to inefficient or even null detections if the mismatch is large enough \citep{Hoeijmakers_2015}. Line-list inaccuracies are likely the cause of why the molecular species TiO has been challenging to detect until recently \citep[e.g.][]{Hoeijmakers_2015,Nugroho_2017_DaysideTiO,Sedaghati_2021,Prinoth_2022}. Therefore, accurate line lists that are representative of the atmosphere are required \citep{Kitzmann_2021_Template_Paper}. Combining lines that would not be present in a spectrum, or assigning too much weight to a specific line reduces the efficiency of the cross-correlation process. In all cases, the result is a noisier signal with a high risk of non-detection for faint signals. 

Another inhibiting factor that has a tangible effect on the cross-correlation signal is the wavelength range of the spectrograph. Typically, studies only analyse data from one instrument \citep[e.g.][]{Yan_2018,Hoeijmakers_2019_Fe_Detections,Ehrenreich_2020,Merritt_2021_Chemical_Inventory_WASP_121b,Prinoth_2022,Stangret_2022,Bello-Arufe_2022_Iron_NOT}, as it is relatively easy to combine observations from the same instrument. Techniques that combine spectrographic observations from different high-resolution instruments are still in their infancy, but they have been used in the past \citep[e.g.][]{Carmenes_Paper_2_Method,Asnodkar_2022_Single_Lines,Pino_2022_K9_Dayside}. As archives collect multiple transit observations of the same target, developing robust combination techniques becomes important. Combining the information from multiple spectrographs enhances the quality of observations beyond the capabilities of a single instrument, as it allows for a more extensive coverage of wavelength ranges that may be more conducive to detecting certain species over others.

An important topic discussed less in cross-correlation studies is aliases. Aliases are systematic artefacts of cross-correlation functions, either from auto-correlation with the search species \citep{Hoeijmakers_2018_Has_autocorrelation} or correlations with alternative species in the spectrum \citep{Hoeijmakers_2018_Fe_Detection,Hoeijmakers_2019_Fe_Detections}. The systematic variation distorts the noise in the baseline of the cross-correlation function and can, in turn, severely undermine the uncertainty analysis. This problem becomes more apparent when searching for less abundant species or species with fewer lines, as interference from species with stronger lines becomes more likely.

The growing number of observations of KELT-9\,b leads to a relative decrease in what is achievable in a single night of observations, while  increasing the motivation to combine multiple spectroscopic observations. This study combines archival data from the HARPS-N and CARMENES spectrographs to detect exotic species in the atmosphere of KELT-9\,b. The defined approach combines cross-correlation functions from multiple observations, and constrains the predicted alias contributions to yield more statistically robust detections. First searches for the same elements found in the publications of \cite{Hoeijmakers_2019_Fe_Detections}, \cite{Yan_2018}, and \cite{Carmenes_Paper_Method} serve as verification before searching for new undetected species in the combined data sets. The aim is to show that signals increase in their S/Ns when multiple spectrographs are combined, and also begin to resolve alias structures, while constraining these structures can lead to new detections of atoms in KELT-9\,b's atmosphere.

This paper is structured as follows: in Sect. \ref{DataProcessesing} we summarise the data and processing steps. In Sect.~\ref{CCFApproach} we introduce the cross-correlation technique and our new method for extracting the planetary signals from the data; additional supplementary information is provided in Appendix~\ref{Weight Results}. We present the results of our attempts to detect new species in KELT-9\,b's atmosphere in Sect.~\ref{results}, with additional results discussed in more detail in Appendix~\ref{DetectionStatistics}, and validating these detections in Appendix~\ref{A:ModelPlots}. Finally, we discuss the implications of the results in Sect.~\ref{Discussion}, and draw conclusions in Sect.~\ref{Conclusion}.

\section{Data collection \& processing}
\label{DataProcessesing}

The first two of the four analysed data sets originate from observations taken with the HARPS-North spectrograph \citep[HARPS-N,][]{HARPS_Mayor_2003} located on the Telescopio Nazionale Galileo. The original studies that used these data were \cite{Hoeijmakers_2018_Fe_Detection,Hoeijmakers_2019_Fe_Detections}. HARPS-N is an optical spectrograph spanning a wavelength range between 387.4 and 690.0\,nm, at a spectral resolution of 115,000. The two transit observations were conducted on 31 July 2017 and 20 July 2018, respectively, with 600\,s exposures. Both observations cover the 3.9\,h transit and additional baseline measurements totalling 49 and 46 exposures, with median S/N values over the echelle orders of 74 and 71, respectively. The primary science products obtained from these observations are the individually extracted echelle orders from the 2D echellogram and the blaze-corrected, stitched and resampled 1D spectra.

The two additional data sets originate from the CARMENES spectrograph \citep{CARMENES_2014} from the Calar-Alto Observatory. CARMENES is an optical and near-infrared spectrograph with the optical arm covering a wavelength range between 520 and 960 nm and an infrared wavelength between 960 and 1,710\,nm at a resolution of 94,600 and 80,400. The first publications to analyse this data were \cite{Yan_2018} and \cite{Carmenes_Paper_Method}. The observations were taken on 6 August 2017 and 16 June 2018, covering the entire transit. Both observations cover the entire transit with 300\,s exposures for the first night and 111\,s for the second. The present study uses the optical arm of the instrument with median S/N values of 35 and 34. The primary science products drawn from the archive are the individually extracted and blaze-corrected echelle orders, estimated errors and blaze profiles. The 2D orders were stitched together using a weighted average to construct a 1D spectrum of each exposure. Cross-correlation is more efficient when the blaze profile has not been removed, as this eliminates the need to re-weigh lines in the template by the effective throughput of the spectrograph (i.e. near order edges). Therefore, the profiles are divided back into spectral orders. This information is summarised in Table \ref{tab:ObservationalSummary}.

\begin{table*}
    \centering
    \caption{Summary information of the transit observations used in this study}
    \label{tab:ObservationalSummary}
    \begin{tabular}{c|c|c|c|c|c|c}
        Observation & Spectrograph & Wavelength Range (nm) & Resolution & S/N & Exp-Time (s) & Exp-Number\\ \hline
        1 & HARPS-N & 387.4 - 690.0 & 115 000 & 74 & 600 & 49 \\
        2 & HARPS-N & 387.4 - 690.0 & 115 000 & 74 & 600 & 46 \\
        3 & CARMENES & 520 - 960 & 94,600  & 35 & 300 & 54 \\
        4 & CARMENES & 520 - 960 & 94,600  & 34 & 111 & 140\\
    \end{tabular}
\end{table*}

In order to optimise the cross-correlation search, we removed any systematic signals that do not belong to the planet itself. The adopted process follows the same methodology applied in \cite{Jens_WASP_121b_Transit_Detections_of_Atoms_2020}. This reference provides an in-depth walkthrough of the data preparation. The systematics primarily consists of atmospheric telluric contamination, velocity offsets, the stellar spectrum, and any additional effects such as interstellar medium (ISM) lines or residual continuum offsets.

The telluric lines produced by the $\rm{H_2O}$ and $\rm{O_2}$ absorption bands in the atmosphere are removed using \texttt{molecfit} \citep{molecfits_puplication_paper}. \texttt{molecfit} creates a model for the telluric transmission spectrum over the entire wavelength of the spectrograph range using the 1D spectra of each observation. The models are then interpolated onto the wavelength solutions of the 2D echelle orders and divided out. This process applies to spectra without blaze correction since the operation is multiplicative. The correction is enough for HARPS-N and the CARMENES visible arm to bring the telluric corrections to the noise level of the spectra.

In addition to the telluric removal, the stellar spectra require two Doppler corrections to correctly place each spectral order into the host star's rest frame. The first is the centre of mass motion of the Earth-Sun system, the barycentric velocity, while the second corrects the star's radial velocity induced by the planet's gravitational pull. The correction leaves the spectra at a constant radial velocity shift set by the systemic velocity of approximately $-18$\,\kms  \citep{Gaudi_2017_KELT9b,Borsa_2019,Hoeijmakers_2019_Fe_Detections,Asnodkar_2022_Single_Lines}, all other system parameters were taken from \cite{Gaudi_2017_KELT9b}. We removed the stellar signal from the time series spectra by calculating an out-of-transit baseline of the star and dividing each spectrum by it. The spectral orders were then flattened using a median filter in order to remove broadband continuum variations, and with a running standard deviation applied in suit. Values greater than 5$\sigma$ were flagged as NaNs. The process was then repeated instead using a Gaussian high-pass filter of 80,\kms, with any remaining 5$\sigma$ outliers flagged as NaNs. The resultant NaNs created an outlier mask which avoided problematic pixels in the cross-correlation process \citep{Hoeijmakers_2018_Fe_Detection}.

\section{Cross-correlation analysis for transit spectroscopy}

\label{CCFApproach}
Cross-correlation combines the expected contributions of the absorption lines for a particular species in spectrum. The procedure requires using a spectral template $T$ of a specific species over a defined wavelength axis $i$ ($T_i$), which predicts the transmission spectrum of a specific species. The template is converted to a set of weights by normalisation:
\begin{equation}
    T = \frac{T_i(v)}{\Sigma T_i(v)},
\end{equation}

where $T_i(v)$ represents the template species Doppler shifted to a specific velocity, and $v$ represents the Doppler shift of the template. Cross-correlation is implemented using the following equation:
\begin{equation}
    c(v) = \sum_{i=0}^{N} x_i T_i(v),
    \label{eq:ccf}
\end{equation}

where $c(v)$ represents the weighted average of line positions at a given Doppler shift, $N$ represents the total number of wavelength bins, and $x_i$ is the spectral data which shares the same wavelength range $i$ as the template. The template and cross-correlation functions are velocity-dependent. During transit, a planet's signal Doppler shifts to its instantaneous radial velocity and a cross-correlation will form a maximum at this point, assuming the atmosphere's velocity profile is comparable to the planet's. With a radial velocity step of 1 \kms, we sample a radial velocity range of -1,000\,\kms\, to 1,000\,\kms. This option covers more aliasing lines in the cross-correlation function, which allows for tighter constraints on the alias-regression models without compromising computational efficiency. We introduce this in Section \ref{Aliases}.

During transit, the stellar disk becomes partially obscured and this creates a Doppler shift due to the rotation of the star \citep{Rossiter_RM,McLaughlin_RM}. In a cross-correlation map, this results in a bright apparent emission feature called the Doppler shadow that requires removal. We remove the Doppler shadow by masking where the planet's signal is expected to lie on the cross-correlation plane, assuming a Keplerian circular orbit:
\begin{equation}
    v_{rv} = v_{\rm{orb}}\sin{2\pi\phi}\sin{i} + v_{\rm{sys}}
    \label{orbital_equation}
\end{equation}

where $v_{rv}$  is the planetary radial velocity during transit, $v_{orb}$ is the orbital velocity of the system, $\phi$ is the orbital phase of the planet, $i$ is the system inclination, and $v_{(sys)}$ is the star's systemic velocity. We use this expression to mask the planet's signal along the trace by masking pixels in the cross-correlation function that lie within 20 \kms of the expected planet signal. Removal of the shadow follows the same procedure as \cite{Jens_WASP_121b_Transit_Detections_of_Atoms_2020}, which fits an empirical model consisting of two Gaussians as a profile that has a positive central core, negative side-lobes and a free scaling factor. 

\begin{figure*}
    \centering
    \includegraphics[scale=0.7]{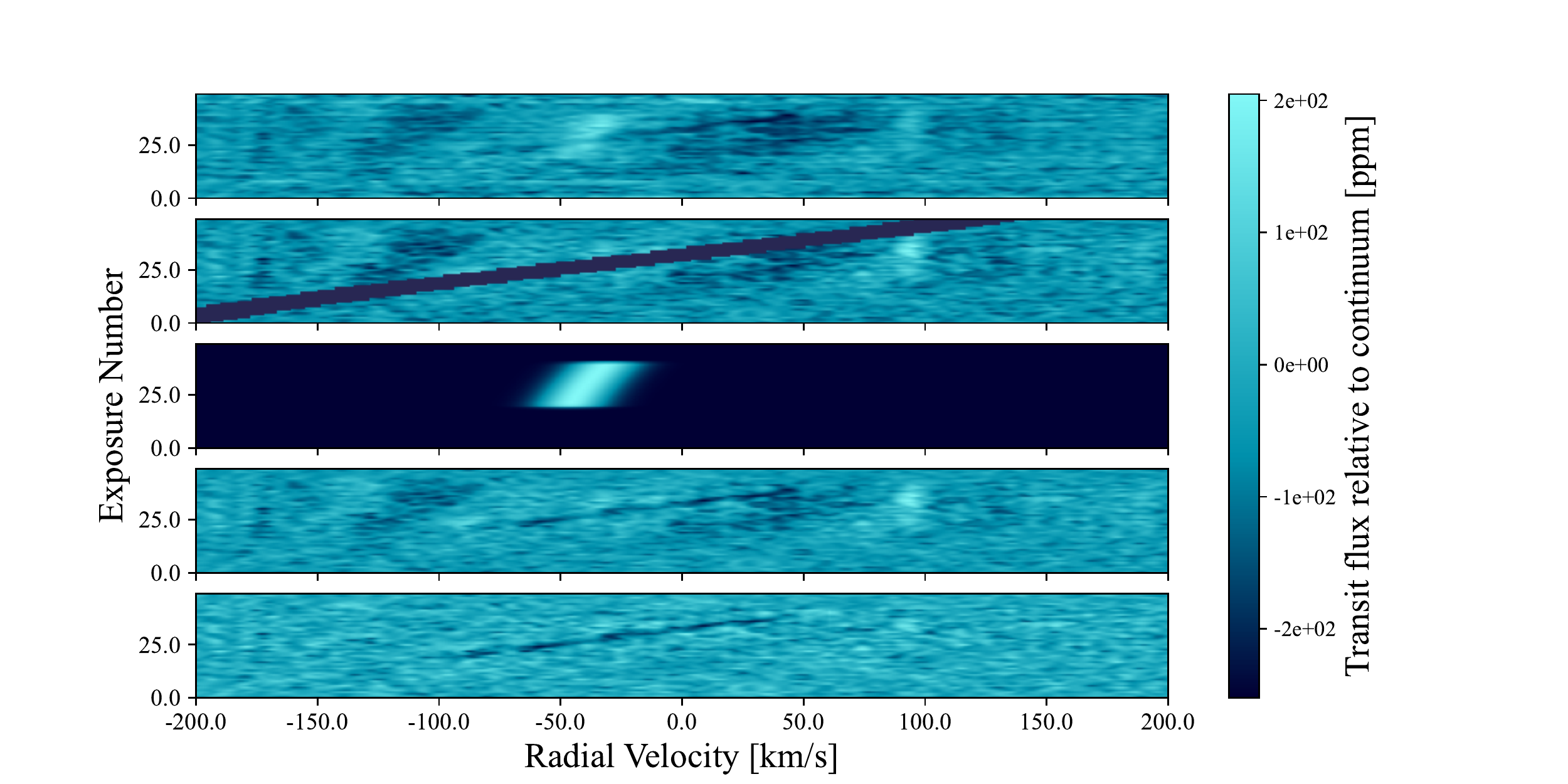}
    \caption{Overview of the cross-correlation cleaning process. \textit{Panel A:} The raw cross-correlation function time-series applied after the out-of-transit spectra have been divided and the residual broadband features removed. Both the Doppler shadow and the alias structures are present. \textit{Panel B:} The planet mask covers the expected planet signal, the half-width of the planet mask is 20 \kms. \textit{Panel C:} The fitted Doppler-shadow profile is a two-component Gaussian. \textit{Panel D:} The cross-correlation function after removal of the Doppler shadow, residual stellar aliases are still present. \textit{Panel E:} The resulting cross-correlation map after removing the stellar aliases. The signature of the planet appears as the dark slanted trace. In this plot, the sign is negative; however, it is flipped in preceding figures to denote absorption.}
    \label{fig:cleaning_steps}
\end{figure*}

Furthermore, correlated structures caused by the incomplete removal of strong stellar and telluric lines can remain in the cross-correlation function. These stellar aliases create near-vertical structures. Removing these artefacts follows a modified approach to \cite{Prinoth_2022}, which fits a 1D polynomial from each velocity value and subtracts the fit. Instead, rather than fitting for a single velocity value, a fit is applied to a sliding average on a column of ten velocity values, and subtraction is applied to the central column. The column window is then moved across the cross-correlation map, subtracting over the whole radial velocity range. The modified approach utilises the partial correlation between neighbouring pixels to better sample vertical trends. Fig. \ref{fig:cleaning_steps} illustrates the applied cleaning steps.

The exoplanet's signal can be strengthened further by co-adding all pixels in the planet's trace. To do this, the individual exposures are shifted to the time-independent rest frame of the planet. Since the planet's orbital phase and inclination are well known, it is possible to apply Eq. (\ref{orbital_equation}) by assuming a range of orbital velocities, in this case spanning 0 to 400\,\kms and shifting the pixels in the cross-correlation map accordingly. Co-adding in-transit cross-correlation functions creates a 1D cross-correlation function for every potential orbital velocity \citep{Nugroho_2017_DaysideTiO,Prinoth_2022}. Stacking these 1D cross-correlation functions produces a \kpvsys map \citep{Brogi_Welch_ttest}. If an absorption spectrum of the search species exists then the resulting map should show a peak at the orbital velocity of the planet and the stellar systemic velocity.

\subsection{Cross-correlation templates}
Cross-correlation requires a template mask to define the weights with which pixels are co-added together, as specified in Eq. (\ref{eq:ccf}). We constructed masks through forward modelling KELT-9\,b's transmission spectrum, assuming that the planet contained the absorption lines of the search species. The calculation of these templates primarily follows that of \cite{Hoeijmakers_2019_Fe_Detections}. We used \textsc{HELIOS-K}\footnote{\url{https://github.com/exoclime/HELIOS-K}} \citep{HELIOS_K_Grimm_2015} to compute the opacity functions for each species, drawing from the transmission tables of the VALD \citep{VALD_Database} and Kurucz \citep{Kurucz_2018} databases. The continuum opacity function for the mask comprised H-He, $\rm{H_2}$-$\rm{H_2}$ and $\rm{H_2}$-He collision induced absorption obtained from HITRAN \citep{HITRAN}, while the $\rm{H^-}$ bound-free and free-free absorption contributions were taken from \cite{John_Continuum_Opacity}. Cross-correlation templates are computed following the method described in \citet{Kitzmann_2021_Template_Paper}. The chemical composition is determined 
with the equilibrium chemistry code \textsc{FastChem}\footnote{\url{https://github.com/exoclime/FastChem}} \citep{Stock2018MNRAS.479..865S, Stock2022MNRAS.tmp.2449S} assuming solar element abundances and an isothermal temperature profile with 4,000\,K in the pressure range 10 - $\rm{10^{-15}}$\,bar. These high temperatures create partially ionised atoms and a low abundance of molecules. The templates for the detected species can be found in Fig. \ref{fig:LineListPlots}.

\subsection{Stacking observations}
Combining spectroscopic observations from multiple nights with different spectrographs is non-trivial. When using a single spectrograph, it is possible to shift the observed spectra to the planetary rest frame for each night, and compute an average spectrum before applying the cross-correlation function \citep{Jens_WASP_121b_Transit_Detections_of_Atoms_2020}. Alternatively, observations can be averaged in \kpvsys space as the process projects the cross-correlation functions on the same grid for each night. In either case, an average dampens the noise leading to clearer signals. However, a standard mean does not consider variations in data quality caused by weather and air-mass variations. Additionally, when using different instruments, each spectrograph's unique wavelength coverage also creates differences in results. Therefore, additional steps are required to take these aspects into consideration.

\begin{figure*}
    \centering
    \includegraphics[scale=0.485]{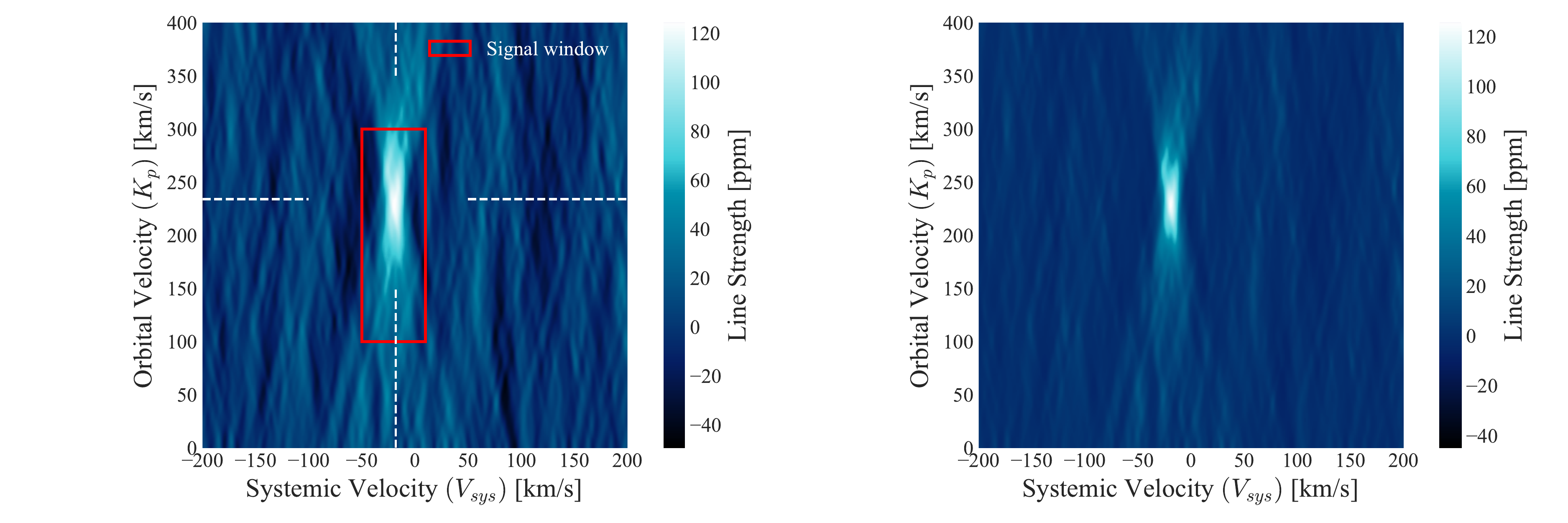}
    \caption{Comparison plot demonstrating the improved signal from applying the combination method for the species Fe\,I. \textit{Left Panel:}  \kpvsys map of a single transit observation, this specifically is the first HARPS-N transit. The red border marks the window where the average signal strength is computed, and the region outside where the standard deviation is computed. \textit{Right Panel:}  \kpvsys map when combining all four transit observations. The result is a much higher S/N.}
    \label{fig:ImprovementFromCombination}
\end{figure*}

To deal with the variations in data quality from the same instrument, we were able to mask out the planet's signal in the \kpvsys map. This was done by constructing a window of values surrounding the expected area of the signal. In our case, the window ranged from $-50$ \kms to $10$ \kms for the systemic velocity, and $100$ \kms to $300$ \kms for the orbital velocity (see Fig. \ref{fig:ImprovementFromCombination}). The window selected deliberately encompassed a wide range of velocity points to ensure any signal which may be offset, due to different atmospheric dynamics \citep{Prinoth_2022}, was captured (see Fig. \ref{fig:Categorical}). We then masked-out the signal, and took the standard deviations of the remaining pixels. Taking the standard deviation of the masked \kpvsys map measured the noise and any systematics that remained in the data without contribution from the signal of the planet. It was then possible to invert the mask and apply it to the \kpvsys map such that only the expected region where one would expect the signal to lie was extracted. If a signal existed in this region taking the mean would lead to a value which was larger on average than the background, if no signal existed the mean would average the noise and any systematics that lay in that region. A pseudo-S/N value could be constructed by taking the ratio of the mean and the standard deviation, whose value was dependent on the night's quality.

The pseudo-S/N calculation to \kpvsys maps is not sufficient to compare the quality of observations between different spectrographs. Different instruments cover different wavelength ranges and thus cover different lines. Since cross-correlation uses the line position of all lines across the wavelength range, this will create a difference in the result even if the data quality is identical. It is possible to account for this difference through the templates applied in cross-correlation. Applying a simple sum of the total absorbed flux of the template over the wavelength range achieves an approximate measurement for this value. This number will be the same for the same spectrograph but will differ for other instruments. Instruments that contain deeper more readily observed lines obtain a higher value.

With the sum of the line depths of the instrument and the pseudo-S/N of the masked \kpvsys map, it is possible to compute a statistic that enables comparison between nights and normalise the values to create a set of weights,
\begin{equation}
    \textbf{w} = \frac{\Sigma_{T}/\rm{(S/N)}_{\rm{pseudo},i}}{\sum_i^N (\Sigma_{T}/\rm{(S/N)}_{\rm{pseudo},i})},
\end{equation}

where i refers to the night of observation, $\Sigma_{T}$ is the sum of the template flux, and $\rm{(S/N)}_{pseudo,i}$ is the pseudo-S/N of the \kpvsys map. Fig. \ref{fig:Weight_Process} outlines the weight extraction process for Fe\,I and Na\,I. For Fe\,I, higher preference is given to the HARPS-N nights. This is likely due to its higher quality produced by the telescope, its high efficiency in resolving blue wavelengths and greater abundance of deeper lines. For Na\,I, however, the weights the CARMENES share a larger proportion of the total sum, as the combined flux from both spectrographs for this species is comparable. However, the data quality is still of better quality for the HARPS-N nights leading to stronger weight values.

The last step required before the \kpvsys maps can be combined is to normalise each map by subtracting the mean of the masked map and dividing by the standard deviation for each \kpvsys map. The transformation creates S/N maps commonly used in the literature \citep{Giacobbe_2021,Guilluy_New_Molecules_Detections_Gaicobbe_2022}. The act of scaling removes the implicit dependence of the line depth, which vary with different instruments. It is then possible to apply a weighted average using the pre-calculated weights to all nights to create one master S/N map. Finally, to reconvert the master map to one with a line strength: the master S/N map is normalised to a maximum value of one. This array of values is multiplied with the \kpvsys map which produces the most significant weight. At this point, all maps have been combined to produce a signal that mimics the profile of the best observation but at a larger S/N. Fig. \ref{fig:ImprovementFromCombination} illustrates the improved signal of Fe\,I, moving from the best HARPS-N night to the combined observation.

\subsection{Aliasing in cross-correlation}
\label{Aliases}
A potential vulnerability with the cross-correlation technique is the presence of aliases, which occur when lines in a template correlate with the lines of other species. In a 1D cross-correlation function, this appears as an apparent signal located away from the systemic velocity, which is easy to spot by eye \citep{Hoeijmakers_2019_Fe_Detections, Jens_WASP_121b_Transit_Detections_of_Atoms_2020}. However, peaks that lie close to the systemic velocity are more difficult to untangle, given that the dynamic effects of an UHJ's atmosphere can also shift signals of the order of a few \kms \citep{Ehrenreich_2020, Prinoth_2022, Asnodkar_Super_Sonic_Winds_2022}.

A simple toy model of a spectrum illustrates how aliases may cause false detections. In the following example, presume that a transmission spectrum consists of three different species over 400–420\,nm. One species has many absorption lines (species A), for example, Fe\,I, which has approximately 500 lines in this region. The other two species (species B \& C) have 100 and 20 lines, respectively. A summation of a uniform distribution of Gaussian peaks, with random height, width, and centre positions creates the transmission spectrum:
\begin{equation}
    F = 1 - \sum_{n=0}^i \sum_{n=0}^l U_1(A_{\rm{min}},A_{\rm{max}})e^{\frac{-(\lambda-\lambda_{c}(U_2(\lambda_{\rm{min}},\lambda_{\rm{max}})))^2}{\sigma_c^2(U_3(\sigma_{\rm{min}},\sigma_{\rm{max}}))}}
\end{equation}

where $i$ is the total number of species, $l$ is the total number of lines that each species has, and $A$, $\lambda$ and $\sigma$ represent the wavelength amplitudes, centroid position and standard deviation, respectively.

\begin{figure*}
    \centering
    \includegraphics[scale=0.4]{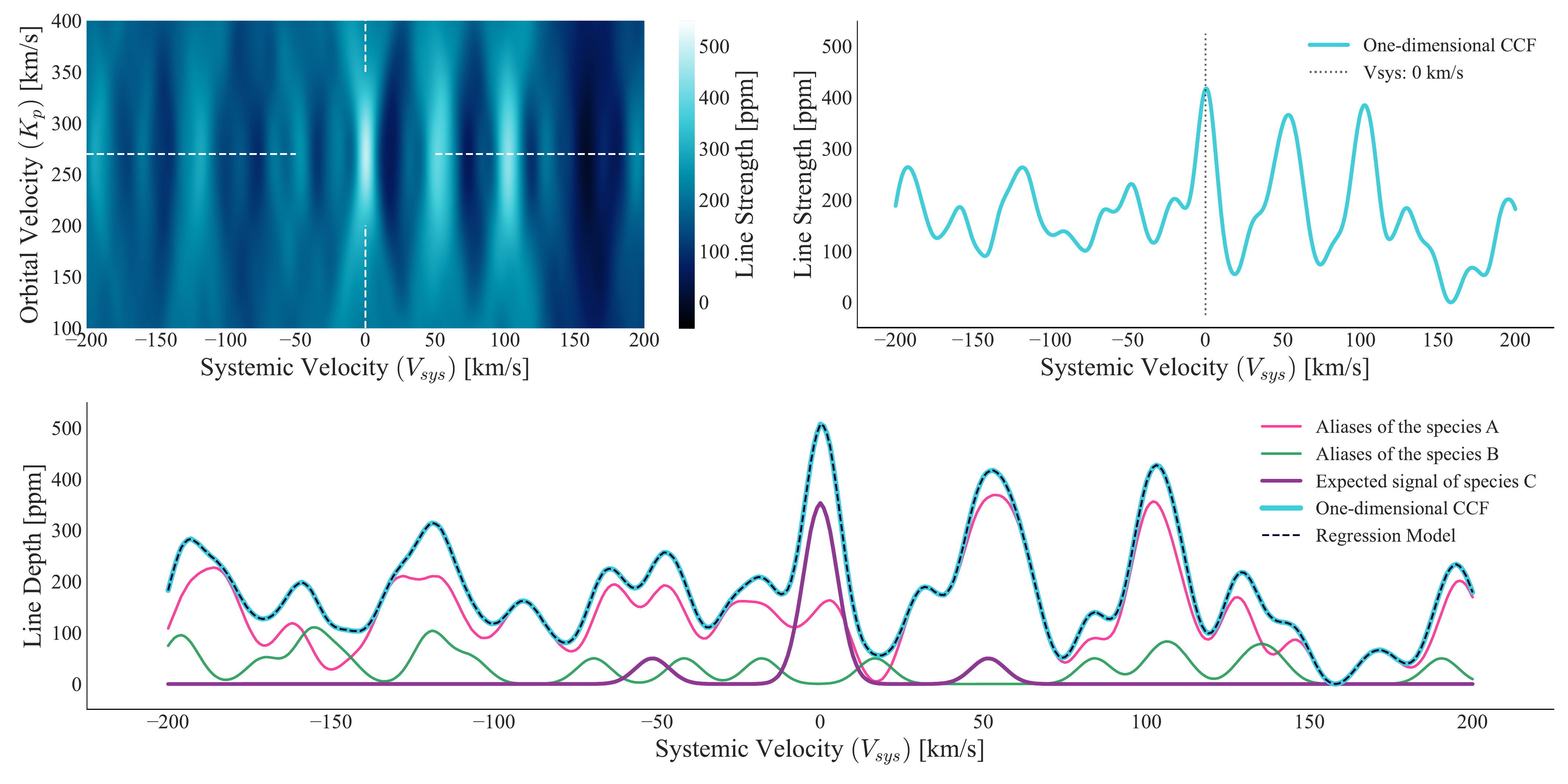}
    \caption{Cross-correlation results from correlating species C with Species A, Species B, and the whole toy spectrum. \textit{Top left panel}: The generated \kpvsys map using the toy spectrum. In the absence of noise, clear signals occur at the same orbital velocity but various systemic velocities, which can not be related to the search species signal. \textit{Top right panel}: The extracted 1D CCF, while a peak has formed at the expected location of the signal, the resulting cross-correlation function is highly structured with two large peaks forming close to 50 \kms and 100 \kms. \textit{Bottom panel}: The decomposition of the 1D cross-correlation into its signal and alias components, we see that species A is largely responsible for the signal structure away from the planet's signal, including the two offset peaks. The regression model specified in Eq. \ref{modelregresionequation} has successfully reconstructed the 1D cross-correlation function.}
    \label{fig:predicted_alises_toy}
\end{figure*}

The amplitude of each line varies from 0.001 to 0.005 of a transit depth, with a standard deviation of 0.003 to 0.005, with the exception of species C which has twenty lines all with the same line depth of 0.005. The produced data sets represent a simplified transmission spectrum with the planet centred at 0 \kms. The different species can be extracted with the cross-correlation technique. Fig. \ref{fig:predicted_alises_toy} displays the cross-correlation results of searching for species C in the transmission spectrum.

In the \kpvsys map, we see a central peak where we expect to see the planetary signal. However, two additional peaks lie at 50 \kms and 100 \kms. Converting to a 1D cross-correlation function, we see that the central peak barely sits above the other two. Peaks of the cross-correlation function no longer represent random noise but are artificial aliases caused by cross-correlating with the other species that reside within the spectrum. Strong aliases thus carry the potential to bury signals of fainter species. This situation is most potent in two cases: for species with few lines and species with weak lines. In the case of only a few lines, each line position is given more weight in the cross-correlation function, making the miss-correlation with other lines in the spectrum more significant. For faint species, the alias signals become more substantial since the more dominant species add a stronger contribution to the overall alias structure. For more dominant species with many lines, aliases are expected to be less pronounced. Since many lines contribute to the cross-correlation function, the lines do not correlate as well with other absorption species in the spectrum, leading to a noise that much better approximates the theoretical photon noise. While all species have alias profiles, the contribution is more negligible for species with strong lines.

Observations have routinely demonstrated that signals from different species detected through cross-correlation can have offsets in systematic velocity \citep{Asnodkar_2022_Single_Lines}. Peaks are known to deviate of the order of a few \kms, typically explained through atmospheric dynamics \citep{Ehrenreich_2020, Prinoth_2022}. This effect is slightly noticeable in Fig. \ref{fig:predicted_alises_toy}, as species A has a peak close to the planetary rest frame at 2.8 \kms, which has increased the signal-peak height and produced a positive offset of 0.3 \kms. This result implies that other species can potentially contribute to the cross-correlation signal and, if a signal is not present, appear to be one. This scenario outlines the need for an analysis method when applying cross-correlation, especially for line lists with relatively few lines. 

The root of the alias problem also betrays its solution. It is possible to predict their positions and relative strengths, and leverage those predictions to more accurately determine if an expected peak is, in fact, a signal. This can be done by cross-correlating the model spectrum of the species suspected of producing aliases with the search-species templates to produce models of alias profiles. Additionally, correlating the search species template with itself creates an expected auto-correlation function. The sum of the aliases and the expected signal comprises the entire cross-correlation function. Fig. \ref{fig:predicted_alises_toy}'s final panel contains the alias profiles of species A and B.

Assuming that the cross-correlation function is a linear sum of the aliases and the true signal, we propose the following numerical model:
\begin{equation}
    Y_i = \beta_0 + \beta_1\textbf{X}_{1i} + ... + \beta_{k}\textbf{X}_{ki} + \epsilon_i,\;\;\;\; i = 1,...,N 
\end{equation}

where $Y_i$ is the response variable and represents each data value $i$, $k$ is the total number of predictor variables, $\textbf{X}_{ki}$ are the predictor variables, $\beta_0$ is a numerical intercept of the model, and $\beta_k$ is a scaling factor, $N$ is the total number of data values allowed into the model, and $e_i$ is the error on each measurement. If $e_i$ measurements are independent and identically distributed, with a zero mean $E(e_i)=0$, and variance $VAR(e_i) = \sigma^2$, then regression can be used to estimate the $\beta_i$ parameters. The alias-regression model is estimated by multiple least squares \citep[MLS,][]{Regression}.

Applying multiple least squares regression to this data provides estimates for the coefficient values whose standard errors can be measured through jack-knife resampling \citep{Jack_Knife_Resampling}, and creates an empirical model to compare with the final results. The division of any coefficient value by its standard deviation provides a measure of the detection significance. Under idealised conditions, the error distribution follows a Student's t-distribution and can therefore be subjected to a Wald t-Test \citep{Regression}. The use of t-statistics is routine in the verification of cross-correlation signals \citep{Brogi_Welch_ttest,Nugroho_2017_DaysideTiO}. Additionally, a statistic which describes the overall constraining power of the model is the number of degrees of freedom: 
\begin{equation}
    df = n - p
    \label{degreesoffreedom}
\end{equation}

where $df$ is the degrees of freedom of the model, $n$ is the total number of data points, and $p$ is the total number of predictor variables. If $df\geq30$, a t-distribution can be accurately approximated to follow a Gaussian distribution $N(0,1)$ \citep{Regression}. As is currently conducted in the literature, claiming a detection requires the t-statistic to be $\geq5\sigma$, weaker signals lower than this value are usually classed as `tentative' detections.

It is essential to be aware that 1D cross-correlation functions can strongly correlate with neighbouring data points depending on one's choice of radial velocity step, and this can over-constrain any fit applied. Under the statistical assumptions applied in MLS, it is expected that the data points are independently distributed \citep{Regression}. Therefore, we mitigate this issue by only fitting every fourth sample of the 1D cross-correlation function \citep{Hoeijmakers_2019_Fe_Detections,Jens_WASP_121b_Transit_Detections_of_Atoms_2020}, corresponding to a step-size of 4 km/s.

\subsubsection{Applying alias-regression to the theoretical case}
Following the outlined methodology, the following MLS regression model is constructed for the hypothetical sample spectrum for species A, B, and C:
\begin{equation}
    CCF_{\rm{toy}} = \underset{(0.05)}{1.0} X_{\rm{A}} +  \underset{(0.16)}{1.0} X_{\rm{B}} + \underset{(0.10)}{1.0} X_{\rm{C}} - \underset{(0.0001)}{0.0001}.
    \label{modelregresionequation}
\end{equation}

In this ideal case, the estimated alias profiles match the output of the toy model exactly. Implying that the coefficient values have a value of unity and do not need to be scaled, note the values in the parenthesis denote the standard error measurements of the coefficients. Fig. \ref{fig:predicted_alises_toy} illustrates this model as a dashed line and demonstrates that the regression approach re-creates the cross-correlation function, and the significance of each coefficient in the model is summarised in Table \ref{tab:toy_detections}. Template C has a t-statistic of $t=9.8$, which passes the detection threshold ($\geq5\sigma$). Another result of this fit is the detection significance of the A and B line profiles, both of which are detected, with species A obtaining the highest significance value. This suggests that aliases peaks carry enough information to enable detections by themselves, and an example of such aliases has previously been identified in \cite{Hoeijmakers_2019_Fe_Detections}. 

\begin{table}[ht]
    \caption{Summarised t-scores for the constructed toy model.}
    \label{tab:toy_detections}
    \centering
    \begin{tabular}{c|c|c|c}
        Species & A & B & C \\\hline
        C$_{Y_i}$ & 16.1 & 5.6 & \textbf{9.8}
    \end{tabular}
    \caption*{\textbf{Notes.} The dependent variable of the model is placed in the first column. The detection significance of the search species is in bold.}
\end{table}

Since the model reproduces the aliases; noise, systematics, or omitted alias contributors are likely responsible for deviations. The toy model presumes no stellar or other systematic contributions exist in the transmission spectrum, and it is common for residual stellar signals to be present within cross-correlation maps \citep{Jens_WASP_121b_Transit_Detections_of_Atoms_2020}. Finally, alias-regression also operates under the assumption that all alias contributors are known. As it is challenging to know every species that may contribute to the alias structure ahead of time, deviations from the model should be expected. All detections found in this paper were included in the alias-regression model. Detections follow the same detection criteria ($\sigma$) as for the toy model.

\subsection{Validating detections}
\label{Validating Detctions}
Candidate signals must undergo scrutiny to test their robustness. A good primary step is to confirm that the signals are not produced by systematic artefacts and originate exclusively from the in-transit exposures, which is achieved through bootstrapping the in- and out-of-transit exposures \citep[see][and references therein]{Wyttenbach_2015}. Following this confirmation, the alias-regression models can be statistically examined through an Analysis of Variance (ANOVA) $F$-test and visually examined through response and residual plots. 

\subsubsection{Bootstrapping}
We tested the robustness of detections through a bootstrapping approach applied to the 2D cross-correlation functions as described in \cite{Jens_WASP_121b_Transit_Detections_of_Atoms_2020}. The method tested the distribution of candidate signals caused by correlated noise in the cross-correlation functions to quantify the probability that the signal was caused by systematic noise, by selecting in-transit values of the cross-correlation function away from the expected planet signal and shifting these values to a random radial velocity. It was possible to test for this. Using a random Gaussian probability distribution, we sampled the shifts with 10,\kms and 20,\kms fixed widths. Without systematics, the profiles were expected to be Gaussian-centred at an average line depth of zero, with the same widths specified by the fixed widths used to sample the random shifts. The signal was validated by plotting the measured line strength of the signal as a vertical line, and the line should be distinctly offset from the two Gaussians if a signal was present. Any systematics present could cause an offset in the Gaussian histograms leading to overlapping with the signal, which would suggest that the signal came from this source as opposed to the planet signal.

We conducted an analysis to assess whether the signal originated from the in-transit exposures. Beginning with the cleaned cross-correlation function, the in-transit and out-of-transit exposures were separated into two groups and averaged. The average of the out-of-transit group should have been zero, while the in-transit one should have experienced an offset due to absorption by the planet's atmosphere. Fifty percent of in- and out-of-transit exposures were randomly selected, averaged and divided by the master in and out-of-transit cross-correlation function. Repeating this process 20,000 times created three distributions of in-in, in-out, and out-out residuals. Plotting these as histograms enabled one to identify if the signal indeed did come from the in-transit exposures, as the in-out transit distribution should have been offset from the in-in and out-out distributions, which should have been centered around zero. An offset confirmed that the signal came from the in-transit exposures.

\subsubsection{Testing alias-regression through F-statistics, response, and residual plots.}
A MLS model comprised of the predicted signals and alias functions should describe the 1D cross-correlation function. How well this is achieved can be evaluated through a response plot, residual plot, and an ANOVA $F$-statistic. A response plot is a scatter plot of the predicted values of a model against the measured data values. In this plot, all values should be distributed around the identity line $y=x$, and measure how well the model represents the data, with the gradient representing the overall predictive power of the model \citep{Regression}. The residual plot demonstrates the difference between the two models against the $x_i$ data points. Ideally, these plots should show a random distribution of points along a $y_i$ value of zero \citep{Regression}. 

The ANOVA $F$-statistic examines if the predictors are needed in the MLS model, that is, whether or not the model $Y_i$ offers a better prediction than the sample mean $\Bar{Y}$. In other words, ANOVA tests the coefficients against the null hypothesis that $\beta_1 = ... = \beta_k = 0$ \citep{Regression}. The $F$-statistic is calculated as the ratio of the explained variance and the unexplained variance of the model
\begin{equation}
    \rm{F} = \frac{\rm{explained\ variance}}{\rm{unexplained\ variance}} = \frac{\displaystyle\sum_{i=1}^K \frac{n_i (\Bar{Y}_i - \Bar{Y})^2}{K-1}}{\displaystyle\sum_{i=1}^K\displaystyle\sum_{j=1}^{n_i} \frac{(Y_{ij}-\Bar{Y}_i)^2}{N-K}},
\end{equation}

where $\Bar{Y_i}$ denotes the sample mean of a specific predictor variable group, $Y_{ij}$ is the $j^{th}$ observation out of the predictor variable groups $K$, $n_i$ is the number of observations in the $i$-th group, $\Bar{Y}$ denotes the overall mean of the data, and $N$ is the overall sample size. The $F$-statistic follows an $F$-distribution, which does not approximate a normal distribution with higher degrees of freedom. The statistic will be significant if the explained variance is large compared to the unexplained variance, which is unlikely if the null hypothesis is true \citep{F-test}. 

A one-tailed hypothesis test with a 5$\sigma$ cut-off under a Gaussian assumption has a probability of $ P(z\geq5\sigma) \leq 5.724 \times 10^{-7}$. Software packages such as \texttt{scipy} \citep{2020SciPy-NMeth}, can calculate an equivalent probability value for the $F$-statistic. This value will be used as the cut-off to evaluate its significance.

In addition to the $F$-statistic, it is also possible to test how statistically significant some parameters of the model are in combination through the use of a reduced regression model \citep{Regression}. In this case, we want to evaluate if the alias coefficients, in combination with each other, constrain the cross-correlation function. Computing a second regression model omitting the search-species coefficient enables this. It is then possible to compute an $F$-statistic which evaluates the constraining power of the reduced model using the following formula

\begin{equation}
    F = \frac{(RSS_{Reduced}-RSS_{Full})/p}{R_{Full}/(n-k)},
\end{equation}

where $RSS$ refers to the residual sum of squares for the reduced and full model, p is the number of predictors removed, n is the total number of observations in the data set, and k is the number of coefficients, including the intercept in the full model. The F-statistic follows an $F(d_1,d_2)$ with the degrees of freedom defined as \citep{Regression}

\begin{equation}
    d_1 = df_{Full} - df_{Reduced};\;d_2 = df_{Full}.
\end{equation}

As the alias profiles lie away from the planetary signal, they are better constrained through cross-correlating a wide radial-velocity range. The range $-1,000$\,\kms  to 1,000\,\kms\, was enough to achieve this. 

\section{Results}
\label{results}
The four transit observations of KELT-9\,b were subject to a chemical survey using the procedures outlined in Sect. \ref{CCFApproach}. Detections fell into one of two groups: confirmed detections or new detections. Confirmed detections are species previously found by \cite{Hoeijmakers_2019_Fe_Detections} or \cite{Carmenes_Paper_Method,Carmenes_Paper_2_Method}. The purpose was to show that the method reproduces these results while enhancing the S/N. We explicitly go into more detail with the Mg I detection, and the tentative Ba\,II detection to give the reader a clearer view of how the model fits the data. All detections are subject to the statistical verification techniques outlined in Sect. \ref{Validating Detctions} with the figures located in Appendix \ref{A:ModelPlots}.

\subsection{Detection confirmations}
\begin{figure*}
    \centering
    \includegraphics[scale=0.29]{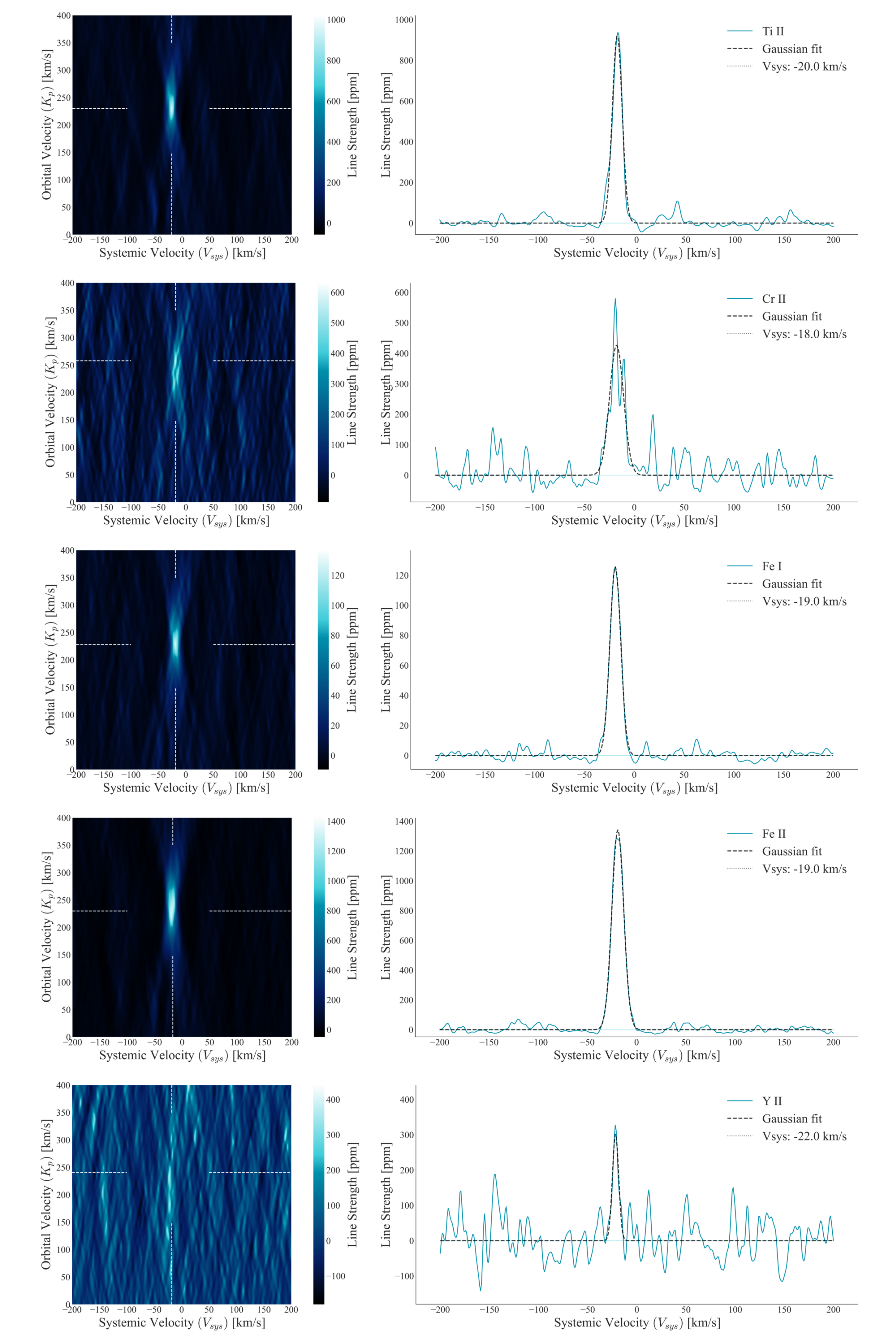}
    \caption{Confirmed species detections in KELT-9\,b. Each panel presents each species' two- and 1D cross-correlation signatures in the following order: H\,I, Na\,I, Mg\,I, Ca\,II, Sc\,II, Ti\,II, Cr\,II, Fe\,I, Fe\,II, and Y\,II. In the \kpvsys maps, the white dashed lines locate the maximum peak of the signal. One-dimensional cross-correlation functions are extracted from the \kpvsys map by selecting the row of pixels at the maximal location of the signal. The dashed dot represents a Gaussian fit to the cross-correlation function, and the vertical dotted lines mark the location of the peak for each example. The blue-shaded region represents the standard deviation of all points away from the peak of each function.}
    \label{fig:ConfrimedDetectionsPlot}
\end{figure*}

The original detections from \cite{Hoeijmakers_2019_Fe_Detections}, \cite{Carmenes_Paper_Method} and \cite{Carmenes_Paper_2_Method} were recovered, with the noted exception of Y\,II which is recovered at a significance of $2.7\sigma$. The \kpvsys maps and 1D cross-correlations for the species H\,I, Na\,I, Mg\,I, Ca\,II, Sc\,II, Ti\,II, Cr\,II, Fe\,I, Fe\,II, and Y\,II are displayed in Fig. \ref{fig:ConfrimedDetectionsPlot}. The centroid velocities for this subset are close to the accepted systemic velocity values quoted in the literature \citep{Asnodkar_2022_Single_Lines} with the notable exception of Ca\,II and H\,I, which have strongly blue-shifted velocities. These signals likely originate from a higher region of the planet's atmosphere. An additional point of interest for these plots is that there appears to be a secondary trough located around $-50$\,\kms; this is likely a residual of the Rossiter-McLaughlin effect \citep{Rossiter_RM,McLaughlin_RM} caused by incomplete removal of the stellar signal. 

Additionally, Sc\,II and Cr\,II detections are verified. While a signal appears present in the \kpvsys plots, their structures are irregular. However, the signals' locations are at the expected velocities. The peak of Cr\,II appears to have some substructure, with a secondary peak at a slightly higher velocity, suggesting aliasing.

\begin{figure*}
\begin{minipage}{1.0\textwidth}

    \centering
    \includegraphics[width=0.853\linewidth]{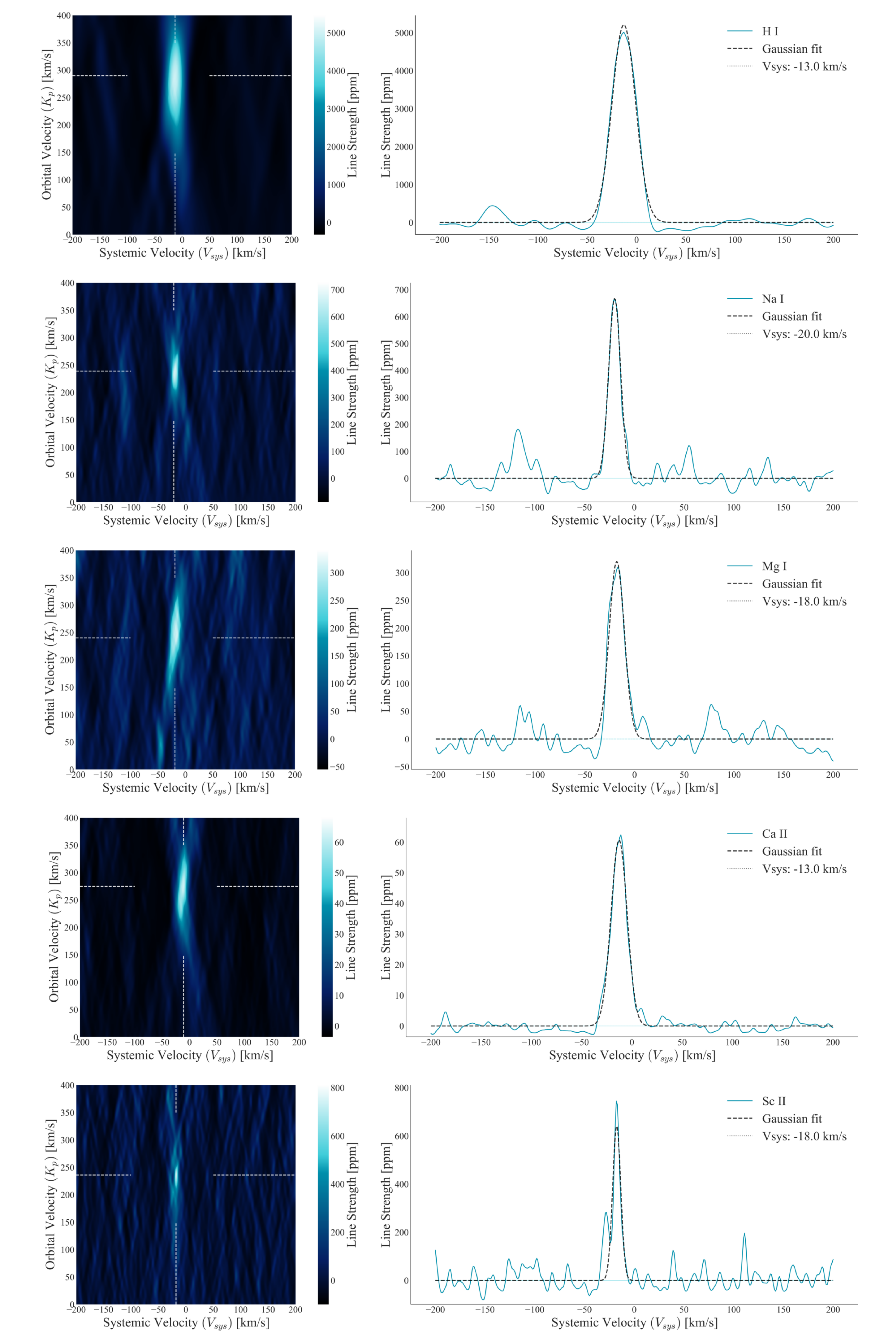}

\end{minipage}
\end{figure*}

\begin{figure*}
\begin{minipage}{1.0\textwidth}
    \centering
    \includegraphics[width=0.853\linewidth]{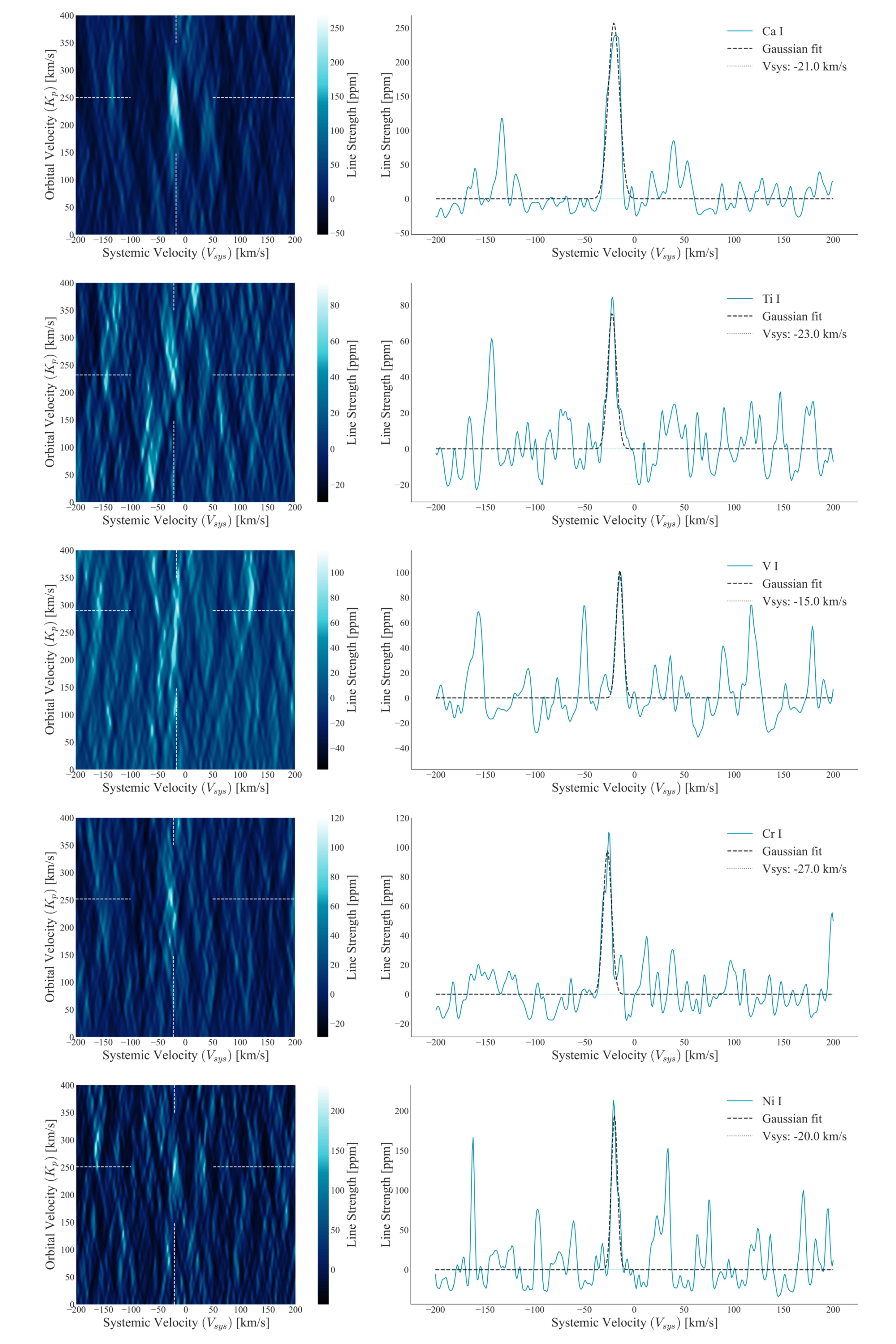}
\end{minipage}
\end{figure*}

\begin{figure*}
    \centering
    \includegraphics[scale=0.31]{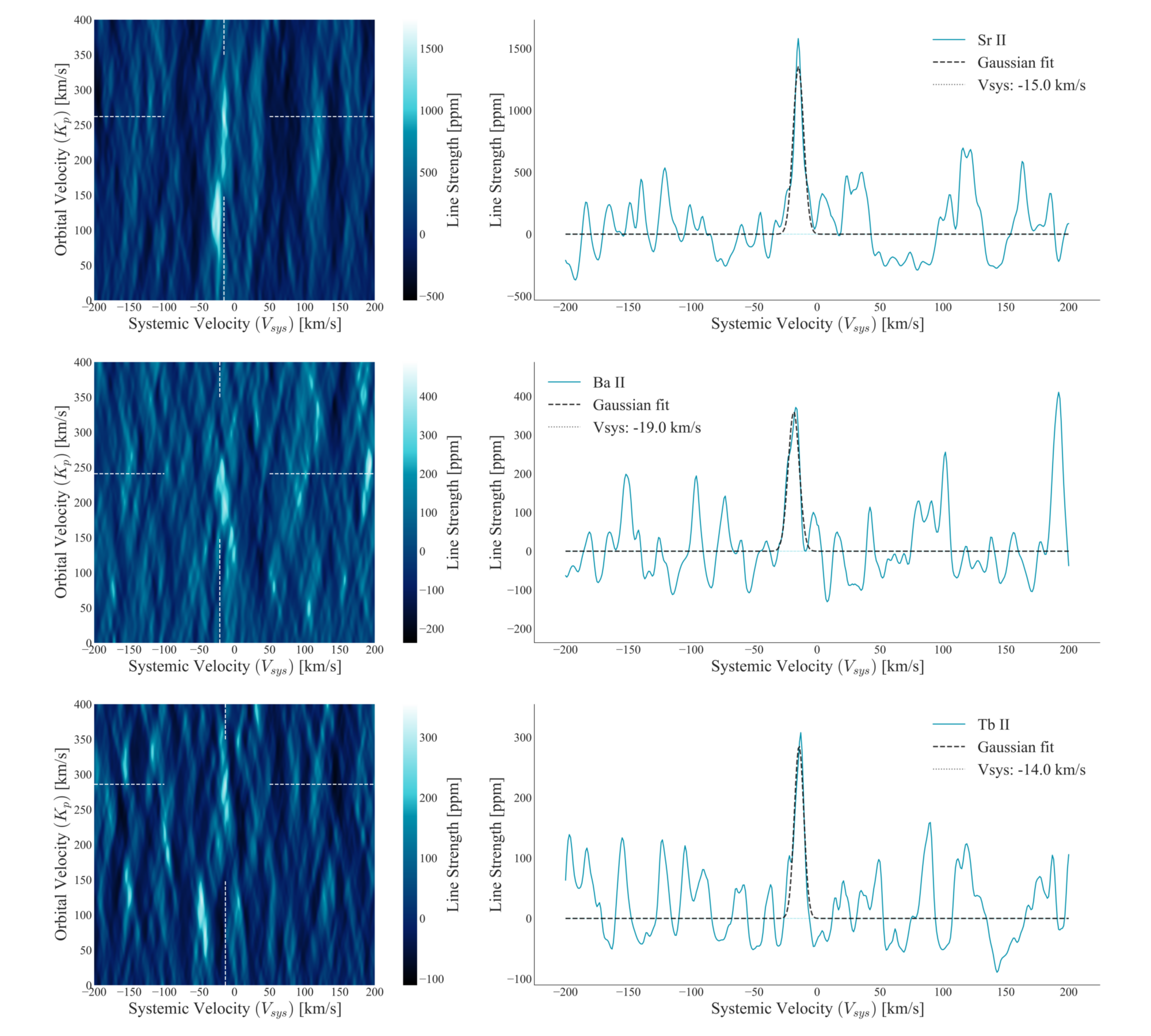}
    \caption{Newly detected species detections in KELT-9\,b. Each panel presents each species' 2D and 1D cross-correlation signatures in the following order: Ca\,I, V\,I, Cr\,I, Ni\,I, Sr\,II, Ba\,II, and Tb\,II. The plots are constructed in the same manner as Fig. \ref{fig:ConfrimedDetectionsPlot}}
    \label{fig:NewDetectionsFigure}
\end{figure*}

\subsection{New detections}
\begin{figure}
    \centering
    \includegraphics[scale=0.65]{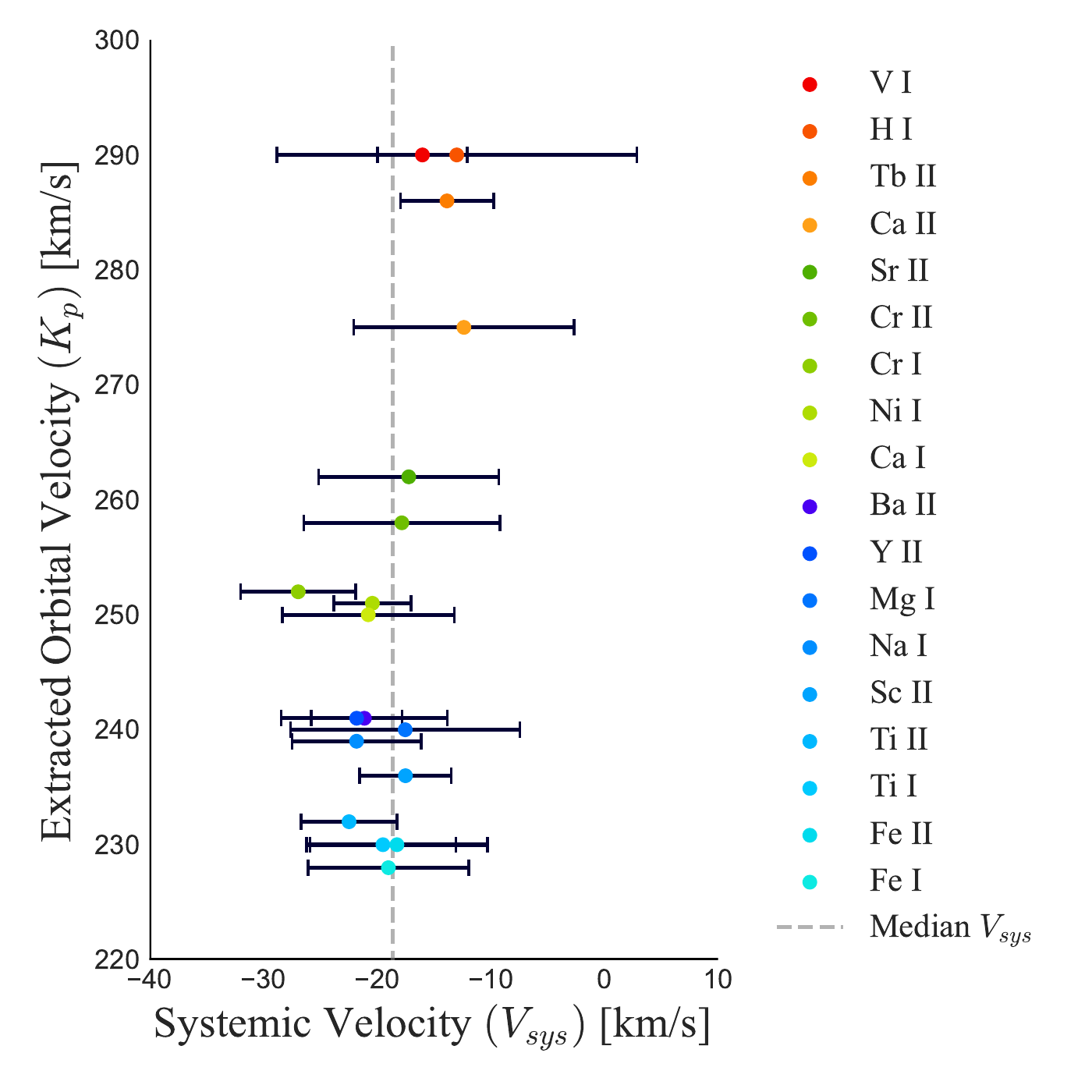}
    \caption{Fitted Gaussian parameters for the systemic velocity centres and widths, obtained from the extracted orbital velocity from the \kpvsys plots. The detections appear to cluster into three groups, suggesting they may be probing different atmospheric layers, which have been colour-coded for easier identification. The median systemic velocity is represented by a grey line on the plot.}
    \label{fig:Categorical}
\end{figure}

In addition to reconfirming the species already detected in the archived data, five new detections above the 5$\sigma$ detection threshold are provided. These are Ca\,I, Cr\,I, Ni\,I, Sr\,II, and Tb\,II, and three tentative detections (above 3$\sigma$) for Ti\,I, V\,I, Ba\,II. There is what appears to be a secondary signal in the Sr\,II detection located at a $K_p$ velocity of 130 \kms. This signal only occurs in the second observation using the HARPS-N spectrograph, while the expected signal appears in both. Suggesting that the feature is specific to that observation and not the planet in general. The detection results are displayed in Fig. \ref{fig:NewDetectionsFigure}. The line lists for these species are plotted in Appendix \ref{Weight Results}. All these species display significant aliasing in their \kpvsys and 1D cross-correlation function, although the aliasing is not severe for Ca\,I. All the signals are located close to the accepted planetary signals with overlapping peak widths. Ionised Tb is the first detection of these species in any exoplanet atmosphere. Other observations have reported detections of the remaining seven, but never in KELT-9\,b. Fig. \ref{fig:Categorical} provides an outline of the measured peak positions and widths for each signal, at each \textrm{$K_p$} value, all values fall well within the averaging window selected in \kpvsys space. There appears to be significant variability in atmospheric signals, with signals clustering into three groups. This may be due to the signals occurring at different atmospheric layers.

\subsection{Verification tests}
\begin{figure*}
    \centering
    \includegraphics[scale=0.45]{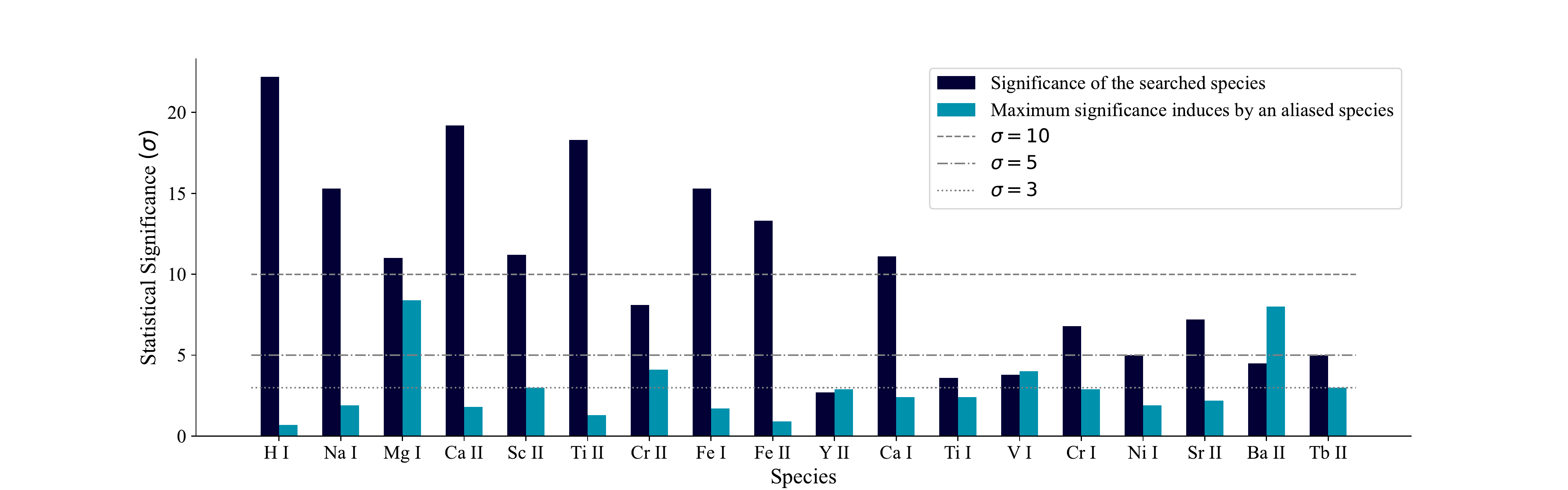}
    \caption{Summary of all detection statistics and the largest $\sigma$ values for the alias coefficients of each alias-regression model. The dashed lines mark the 3, 5, and 10 $\sigma$ detection significance threshold. The values are extracted from Table \ref{tab:Detections}.}
    \label{fig:barplot}
\end{figure*}

Fig. \ref{fig:barplot} provides a summary of the statistical significance of each search species, and the largest t-score of the alias coefficients for each model, while table \ref{tab:Detections} provides a summary of the statistical significance of all coefficient values for the alias-regression models constructed for each species. Most of these alias coefficient values lie below the significance threshold of 5$\sigma$, with the exception of the predicted aliases of Fe\,II with searching for Mg\,I and Ba\,II which are detected at significance's of 8.4$\sigma$ and 8.0$\sigma$, respectively. However, there are multiple occurrences of alias coefficients obtaining significance values above 3$\sigma$, which suggests these artefacts are contributing to the overall signal of the cross-correlation function. The alias modelling appears to have aided in constraining the significance of the detections of each species.

\subsubsection{The aliases of magnesium and ionised barium}
\begin{figure*}
    \centering
    \includegraphics[scale=0.46]{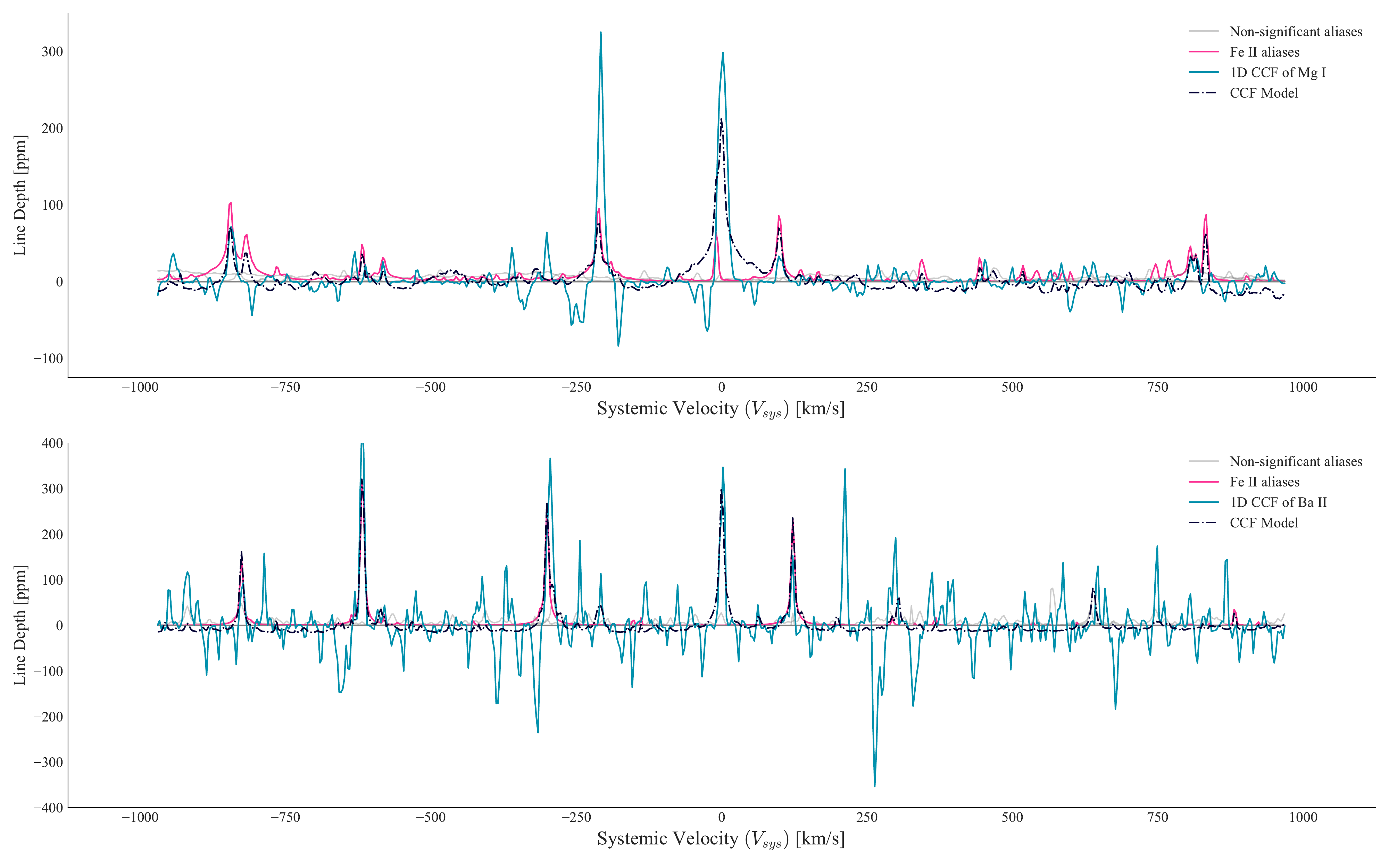}
    \caption{Expected aliases for the 1D cross-correlation function for Mg\,I, those with a $<3 \sigma$ coefficient significance are plotted in grey. The measured 1D cross-correlation function is shown in teal, and the model fit in dashed-black. The significant alias structure of Fe\,II is plotted in pink, and clearly contributes to the 1D cross-correlation function. The remaining predicted non-significant aliases is shown in grey.}
    \label{fig:MgIIAliasBreakDown}
\end{figure*}

Fig. \ref{fig:MgIIAliasBreakDown} illustrates each predicted alias-contribution for the alias-regression model for Mg\,I and Ba\,II, over-plotted with the 1D cross-correlation function and the model itself. The models capture the Mg\,I and Ba\,II signals and multiple Fe\,II aliases through the cross-correlation function. In the panels for Mg\,I and Ba\,II in Fig. \ref{fig:ConfirmedDetectionsModels} we show that the response plots have gradients of 0.406 and 0.278 respectively. This suggests that the model explains a significant proportion of the underlying cross-correlation function, and the residual plot shows that most features have been removed. The other coefficients of the functions do not reach a significance above 3$\sigma$; however, in combination, they constrain the data to a statistically significant degree (see Table \ref{tab:Peak_Information}). The results show that regressing alias profiles adds constraining power to cross-correlation detections.

\subsubsection{Bootstrapping to confirm the location of the signal}
Bootstrapping results for the eight new species are displayed in Fig. \ref{fig:BoostrapConfirmedDetections} and \ref{fig:BoostrapNewDetections}. All detections show that the calculated line strength of the signal is visually offset from the two sampled Gaussians. The offset suggests that any systematics sampled from this bootstrapping approach are not able to explain the signals seen in the \kpvsys maps of the newly detected signals. It is also the case for the confirmed detections seen in Fig. \ref{fig:ConfrimedDetectionsPlot} .

\subsubsection{Verifying the alias-regression models}
The ANOVA $F$-statistic of the alias-regression models are presented in Table \ref{tab:Peak_Information}. All models have $F$-statistics which generate p-values that are more significant than above a $5\sigma$ threshold of the normal distribution. The result suggests that all coefficient values in combination add constraining power to the 1D cross-correlation function. The results presented by the $F$-statistics agree with the results presented in Fig. \ref{fig:NewDectionsModels} and Fig. \ref{fig:ConfirmedDetectionsModels}, all models obtain positive gradients, with lines of worst fit regions away from zero. The strongest gradient of the species is H\,I, which almost has a one-to-one model-to-data relationship, while the weakest is Tb\,II which, while obtaining a positive gradient, leaves much of the residual data left unexplained. Additionally, the residual plots show that the models remove the signals partially (see Fig. \ref{fig:ConfirmedDetectionsModels}). Based on the results of this section and the bootstrapping, it can be concluded that the new detections are possible detections, but the models could be improved to better explain the signals. Table \ref{tab:Peak_Information} also contains reduced model $F$-statistics; in all cases, the alias coefficients in combination remain statistically significant above the equivalent $>5\sigma$ value. Moreover, the statistics are much more significant for the confirmed species than for the newly detected species, which suggests that the cross-correlation functions for these results contain additional variation that the model does not explain.

\begin{table}[ht]
    \caption{ANOVA $F$-statistics for the full alias-regression model and the reduced model.}
    \label{tab:Peak_Information}
    \centering
    \begin{tabular}{c|c|c}
        Species & ANOVA $F$-statistic & Reduced model $F$-statistic \\\hline \hline
        H\,I & 107.5 & 11310.7 \\
        Na\,I & 56.6 & 332.9 \\
        Mg\,I & 96.8 & 256.5 \\
        Ca\,II & 202.8 & 3220.8 \\
        Sc\,II & 234.4 & 324.3 \\
        Ti\,II & 263.2 & 1762.6 \\
        Cr\,II & 435.7 & 193.7 \\
        Fe\,I & 495.6 & 1250.6 \\
        Fe\,II & 355.3 & 1157.9 \\
        Y\,II & 316.5 & 43.7 \\\hline 
        Ca\,I & 235.2 & 224.0 \\
        Ti\,I & 345.7 & 43.2 \\
        V\,I & 342.0 & 42.5 \\
        Cr\,I & 312.6 & 91.6 \\
        Ni\,I & 417.5 & 68.2 \\
        Sr\,II & 156.2 & 73.7 \\
        Ba\,II & 110.2 & 37.2 \\
        Tb\,II & 265.4 & 46.8 \\

    \end{tabular}
    \caption*{\textbf{Notes.} Both the confirmed and newly detected species are included which and are separated with a horizontal line. All statistics obtain a P-value $>$ $5.724 \times 10^{-7}$, which is an equivalent value to a $>5\sigma$ cut-off and can therefore be considered significant. The significantly reduced model statistics imply that the alias-regression functions offer a significant amount of constraining power to the underlying cross-correlation function.}
\end{table}

\section{Discussion}
\label{Discussion}
The purpose of alias-regression is to explain the signals in cross-correlation maps. The methods and validation techniques in this study illuminate a path forwards that can potentially yield detections of such species. Based on the results of this study, new observations should be combined with archival data to increase the S/N. Atmospheric observations of exoplanets are photon starved, motivating the development of techniques that combine existing and new observations to increase the contrast of known signals. This method also has the potential to produce additional detections and reduce the photon noise in the data to the point where the underlying systematics are revealed. 

We conclude that aliases are present in the cross-correlation functions used to detect the ionised, atomic and molecular species in exoplanet atmospheres with high S/N. Their impact is that any fitting procedure that treats the region around the signal of the cross-correlation function solely as noise will bias the detection statistics. \kpvsys maps, if scaled to magnitudes of standard deviations, may in such cases produce detection significance values above the 3$\sigma$ level away from the planetary signal \citep{Giacobbe_2021,Guilluy_New_Molecules_Detections_Gaicobbe_2022}. Instead, it is recommended to present \kpvsys maps in terms of their actual numerical output and employ additional methods to infer a signal's validity. Alias-regression, model injection  \citep{Hoeijmakers_2018_Fe_Detection,Hoeijmakers_2019_Fe_Detections,Jens_WASP_121b_Transit_Detections_of_Atoms_2020}, and bootstrapping approaches \citep{Giacobbe_2021,Prinoth_2022} are ways to achieve this. A method that acknowledges and accounts for the systematic effects will likely hold more constraining power.

A profound implication of aliases is that they cannot be mitigated by using larger telescopes or repeated observations. However, improved S/N observations can provide additional insight. With additional observations of the same target, aliases will begin to appear and dominate the noise in the cross-correlation function. Therefore, this calls for methods such as alias-regression to be applied, in order to discover new species that may be present in the atmospheres of these exoplanets \citep{Kesseli_2022}. 

We have successfully combined four observations of KELT-9 b and used alias-regression to confirm the previous detections of  \cite{Hoeijmakers_2018_Fe_Detection,Hoeijmakers_2019_Fe_Detections,Carmenes_Paper_Method,Carmenes_Paper_2_Method}, with the exception of Y\,II, for which we obtained a significance value of 2.7. The discrepancy in the detection significance of Y\,II is likely due to the fact that \cite{Hoeijmakers_2019_Fe_Detections} constrained their detection significance by a least-squares Gaussian fit without considering the alias component of the cross-correlation function. Thus, they record a detection significance of 6.3 $\sigma$. In a least-squares approach, the omission of an explanatory variable leads to bias in the remaining coefficients. In this specific case, an alias peak for Mg\,I was found to be in the same position as the signal.  Mg\,I has an alias coefficient value of 2.3 indicating that this species is the likely source of contamination. In this situation, the variance of the estimated coefficients decreases, which can lead to inflated significance values. The reduction in the significance of Y\,II suggests that the signal was likely constraining some alias contribution.

Furthermore, we have obtained eight discoveries of species unobserved in KELT-9\,b's atmosphere, with varying degrees of confidence. Notable detections include Tb\,II which has not yet been found on any exoplanet, and the tentative detection of Ba\,II at a significance level of $4.5 \sigma$. Ba\,II has only recently been discovered in the ultra-hot gas giants of WASP-121 b and WASP-76 b \citep{Barium_Detection}. Tb was not predicted to be present in the atmospheres of these planets, nor has detection been proposed. The detection significance of Tb\,II is 5.0$\sigma$, and the extracted orbital velocity is consistent with the positions of the H\,I, Ca\,II, and V\,I detections. Taken together, the significance of the detection and the consistency of the orbital velocities suggest that this detection is sound. However, follow-up observations should verify the repeatability of this detection. Furthermore, Ca\,I and Cr\,I are also detected, increasing the known number of neutral-ion species pairs from one to three. Additional neutral-ion pairs will allow us to potentially unlock the ability to empirically constrain the ionisation rate of the same species, which can be used to determine the temperature profile and diagnose non-LTE effects. Moreover, apart from being detected by its own template, Fe\,II has also been observed via the aliases of Mg\,I and Ba\,II with an 8$\sigma$ level of confidence, this was observed but not statistically verified in \cite{Hoeijmakers_2019_Fe_Detections}. It is important to note that while this is the first confirmed detection of a species via the alias structures in the cross-correlation function, the use of alias-regression does not replace direct signal detections. The implication of the detection of Fe\,II via aliases is that the alias structures are present in cross-correlation functions and must be accounted for when detecting fainter species.

The \kpvsys maps in this study vary considerably from clear bright features such as the Fe\,II and Ca\,II detections to signals that are less clear and contain many substructures such as the Ni\,I and Tb\,II detections. However, there seems to be a discrepancy between what the detection appears as and what the statistics claim. The discrepancy likely lies in the fact that statistical tests perform hypothesis testing under the assumption that there is no underlying systematics at play. This assumption is tenuous with \kpvsys maps, since aliases have been shown to exist. These systematics are likely to compromise the statistical rigour associated with the 3 and 5 $\sigma$ rules-of-thumb used for detection, and do not represent the confidence generally associated with them. However, there is still a clear trend between detection significance and signal clarity, suggesting that we should be more confident with higher statistics. Due to the arbitrary nature of specifying a statistical threshold in the first place, it would be better to use them as a guide rather than the rule when claiming detections, favouring a conservative approach where possible.

The detection of heavier isotopes is important because they can provide a constraint on the age of a planet. This has been shown to be the case for Jupiter in our solar system, which is essential for understanding the present-day architecture of the Solar System \citep{kruijer2017_jupiter_age}. Age determination can constrain the formation and migration scenarios of these planets. For KELT-9\,b, this could lead to a better understanding of the total mass lost due to the atmospheric accretion exhibited by its host star \citep{KELT-9b_Mass_Loss}, and a better age constraint for the star itself. Detections are the first footfall in determining the abundances of these heavy isotopes. With better constraints on the systematics and additional follow-up observations, it may be possible to determine abundances and, by proxy, the age of KELT-9\,b. 

alias-regression works in both theoretical and observational cases, albeit with some limitations. In this study, the models showed some success in constraining the overall detected signals, but do not account for all of the peaks seen in the 1D cross-correlation function. However, as shown in Fig. \ref{fig:MgIIAliasBreakDown}, alias-regression is successful in predicting some alias peaks. Deviations from the model could be caused by a variety of sources, including unaccounted systematics, and not yet included alias species. Additionally, templates are functionally dependent on many different input parameters, with line strength depending on global parameters such as temperature, abundances, and planetary bulk parameters. Therefore, alias profiles relate how well the templates match the transmission spectrum of the planet in question, similar to the atmospheric retrievals of \citet{Brogi_Line_2019, Gibson_2022}.

It is also prudent to consider that the photon noise may still contribute a proportionate amount to the cross-correlation function, adding to the uncertainty when fitting the alias-regression coefficients. For example, species with few lines absorb less flux and therefore sample more noise in their cross-correlation, reducing statistical robustness. However, each model applied successfully constrains the 1D signals. The results obtained with the reduced model support this case. Observing Table \ref{tab:Detections}, we see that the statistical significance of the individual coefficient values of the regression is often small, but in combination, they yield a high significance. Therefore, modelling aliases will constrain the CCF more than if they were omitted when determining the significance of a detection. 

We conclude that combining the four nights of observations only begins to resolve the alias floor, making it difficult to fit the alias-regression model accurately. Repeated observations with larger telescopes and additional consideration of the shape of the alias profiles should improve these fits.

\section{Conclusion}
\label{Conclusion}
This work has provided a new approach to the cross-correlation technique, which has led to the detection of eight new chemical species in the atmosphere of KELT-9\,b. The method combines the information from multiple spectrographs with different wavelength ranges to produce strong signals in high contrast to the background noise. In addition, by attenuating the noise, alias profiles are resolved, revealing a systematic noise floor in the cross-correlation function. alias-regression accounts for these aliases and provides a statistical measure of the detection significance, and has led to novel detections of eight undiscovered species in KELT-9\,b's atmosphere, as well as reconfirmed detections from previous studies.

Despite its strengths, the approach has some limitations. While the alias-regression models have been shown to have the constraining power of the 1D cross-correlation signals, they leave much of the cross-correlation structure unexplained for species with fainter signals. This may be due to the presence of alternative alias-producing species that are not accounted for, or to other sources of systematic error that have yet to be discovered. As with any statistical inference (model fitting) technique, the effectiveness of the alias-regression technique is limited by the selection of all the relevant alias contributors. In reality, relevant species needed to fully constrain the cross-correlation function may have been excluded.

\begin{acknowledgements}
      N.W.B \& B.P. acknowledge partial financial support from The Fund of the Walter Gyllenberg Foundation.
      B.T.\ acknowledges the financial support from the Japan Society for the Promotion of Science as a JSPS International Research Fellow. 
      R.F acknowledge supported by the Göran Gustafsson Foundation for Research in Natural Sciences and Medicine and from the Royal Physiographic Society in Lund through The Märta and Eric Holmberg Endowment.
      D.K. acknowledges financial support from the Center for Space and Habitability (CSH) of the University of Bern. This work has made use of the VALD database, operated at Uppsala University, the Institute of Astronomy RAS in Moscow, and the University of Vienna.
      Additionally, this article is based on observations made in the Observatorios de Canarias del IAC with the Telescopio Nazionale Galileo operated on the island of La Palma by the Fundaci\'{o}n Galileo Galilei - INAF in the Observatorio del Roque de los Muchachos, and observations collected at the German-Spanish Astronomical Center, Calar Alto, jointly operated by the Max-Planck-Institut für Astronomie Heidelberg and the Instituto de Astrofísica de Andalucía (CSIC). We would also like to thank Y. Bokhari-Friberg for their help in editing and typesetting the manuscript, and the anonymous referee for their helpful contributions.
\end{acknowledgements}

\bibliographystyle{aa} 
\bibliography{bib}

\appendix

\section{Weight results}
\label{Weight Results}
\begin{figure*}[ht]
    \centering
        \begin{minipage}[b]{1\textwidth}
          \includegraphics[scale=0.36]{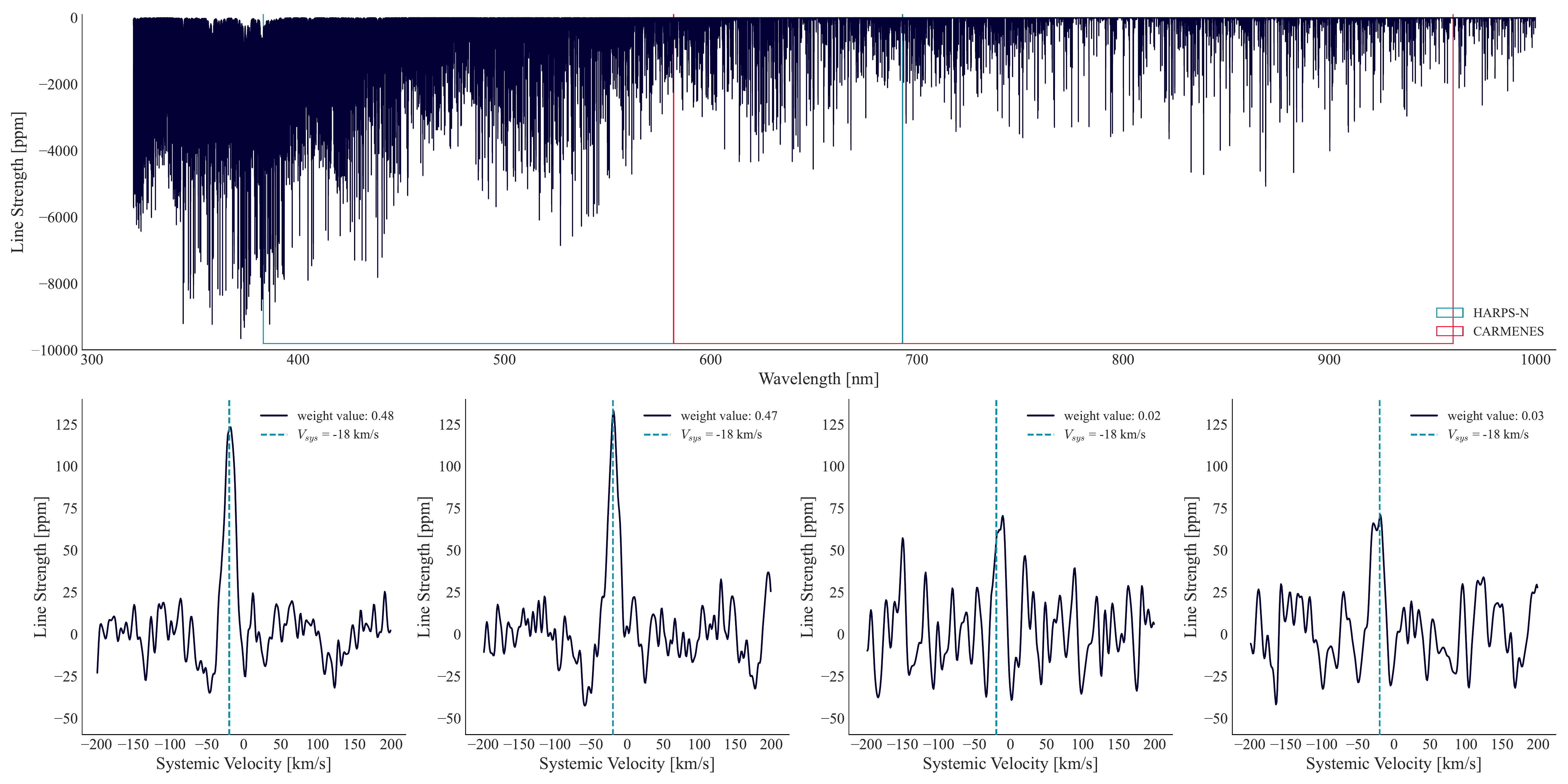}  
        \end{minipage}\qquad
        \newline
        \begin{minipage}[b]{1\textwidth}
          \includegraphics[scale=0.36]{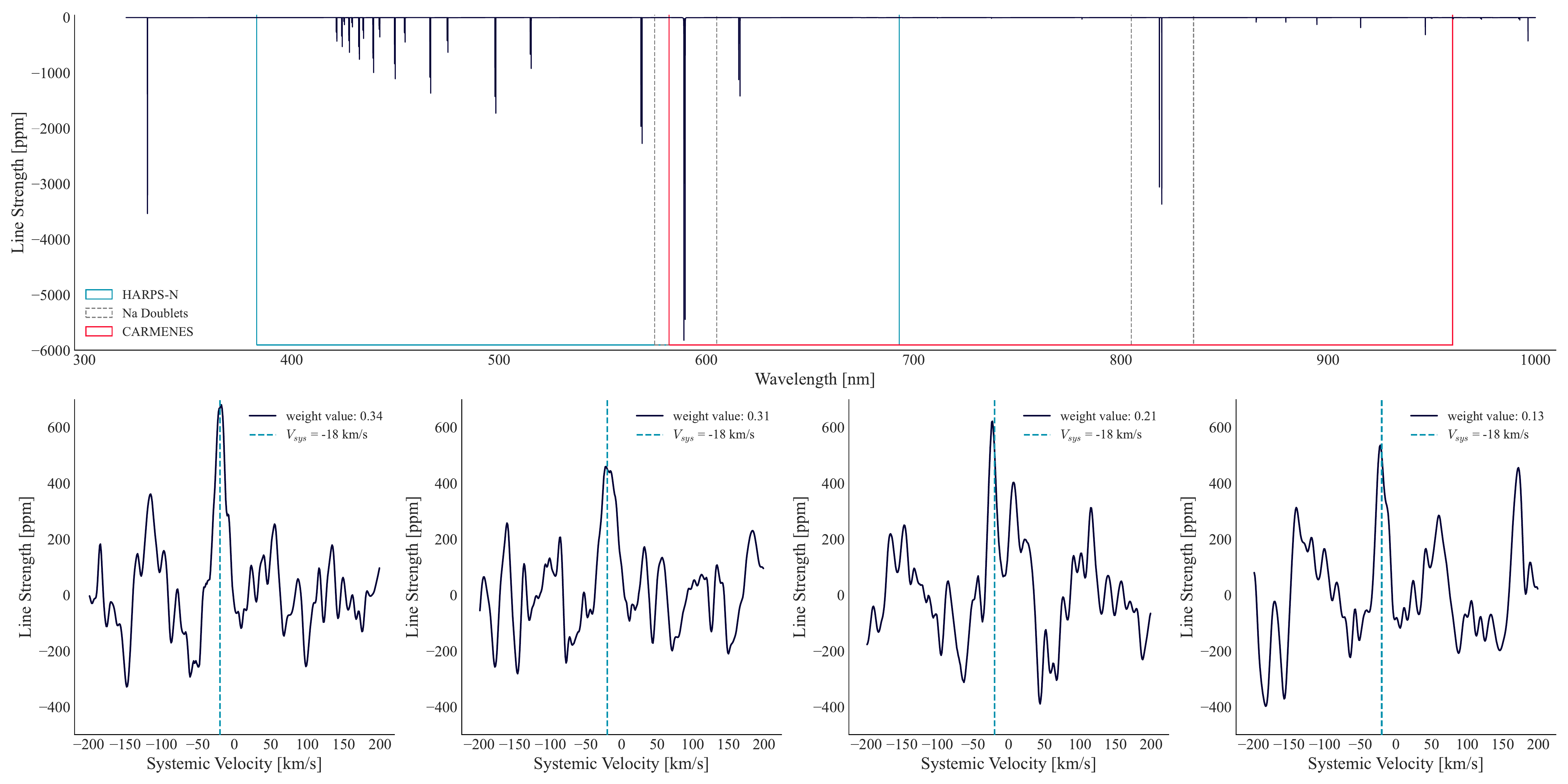} 
        \end{minipage}
    \caption{Summary of the weight extraction process. The first two rows illustrate the weight construction for Fe\,I. \textit{Top panel}: Line list of neutral iron used for cross-correlation. The wavelength ranges of the HARPS-N and CARMENES visible are plotted in boxes to illustrate the difference in wavelength coverage. \textit{Bottom panel}: 1D cross-correlation functions for the four nights of observations. The two HARPS-N nights have much higher S/N and contain less background noise, producing larger weights than the CARMENES observations. The second two rows demonstrate the weight constructions for Na\,I. \textit{Top panel}: Line list for neutral sodium. The proportion of flux obtained by both spectrographs is comparable to each other, therefore data quality dominates the weight calculation when combined. \textit{Bottom panel}: Cross-correlation functions for each night, CARMENES nights obtain a higher proportion in the calculation of weights.}
    \label{fig:Weight_Process}
\end{figure*}

\begin{figure*}
    \centering
    \includegraphics[scale=0.43]{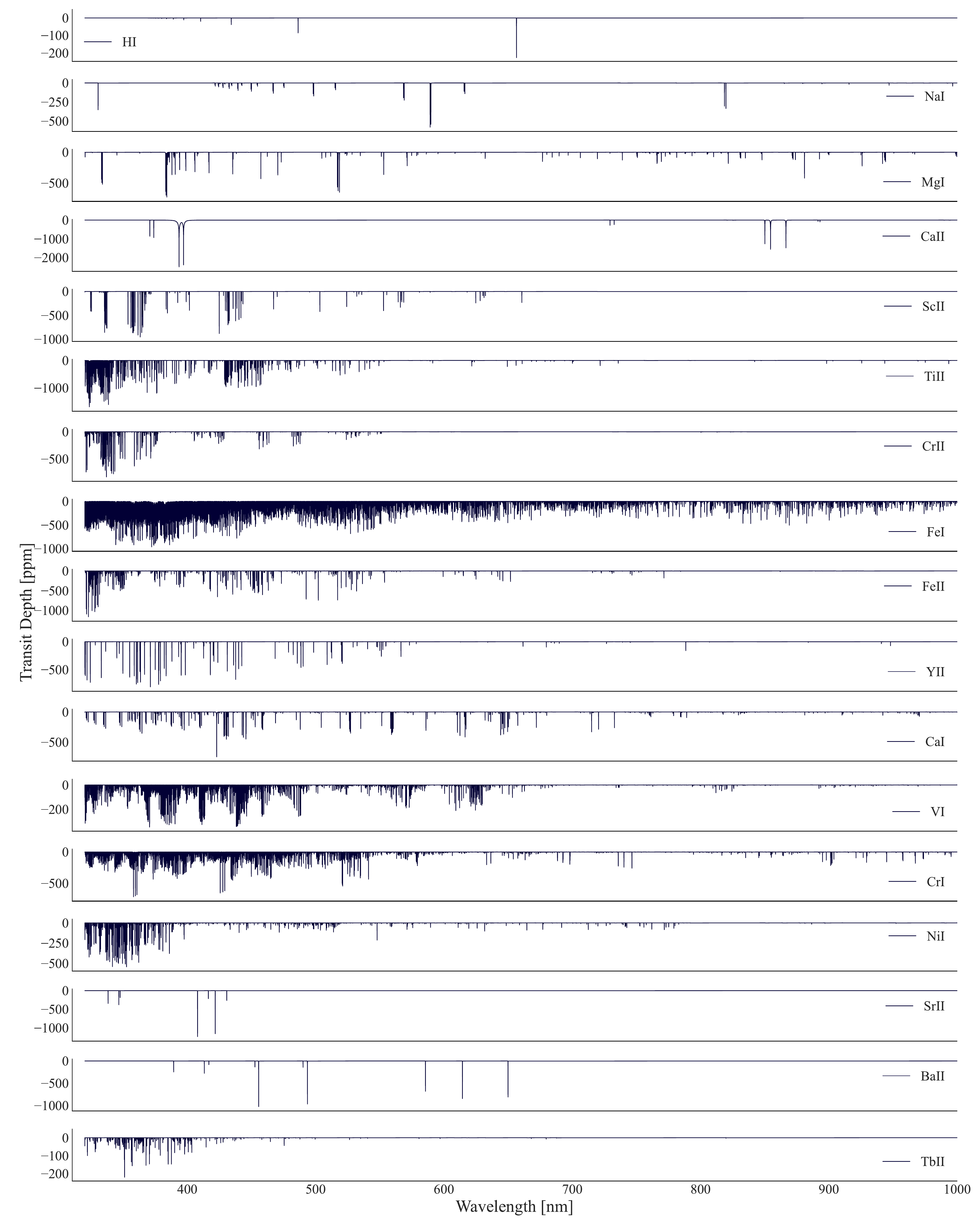}
    \caption{Generated line lists used in the cross-correlation of this report. The $y$ axis denotes transit depth, with zero equal to the stellar out-of-transit flux. The species order ascends first with atomic number, then atoms before ionised species. The included wavelength region span the HARPS-N and CARMENES visible arms.}
    \label{fig:LineListPlots}
\end{figure*}

An overview of the weight calculation process is presented in Fig. \ref{fig:Weight_Process} for two species of Fe\,I (\textit{top panel}) and Na\,I (\textit{bottom panel}). Atomic iron contains many lines, with a larger flux absorption in the blue wavelengths. From Table \ref{tab:ObservationalSummary}, the HARPS-N nights have larger S/N values, and this explains why the cross-correlation peaks are much more significant in the two HARPS-N nights, and has led to the weights favour these nights. For atomic sodium, however, the CARMENES spectrograph captures a deep triplet in the sodium spectrum at a larger wavelength range. In this case, while the CARMENES data is still poor compared to HARPS-N, it contributes a proportional amount to the signal. Table \ref{tab:WeghtExtractionSumaryFe} and Table \ref{tab:WeghtExtractionSumaryNa} decomposes the weights of each night into the pseudo-S/N values and the relative flux summation to better clarify how the weights are constructed. Fig. \ref{fig:LineListPlots} contains the line lists for all the species detected in this report to give the reader a better idea of which spectrograph performs better with each species.

\begin{table*}[ht]
    \caption{Computed values deriving the weights for combining the flux of each spectral observation for the species Fe\,I}
    \label{tab:WeghtExtractionSumaryFe}
    \centering
    \begin{tabular}{c|c|c|c}
        Instrument & Pseudo-S/N & Relative Flux Summation & Normalised Weight Value\\\hline
        HARPS-N (Night 1) & 2.42 & 220 & 0.478\\
        HARPS-N (Night 2) & 2.41 & 220 & 0.467\\
        CARMENES (Night 1) & 0.95 & 21.9 & 0.019\\
        CARMENES (Night 2) & 1.31 & 21.9 & 0.026\\
    \end{tabular}
\end{table*}

\begin{table*}[ht]
    \caption{Computed values deriving the weights for combining the flux of each spectral observation for the species Na\,I}
    \label{tab:WeghtExtractionSumaryNa} 
    \centering
    \begin{tabular}{c|c|c|c}
        Instrument & Pseudo-S/N & Relative Flux Summation & Normalised Weight Value\\\hline
        HARPS-N (Night 1) & 1.57 & 1.65 & 0.342\\
        HARPS-N (Night 2) & 1.42 & 1.65 & 0.310\\
        CARMENES (Night 1) & 1.36 & 1.19 & 0.213\\
        CARMENES (Night 2) & 0.86 & 1.19 & 0.134\\
    \end{tabular}
\end{table*}

\section{Detection attributes}
\label{DetectionStatistics}
Table \ref{tab:Detections} presents a summary of all t-scores of the statistically significant searches and alias-regression coefficients. The majority of the alias coefficients lie below a significance of $3 \sigma$; however, a few obtain values above this threshold. The species which include significant detections above this threshold are Mg\,I, Sc\,II, Cr\,II, V\,I, and Ba\,II.

\begin{table*}
    \centering
    \caption{Individual t-statistics for the coefficient values for each multiple-least squares alias-regression function.}
    \label{tab:Detections}
    \scalebox{0.9}{
        \begin{tabular}{c|cccccccccccccccccc}
        Search Species & HI & NaI & MgI & CaII & ScII & TiII & CrII & FeI & FeII & YII & CaI & TiI & VI & CrI & NiI & SrII & BaII & TbII \\\hline \hline
        H\,I$_{Y_i}$ & \textbf{22.2} & 0.0 & 0.0 & 0.7 & 0.0 & 0.0 & 0.0 & 0.0 & 0.0 & 0.3 & 0.2 & 0.0 & 0.0 & 0.0 & 0.0 & 0.0 & 0.0 & 0.2 \\
        Na\,I$_{Y_i}$ & 0.0 & \textbf{15.3} & 0.6 & 0.0 & 0.0 & 0.0 & 0.5 & 0.0 & 0.0 & 0.0 & 1.0 & 0.0 & 1.9 & 0.0 & 1.8 & 0.0 & 0.0 & 0.5 \\
        Mg\,I$_{Y_i}$ & 0.0 & 0.0 & \textbf{11.0} & 0.9 & 0.0 & 1.2 & 0.0 & 1.4 & \underline{8.4} & 0.0 & 0.9 & 0.0 & 0.0 & 0.0 & 0.0 & 0.0 & 1.0 & 2.7 \\
        Ca\,II$_{Y_i}$ & 0.0 & 1.5 & 0.0 & \textbf{19.2} & 0.7 & 0.3 & 0.0 & 1.2 & 0.1 & 0.0 & 0.0 & 0.7 & 1.8 & 0.0 & 0.9 & 1.2 & 0.4 & 1.1 \\
        Sc\,II$_{Y_i}$ & 1.6 & 0.0 & 0.0 & 0.0 & \textbf{11.2} & 0.0 & 0.7 & 0.0 & \underline{3.0} & 0.0 & 1.1 & 0.5 & 0.6 & 1.8 & 1.3 & 0.0 & 0.0 & 0.5 \\
        Ti\,II$_{Y_i}$ & 0.8 & 0.0 & 0.0 & 0.0 & 0.0 & \textbf{18.3} & 0.6 & 0.0 & 0.6 & 0.0 & 1.3 & 0.3 & 0.0 & 0.4 & 0.0 & 0.0 & 0.4 & 0.0 \\
        Cr\,II$_{Y_i}$ & \underline{4.1} & 0.0 & 1.4 & 0.2 & 0.0 & 0.0 & \textbf{8.1} & 0.0 & 0.5 & 0.0 & 0.0 & 0.0 & 0.0 & 0.0 & 0.0 & 0.6 & \underline{3.3} & 1.8 \\
        Fe\,I$_{Y_i}$ & 0.0 & 0.1 & 1.1 & 0.0 & 0.0 & 0.4 & 0.0 & \textbf{15.3} & 0.5 & 0.5 & 1.0 & 1.6 & 0.0 & 1.7 & 0.2 & 0.0 & 0.5 & 0.5 \\
        Fe\,II$_{Y_i}$ & 0.5 & 0.5 & 0.1 & 0.9 & 0.0 & 0.0 & 0.0 & 0.7 & \textbf{13.3} & 0.4 & 0.0 & 0.0 & 0.0 & 0.0 & 0.0 & 0.0 & 0.0 & 0.0 \\
        Y\,II$_{Y_i}$ & 2.9 & 0.0 & 2.3 & 0.2 & 0.0 & 0.3 & 0.0 & 1.6 & 1.4 & \textbf{2.7} & 0.0 & 0.3 & 0.4 & 0.0 & 0.0 & 0.0 & 1.2 & 1.0 \\\hline
        Ca\,I$_{Y_i}$ & 2.1 & 0.0 & 0.1 & 1.2 & 0.0 & 1.7 & 0.1 & 0.3 & 1.7 & 0.9 & \textbf{11.1} & 0.0 & 1.1 & 0.1 & 0.1 & 2.4 & 0.0 & 0.0 \\
        Ti\,I$_{Y_i}$ & 0.7 & 1.5 & 1.8 & 1.0 & 0.3 & 0.0 & 0.6 & 0.0 & 1.3 & 1.1 & 0.0 & \textbf{3.6} & 0.0 & 2.4 & 0.0 & 0.0 & 0.0 & 0.0 \\
        V\,I$_{Y_i}$ & \underline{4.0} & 0.0 & 0.3 & 1.4 & 1.9 & 0.0 & 0.0 & 0.0 & 0.0 & 0.4 & 1.1 & 1.4 & \textbf{3.8} & 0.0 & 0.0 & 2.1 & 1.2 & 0.0 \\
        Cr\,I$_{Y_i}$ & 0.0 & 1.1 & 0.9 & 0.5 & 0.7 & 0.0 & 2.9 & 0.0 & 1.5 & 0.7 & 0.7 & 0.0 & 1.1 & \textbf{6.8} & 2.3 & 0.0 & 0.0 & 0.0 \\
        Ni\,I$_{Y_i}$ & 0.0 & 0.0 & 0.0 & 0.5 & 1.4 & 0.0 & 0.0 & 0.0 & 0.0 & 1.5 & 1.8 & 0.0 & 0.0 & 0.6 & \textbf{5.0} & 1.9 & 0.4 & 0.0 \\
        Sr\,II$_{Y_i}$ & 0.0 & 0.3 & 0.5 & 0.0 & 0.3 & 0.0 & 0.0 & 1.7 & 0.3 & 2.2 & 2.0 & 0.0 & 0.0 & 0.0 & 0.6 & \textbf{7.2} & 0.0 & 0.8 \\
        Ba\,II$_{Y_i}$ & 1.5 & 0.0 & 0.0 & 0.5 & 0.1 & 2.6 & 1.4 & 1.6 & \underline{8.0} & 0.5 & 0.0 & 0.0 & 0.0 & 0.0 & 0.8 & 0.0 & \textbf{4.5} & 0.0 \\
        Tb\,II$_{Y_i}$ & 0.0 & 0.0 & 0.0 & 1.8 & 2.5 & 3.0 & 0.7 & 0.8 & 0.0 & 0.4 & 0.0 & 1.2 & 1.1 & 0.0 & 1.1 & 0.0 & 0.0 & \textbf{5.0} \\
        \end{tabular}}
        \caption*{\textbf{Notes.} The reconfirmed species and the new detections are separated with a horizontal line. The first column specifies which search species is the dependent variable (${Y_i}$) of the alias-regression model. The values highlighted in bold indicate the search species' statistical significance based on the coefficient of the regression model for the search species. This is what is used to determine a detection. The values underlined are \underline{significant alias values.}}
\end{table*}

\section{Model verification for known species}
\label{A:ModelPlots}
\begin{figure*}
\centering
\begin{minipage}{1.0\textwidth}
  \centering
  \includegraphics[width=1.0\linewidth]{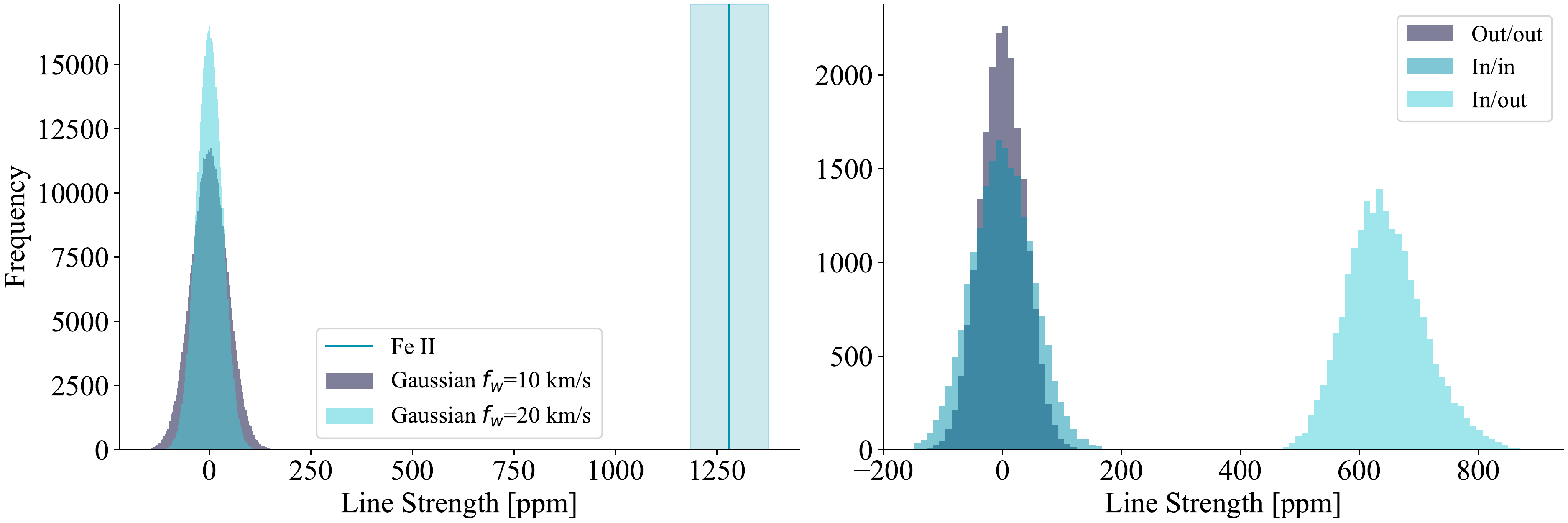}
\end{minipage}%

\begin{minipage}{1.0\textwidth}
  \centering
  \includegraphics[width=1.0\linewidth]{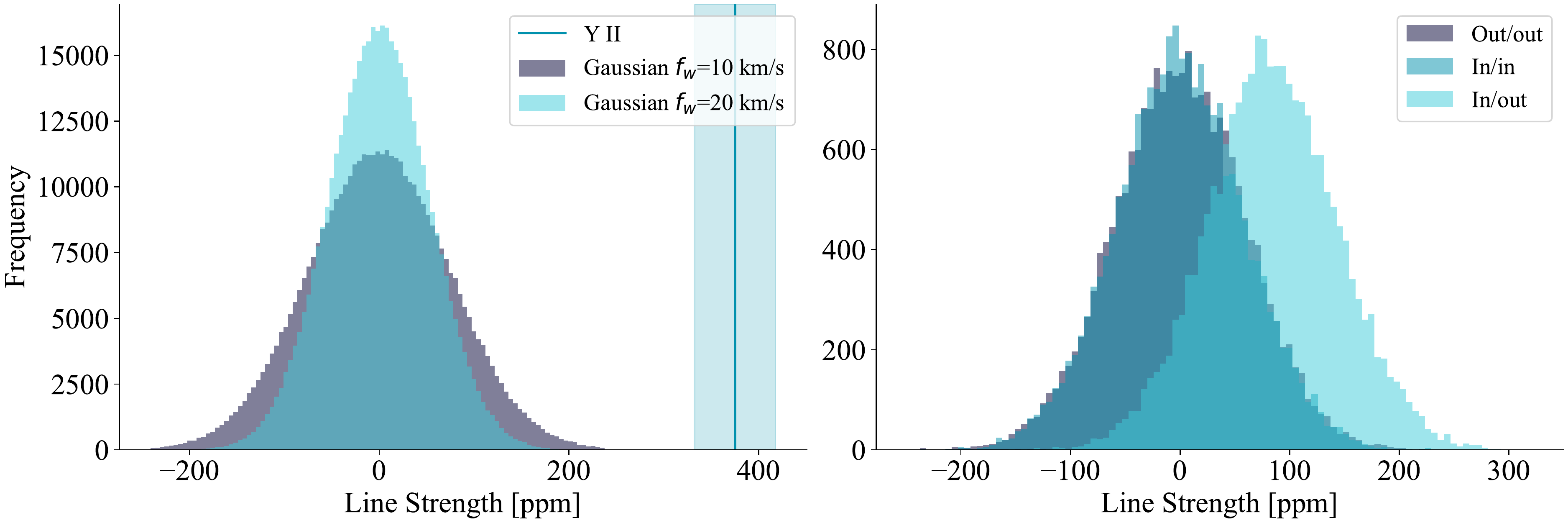}
\end{minipage}
\caption{Bootstrap results for the confirmed detections. Each panel represents the bootstrap results of each new detection in the respective order of: H\,I, Na\,I, Mg\,I, Ca\,II, Sc\,II, Ti\,II, Cr\,II, Fe\,I, Fe\,II, and Y\,II. The dark blue histogram represents the random Gaussian sample obtained with the 10\,\kms width, while the light transparent blue histogram is for the distribution width of 20\,\kms. The solid blue vertical line indicates the measured line depth of the signal, with a standard deviation of the value present as a transparent blue-shaded region. The legend in the right panel plots labels the in-in, in-out and out-out residual distributions calculated during the bootstrapping process.}
\label{fig:BoostrapConfirmedDetections}
\end{figure*}

All species were subjected to verification tests outlined in Section \ref{CCFApproach}. Fig. \ref{fig:BoostrapConfirmedDetections} identifies clear Gaussian profiles which reside well away from the measured line depths, implying the signals come from the planet as opposed to some underlying systematic structure. Fig. \ref{fig:ConfirmedDetectionsModels} depict the response plots and residual plots for the previously detected species in KELT-9\,b's atmosphere. The model's explanatory power varies, from almost obtaining a one-to-one relationship with H\,I, to Tb\,II, which only can explain approximately 8.9\,\% of the variation in the cross-correlation map. However, the error bars on all the gradient values obtain values above zero, implying that these models do have explanatory power.

\begin{figure*}
\centering
\begin{minipage}{1.0\textwidth}
  \centering
  \includegraphics[width=1.0\linewidth]{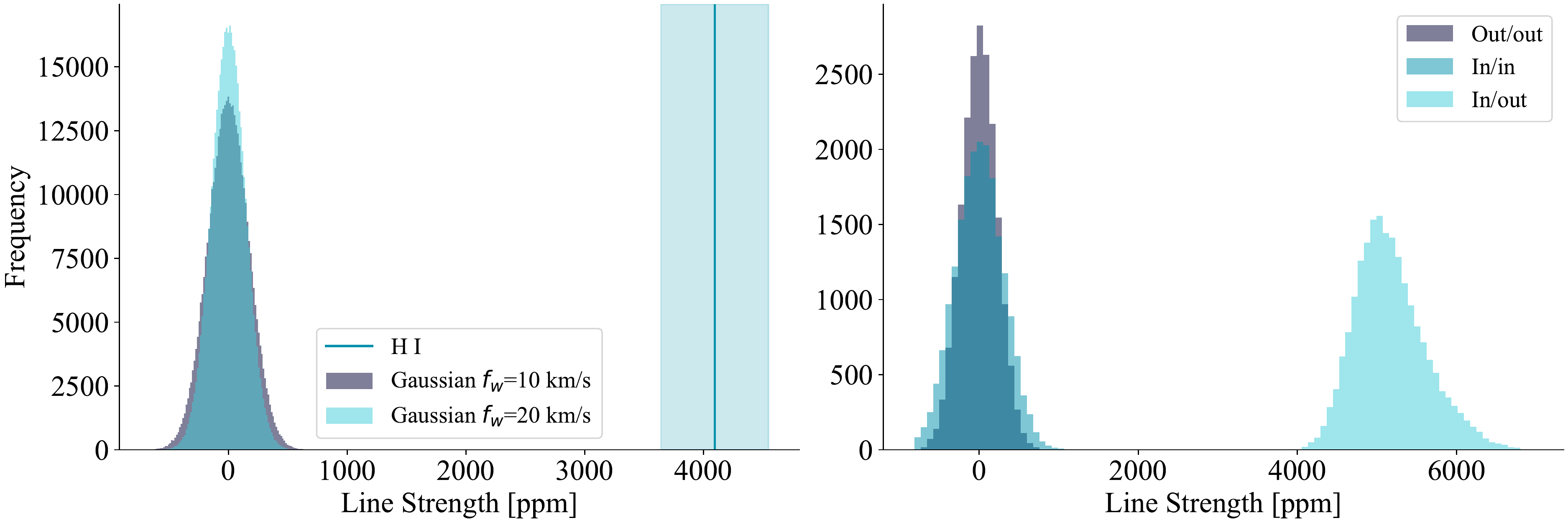}
\end{minipage}%

\begin{minipage}{1.0\textwidth}
  \centering
  \includegraphics[width=1.0\linewidth]{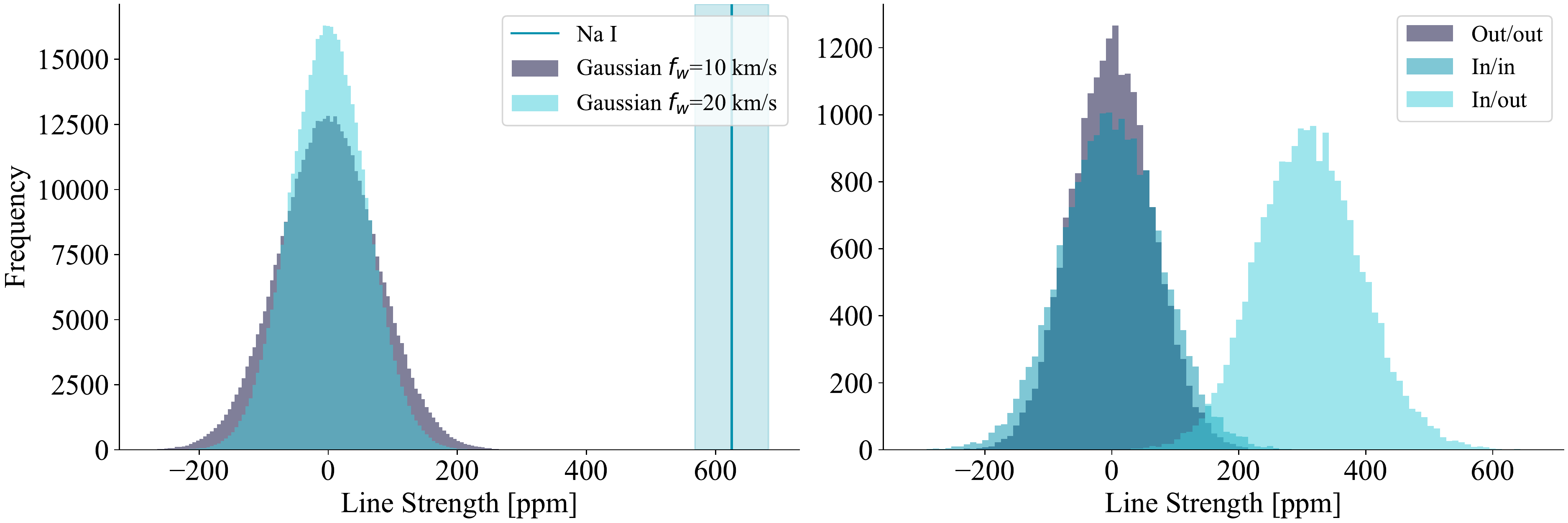}
\end{minipage}
\centering

\begin{minipage}{1.0\textwidth}
  \centering
  \includegraphics[width=1.0\linewidth]{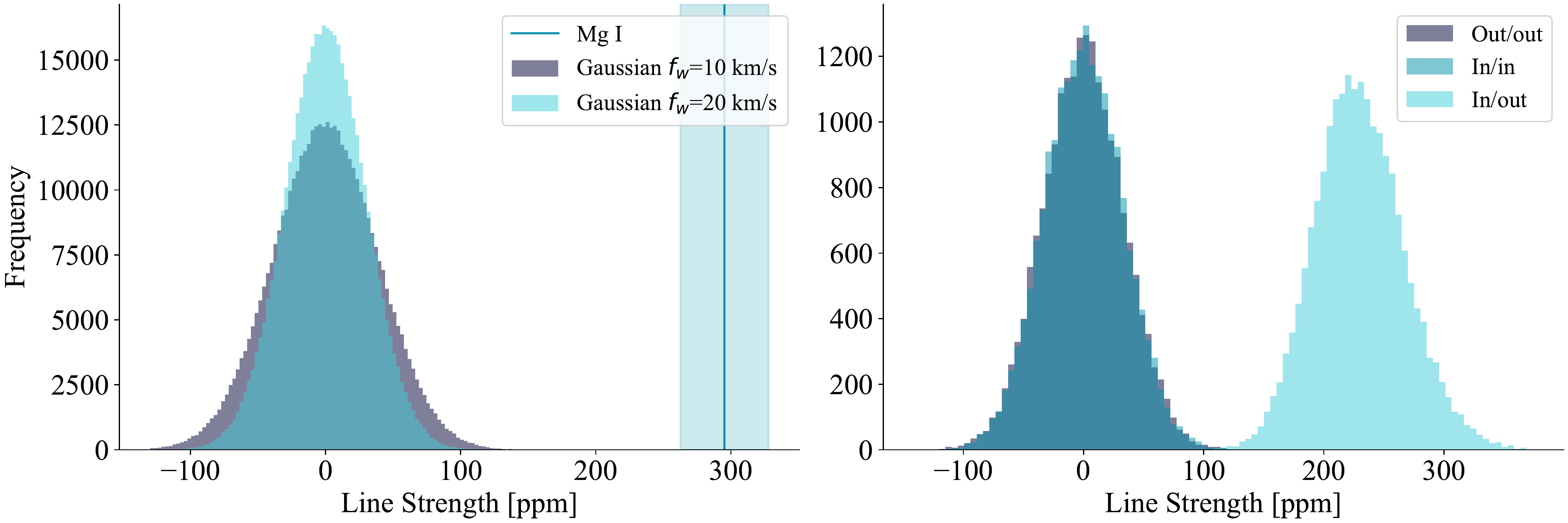}
\end{minipage}%

\begin{minipage}{1.0\textwidth}
  \centering
  \includegraphics[width=1.0\linewidth]{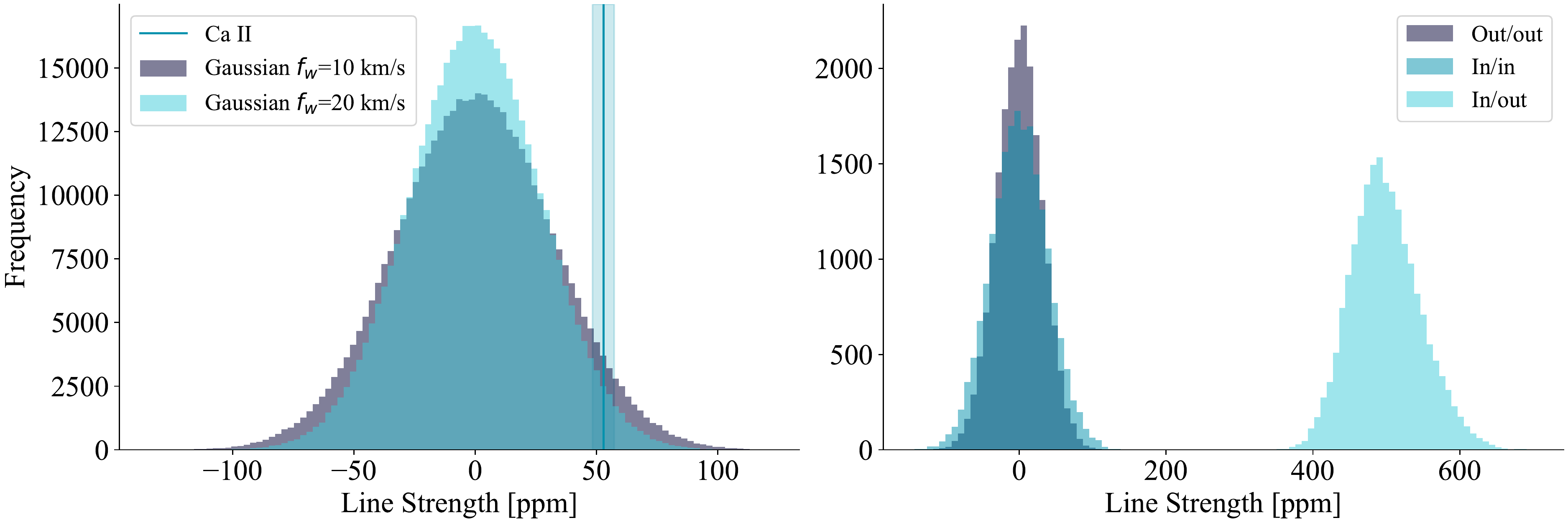}
\end{minipage}
\end{figure*}

\begin{figure*}
\centering
\begin{minipage}{1.0\textwidth}
  \centering
  \includegraphics[width=1.0\linewidth]{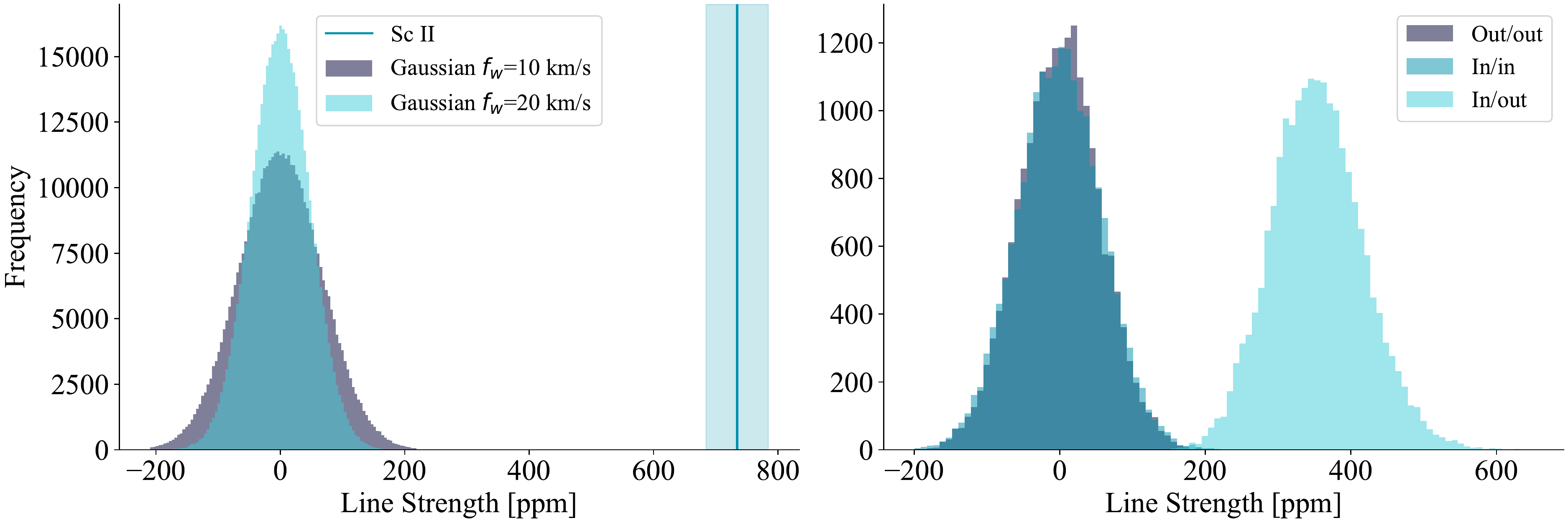}
\end{minipage}%

\begin{minipage}{1.0\textwidth}
  \centering
  \includegraphics[width=1.0\linewidth]{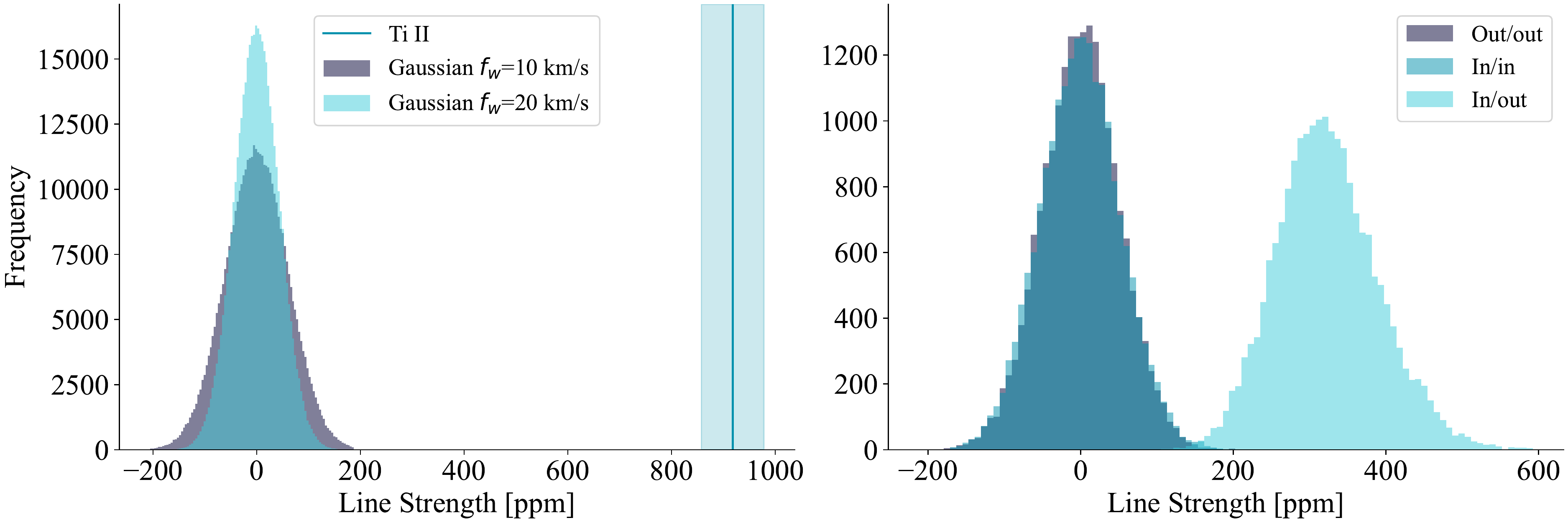}
\end{minipage}

\centering
\begin{minipage}{1.0\textwidth}
  \centering
  \includegraphics[width=1.0\linewidth]{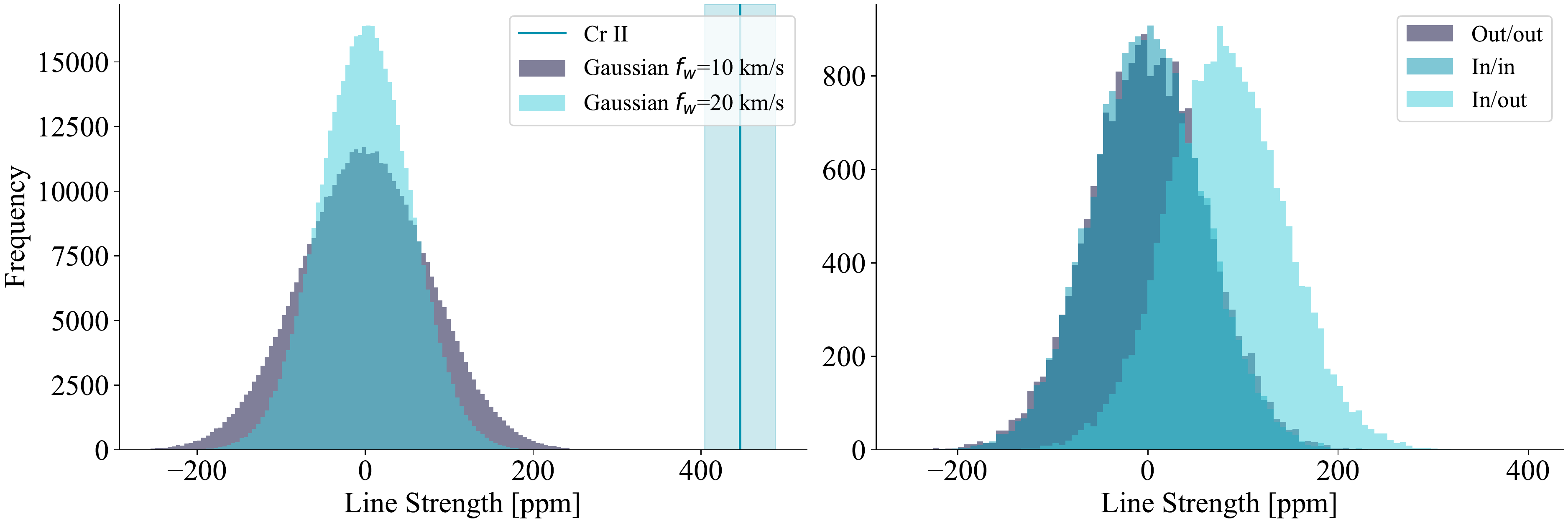}
\end{minipage}%

\begin{minipage}{1.0\textwidth}
  \centering
  \includegraphics[width=1.0\linewidth]{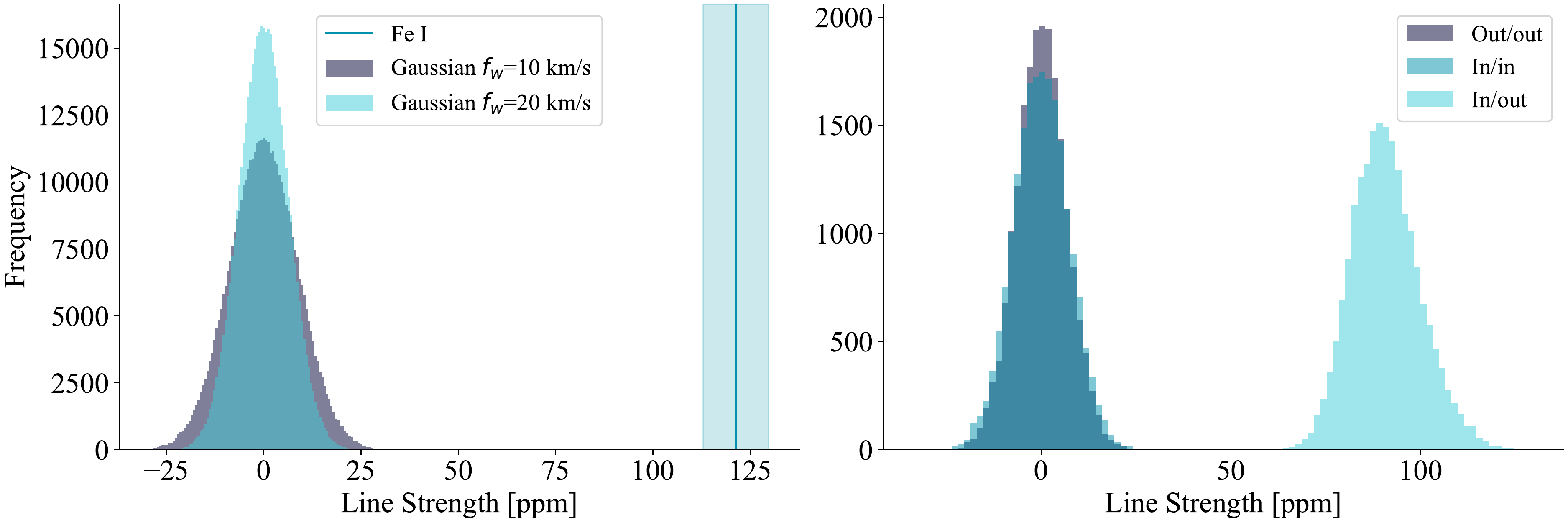}
\end{minipage}
\end{figure*}

\begin{figure*}
\centering
\begin{minipage}{1.0\textwidth}
  \centering
  \includegraphics[width=1.0\linewidth]{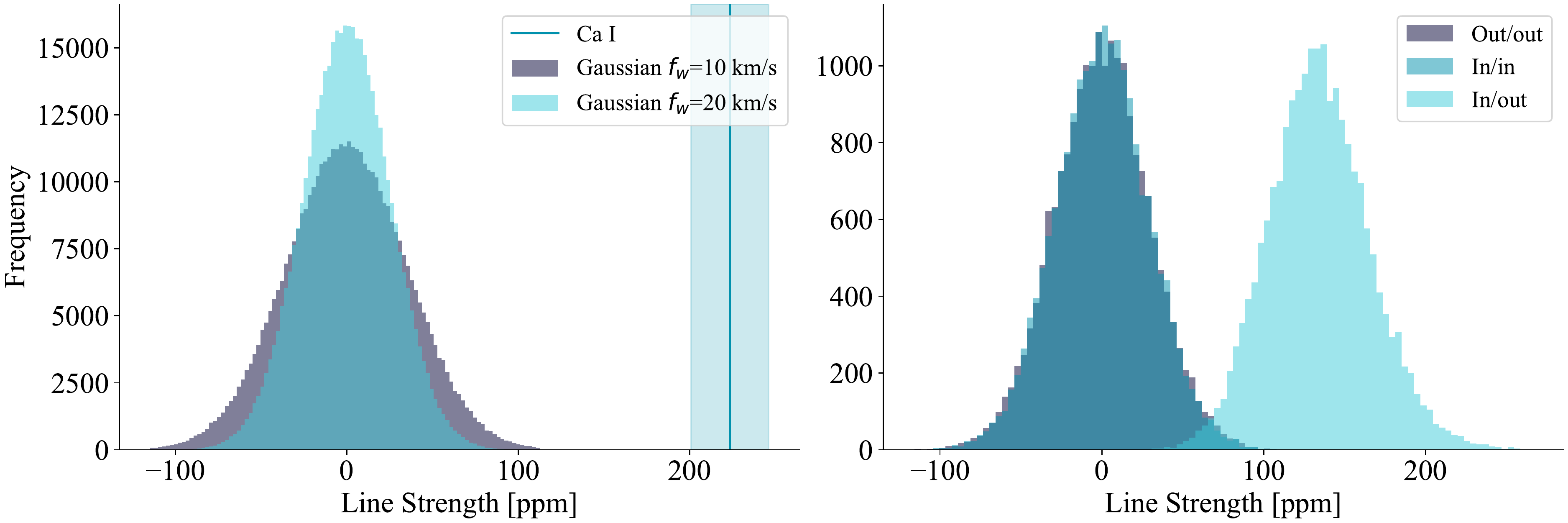}
\end{minipage}%

\begin{minipage}{1.0\textwidth}
  \centering
  \includegraphics[width=1.0\linewidth]{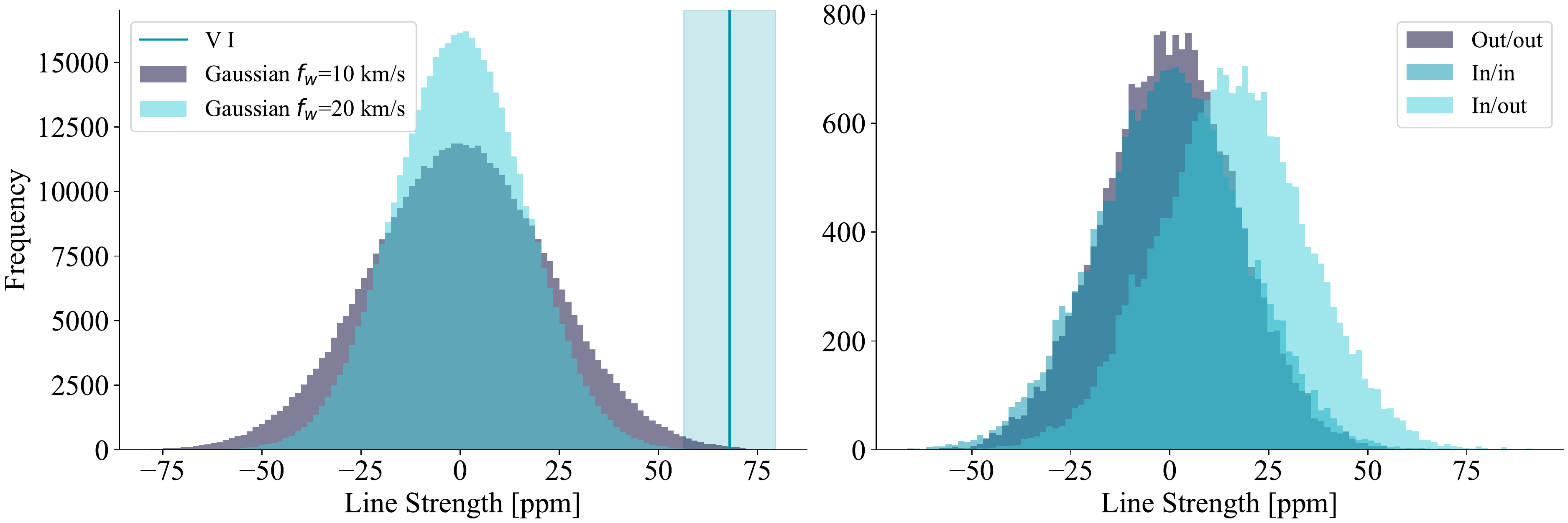}
\end{minipage}

\begin{minipage}{1.0\textwidth}
  \centering
  \includegraphics[width=1.0\linewidth]{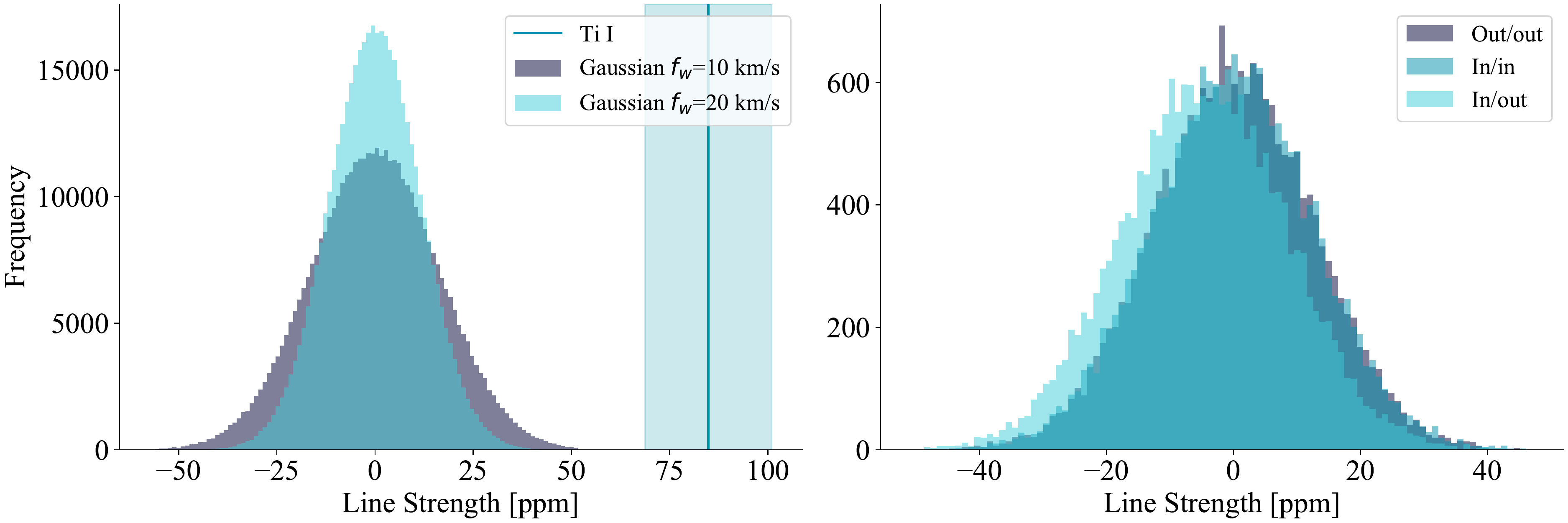}
\end{minipage}

\centering
\begin{minipage}{1.0\textwidth}
  \centering
  \includegraphics[width=1.0\linewidth]{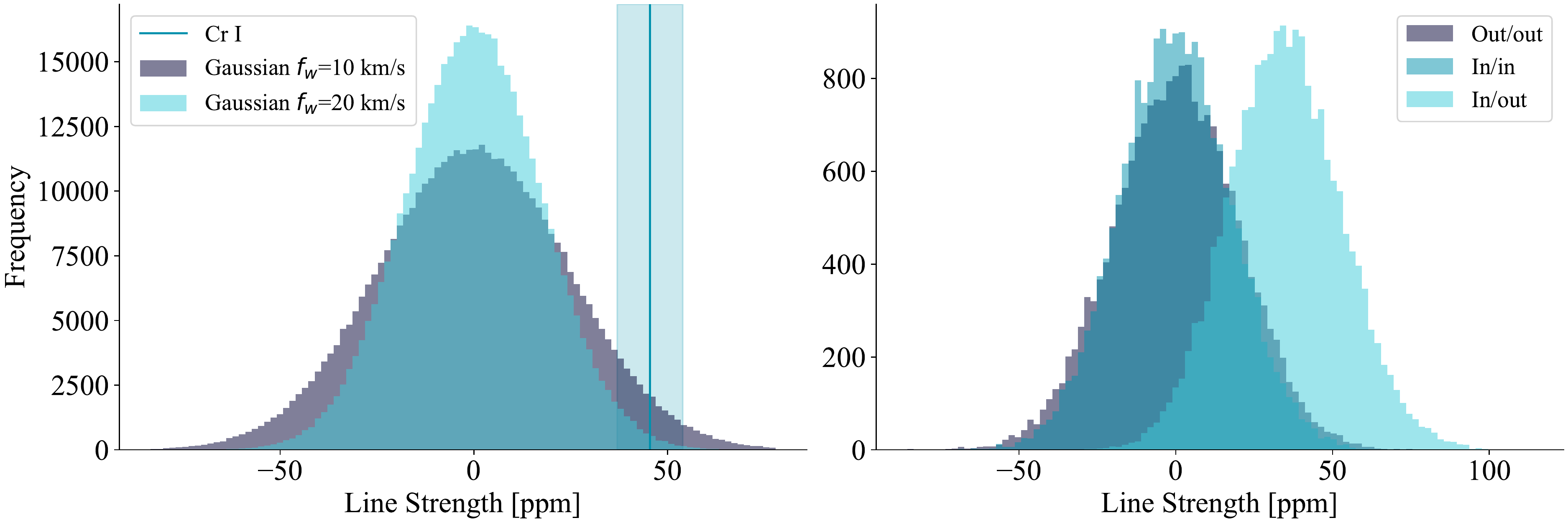}
\end{minipage}%

\end{figure*}

\begin{figure*}
\centering

\begin{minipage}{1.0\textwidth}
  \centering
  \includegraphics[width=0.9\linewidth]{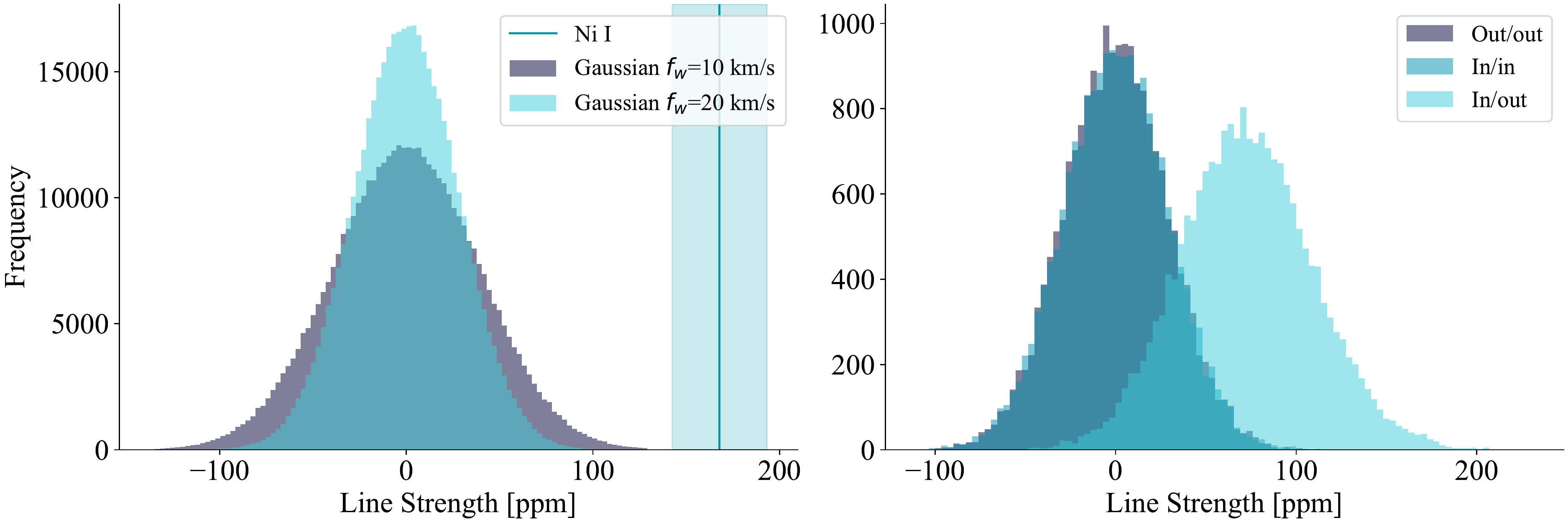}
\end{minipage}

\begin{minipage}{1.0\textwidth}
  \centering
  \includegraphics[width=0.9\linewidth]{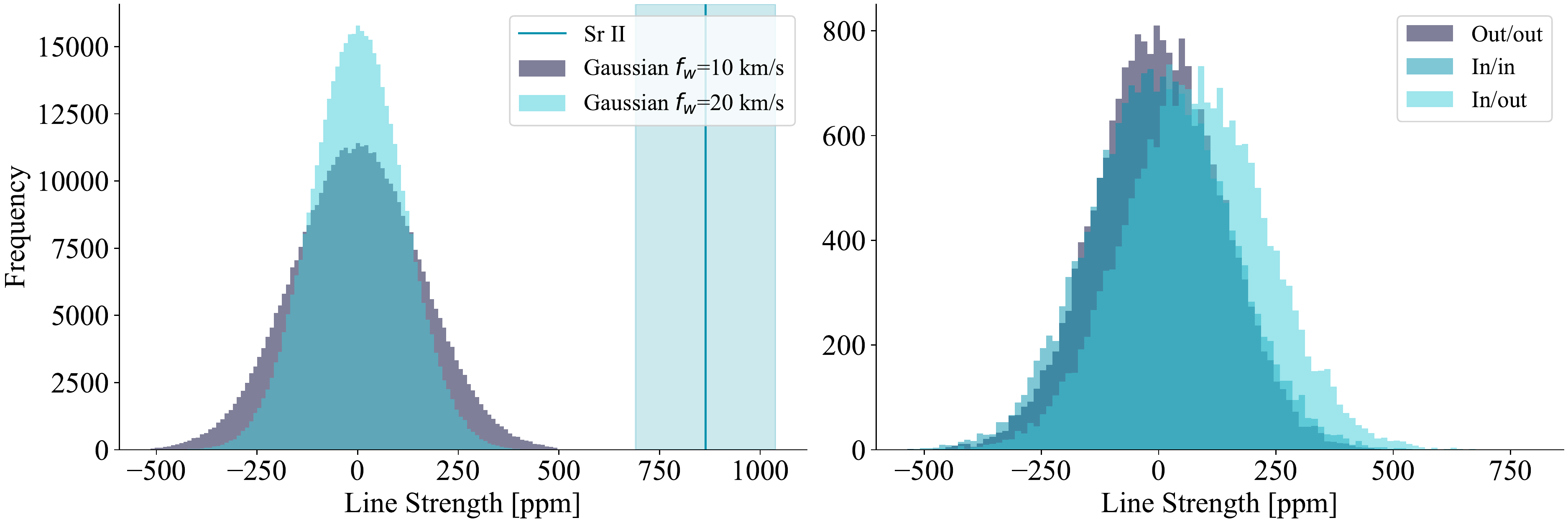}
\end{minipage}%

\begin{minipage}{1.0\textwidth}
  \centering
  \includegraphics[width=0.9\linewidth]{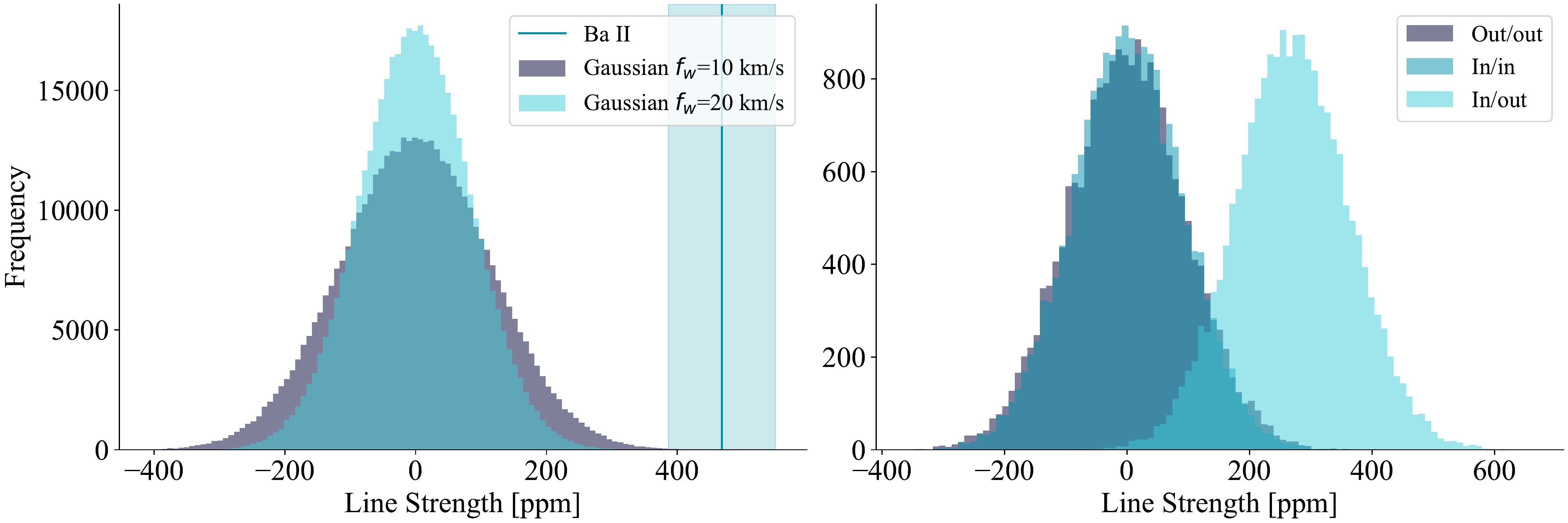}
\end{minipage}

\begin{minipage}{1.0\textwidth}
  \centering
  \includegraphics[width=0.9\linewidth]{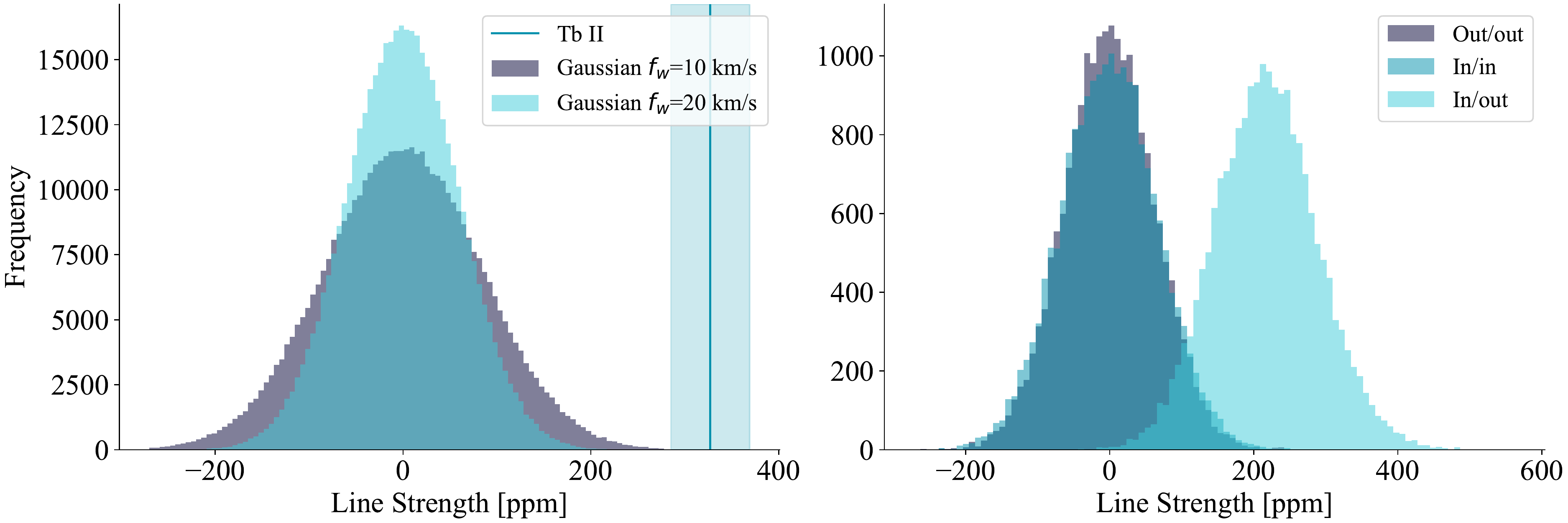}
\end{minipage}%

\caption{Bootstrap results for the new detections. Each panel represents the bootstrap results of each new detection in the respective order of Ca\,I, V\,I, Ti\,I, Cr\,I, Ni\,I, Sr\,II, Ba\,II, and Tb\,II. The plots follow the same construction as Fig. \ref{fig:ConfirmedDetectionsModels}}
\label{fig:BoostrapNewDetections}
\end{figure*}

\begin{figure*}
\centering
\begin{subfigure}{0.35\textwidth}
    \includegraphics[width=\textwidth]{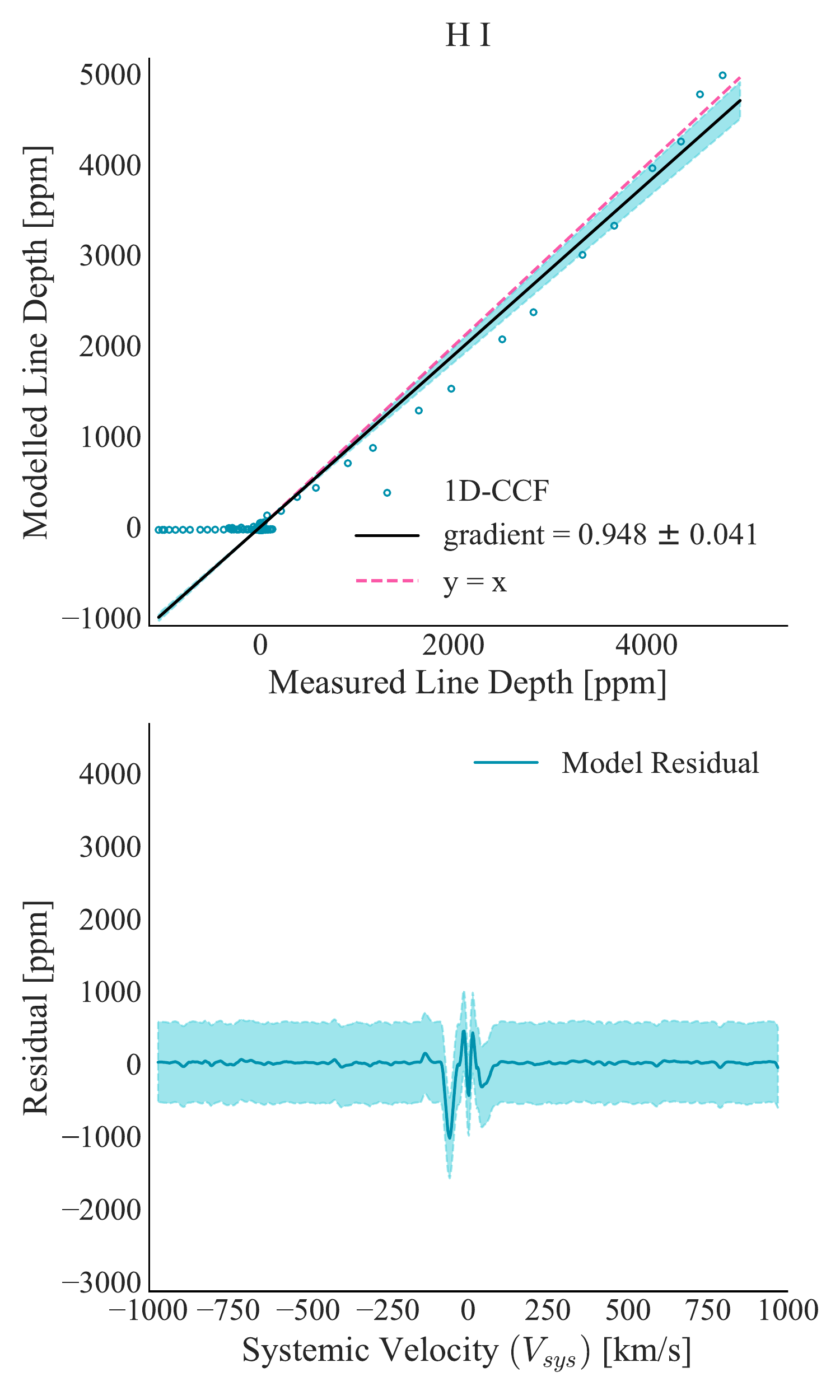}
\end{subfigure}
\begin{subfigure}{0.35\textwidth}
    \includegraphics[width=\textwidth]{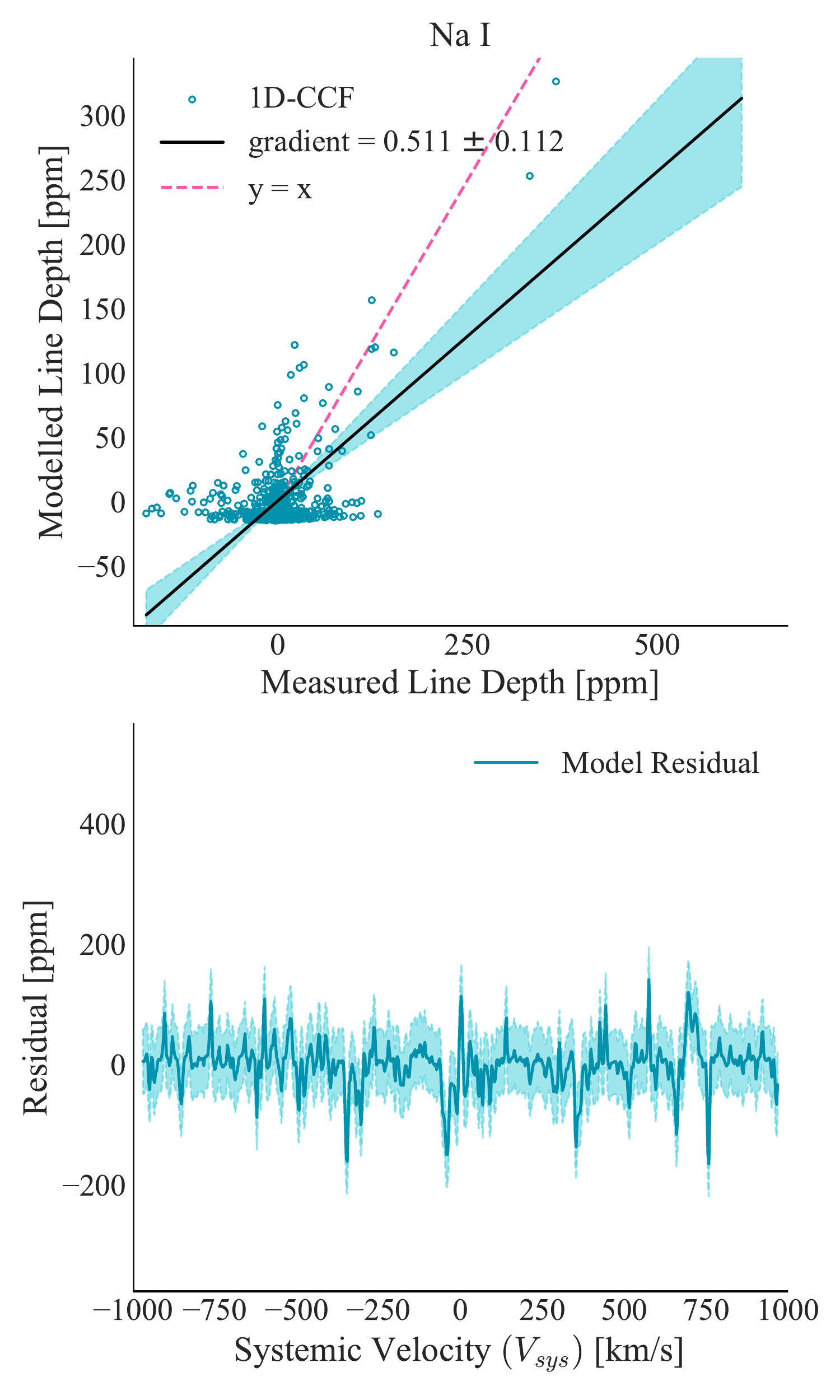}
\end{subfigure}
\hfill
\begin{subfigure}{0.35\textwidth}
    \includegraphics[width=\textwidth]{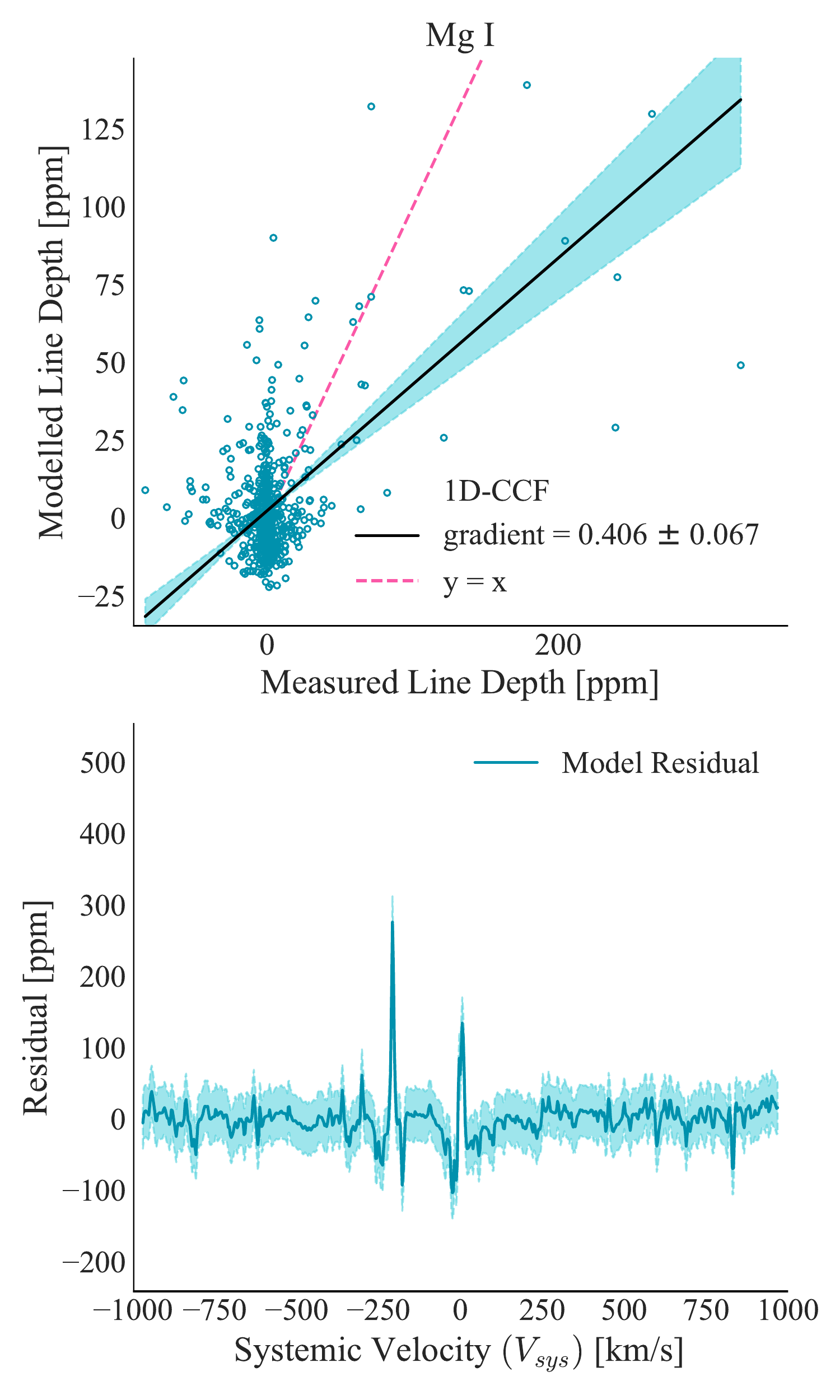}
\end{subfigure}
\begin{subfigure}{0.35\textwidth}
    \includegraphics[width=\textwidth]{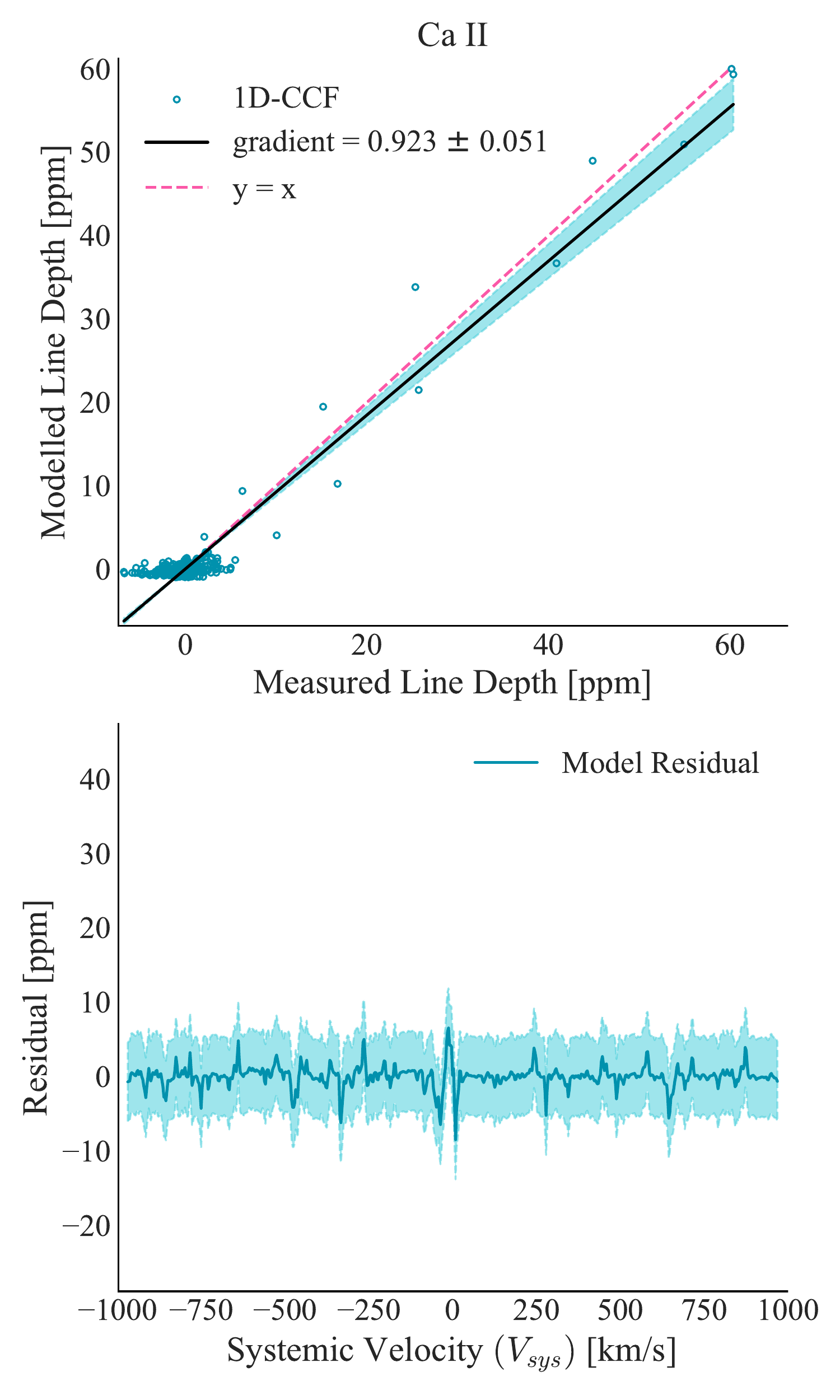}
\end{subfigure}
\caption{Response and residual plots for each confirmed species in KELT-9\,b's atmosphere. Each panel represents a different species. The order of the species is as follows (\textit{left to right}): H\,I, Na\,I, Mg\,I, Ca\,II, Sc\,II, Ti\,II, Cr\,II, Fe\,I, Fe\,II, and Y\,II. Each plot contains a line of best fit plotted through the data of the response plot, spanned by two lines of worst fit. The identity line $y=x$ is plotted to show how closely the model matches the data. The residual plot contains the residuals and the region spanning one standard deviation of the residual data.}
\label{fig:ConfirmedDetectionsModels}
\end{figure*}

\begin{figure*}
\centering
\begin{subfigure}{0.35\textwidth}
    \includegraphics[width=\textwidth]{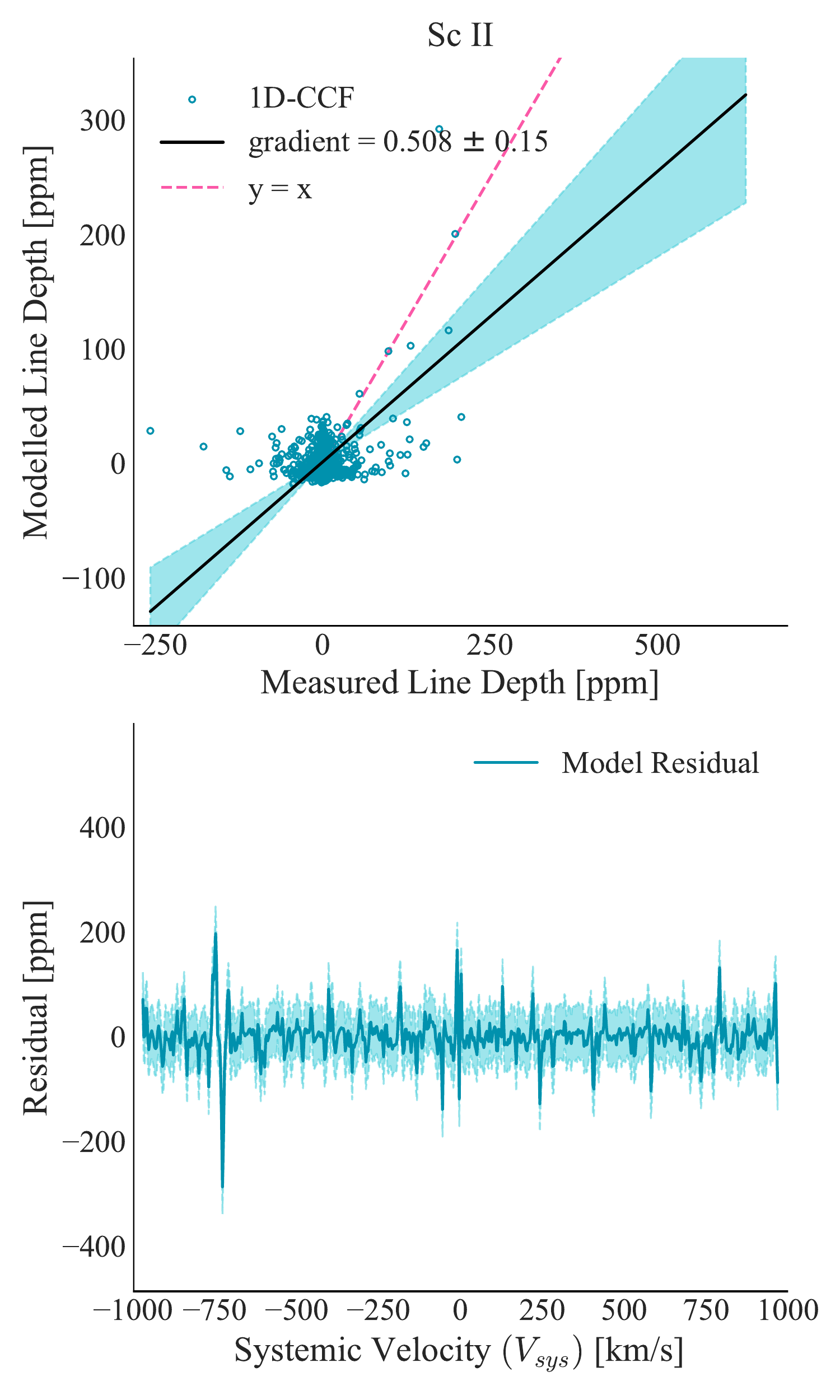}
\end{subfigure}
\begin{subfigure}{0.35\textwidth}
    \includegraphics[width=\textwidth]{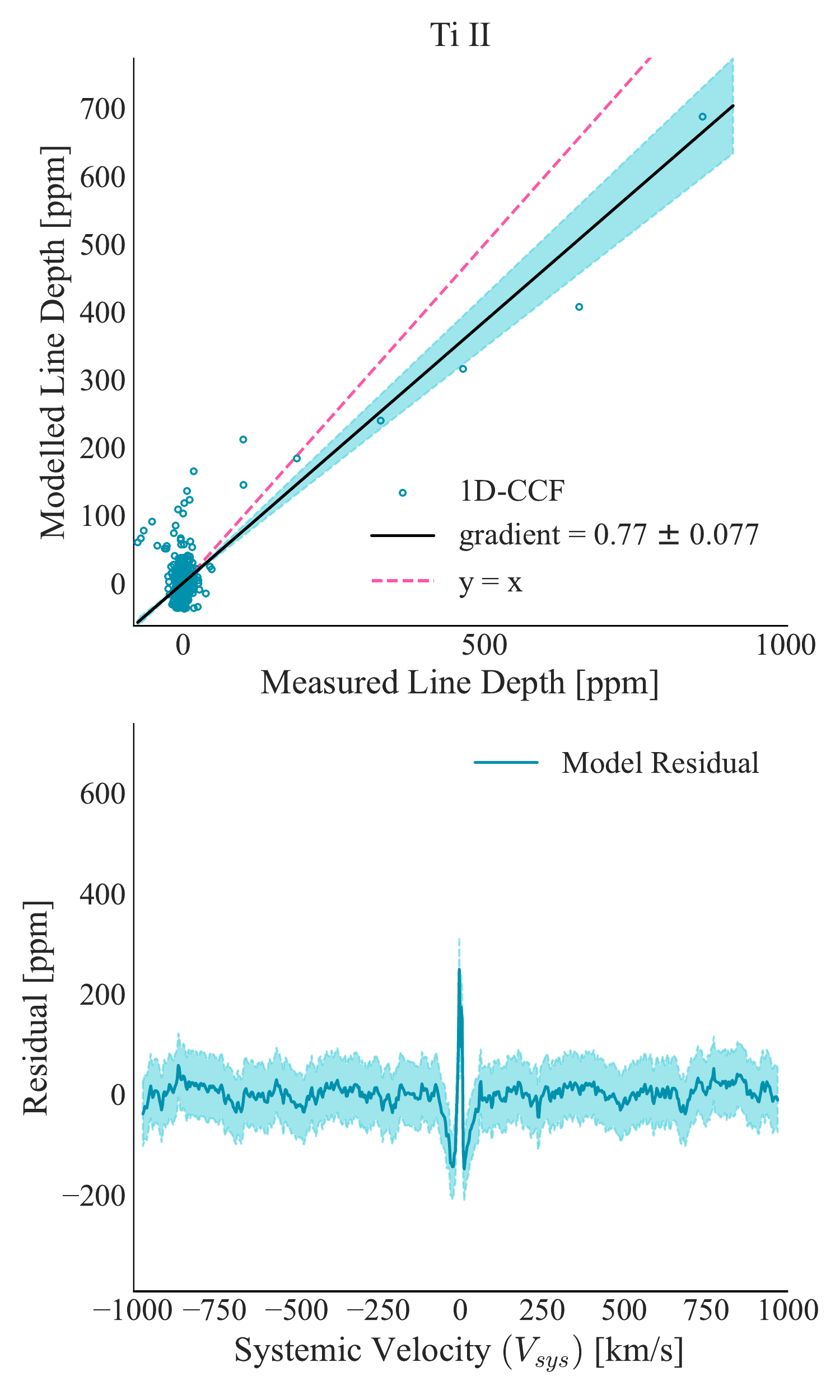}
\end{subfigure}
\hfill
\begin{subfigure}{0.35\textwidth}
    \includegraphics[width=\textwidth]{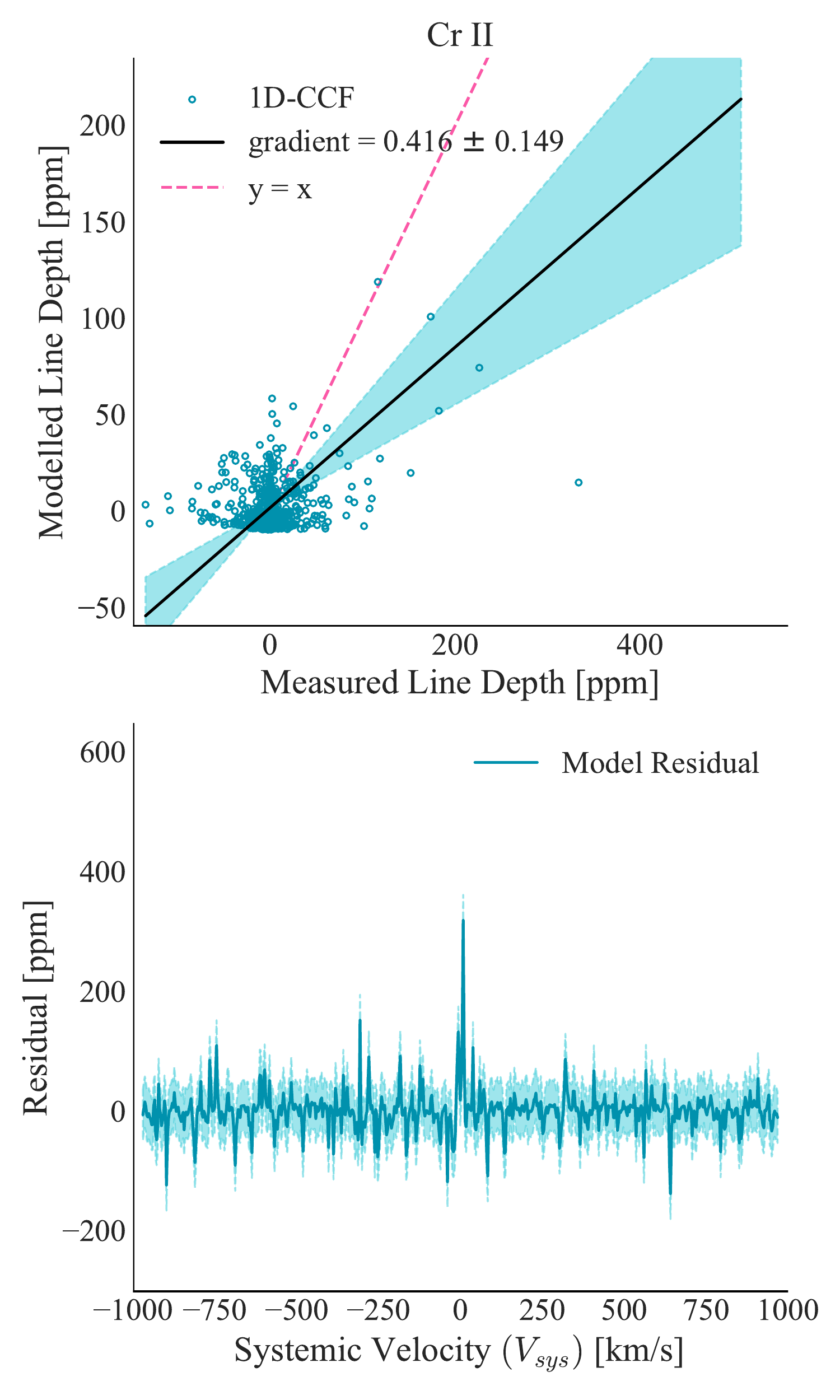}
\end{subfigure}
\begin{subfigure}{0.35\textwidth}
    \includegraphics[width=\textwidth]{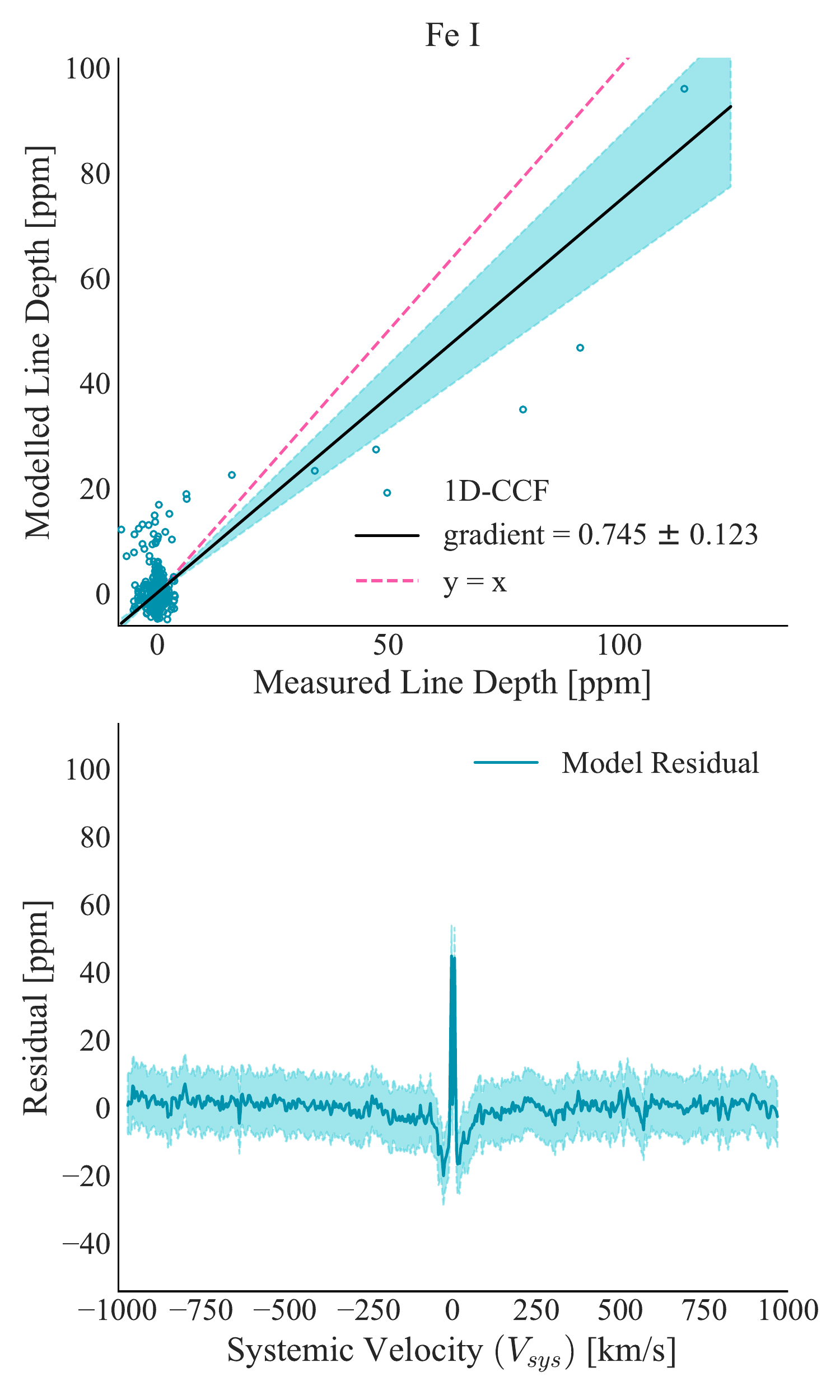}
\end{subfigure}
\caption{\textit{Continued: }Response and residual plots for each confirmed species in KELT-9\,b's atmosphere. The order of the species is as follows (\textit{left to right}): Sc\,II, Ti\,II, Cr\,II, and Fe\,I.}
\label{fig:ConfirmedDetectionsModels2}
\end{figure*}

\begin{figure*}
\centering
\begin{subfigure}{0.35\textwidth}
    \includegraphics[width=\textwidth]{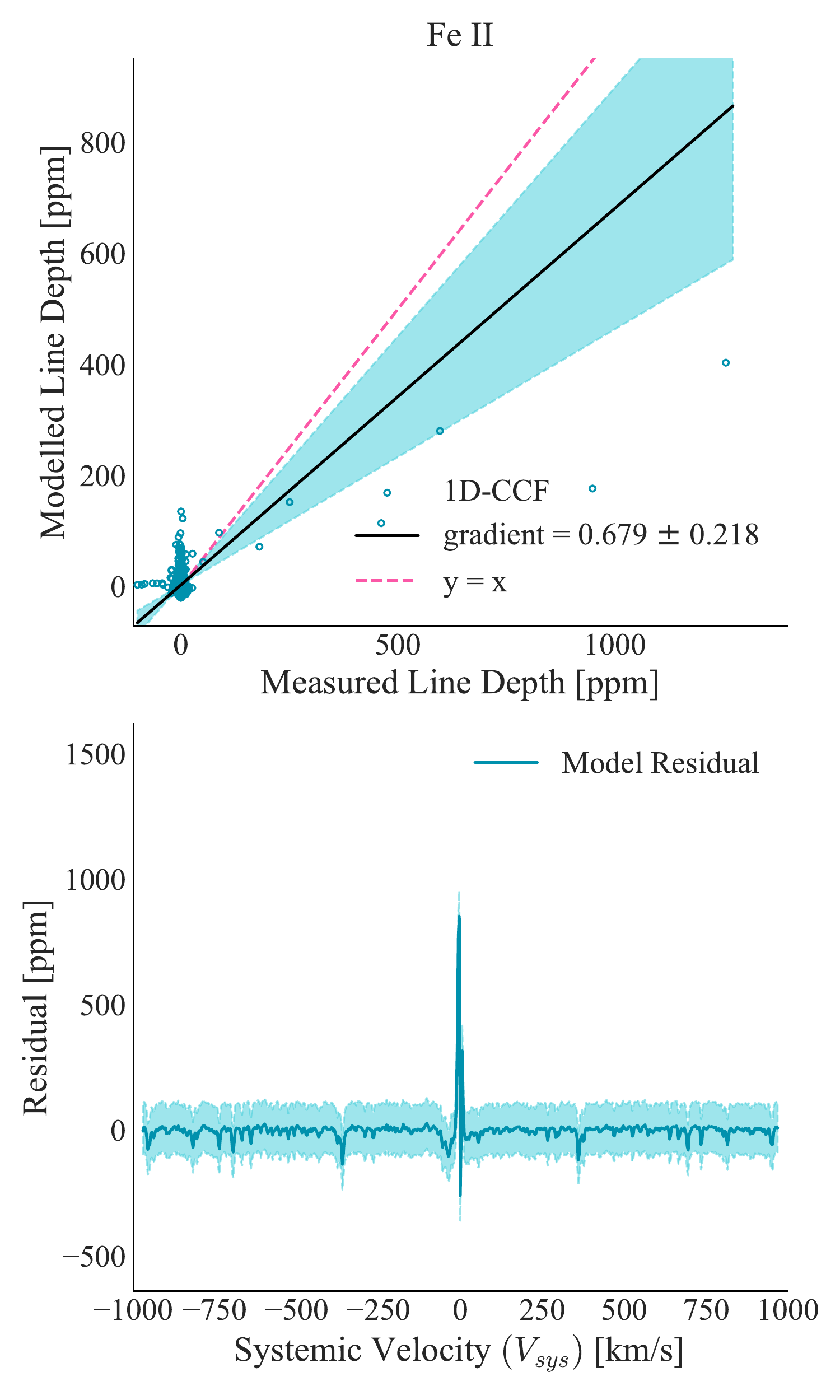}
\end{subfigure}
\begin{subfigure}{0.35\textwidth}
    \includegraphics[width=\textwidth]{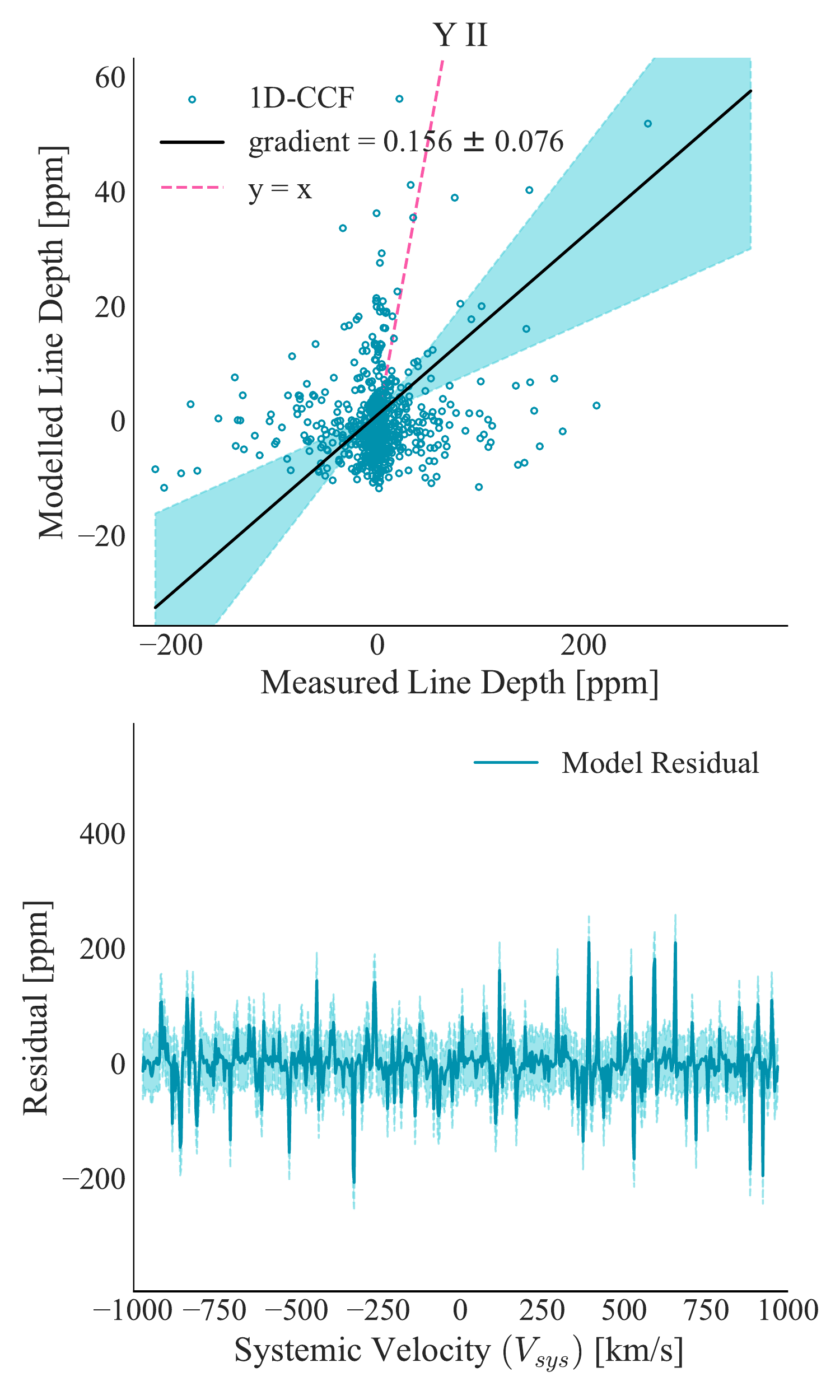}
\end{subfigure}
 \caption{\textit{Continued: }Response and residual plots for each confirmed species in KELT-9\,b's atmosphere. The order of the species is as follows (\textit{left to right}): Fe\,II and Y\,II.}
\label{fig:ConfirmedDetectionsModels3}
        
\end{figure*}

\begin{figure*}
\centering
\begin{subfigure}{0.35\textwidth}
    \includegraphics[width=\textwidth]{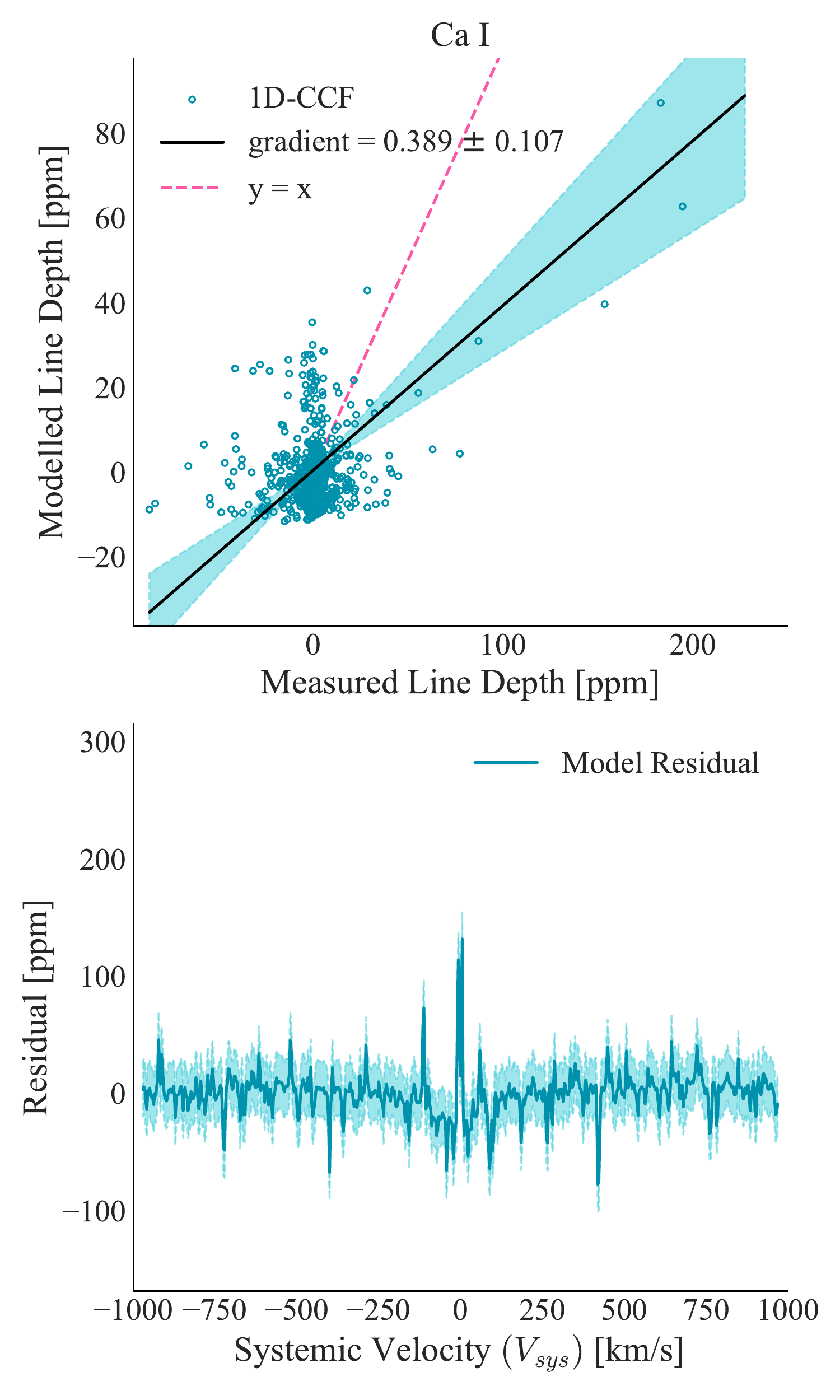}
\end{subfigure}
\begin{subfigure}{0.35\textwidth}
    \includegraphics[width=\textwidth]{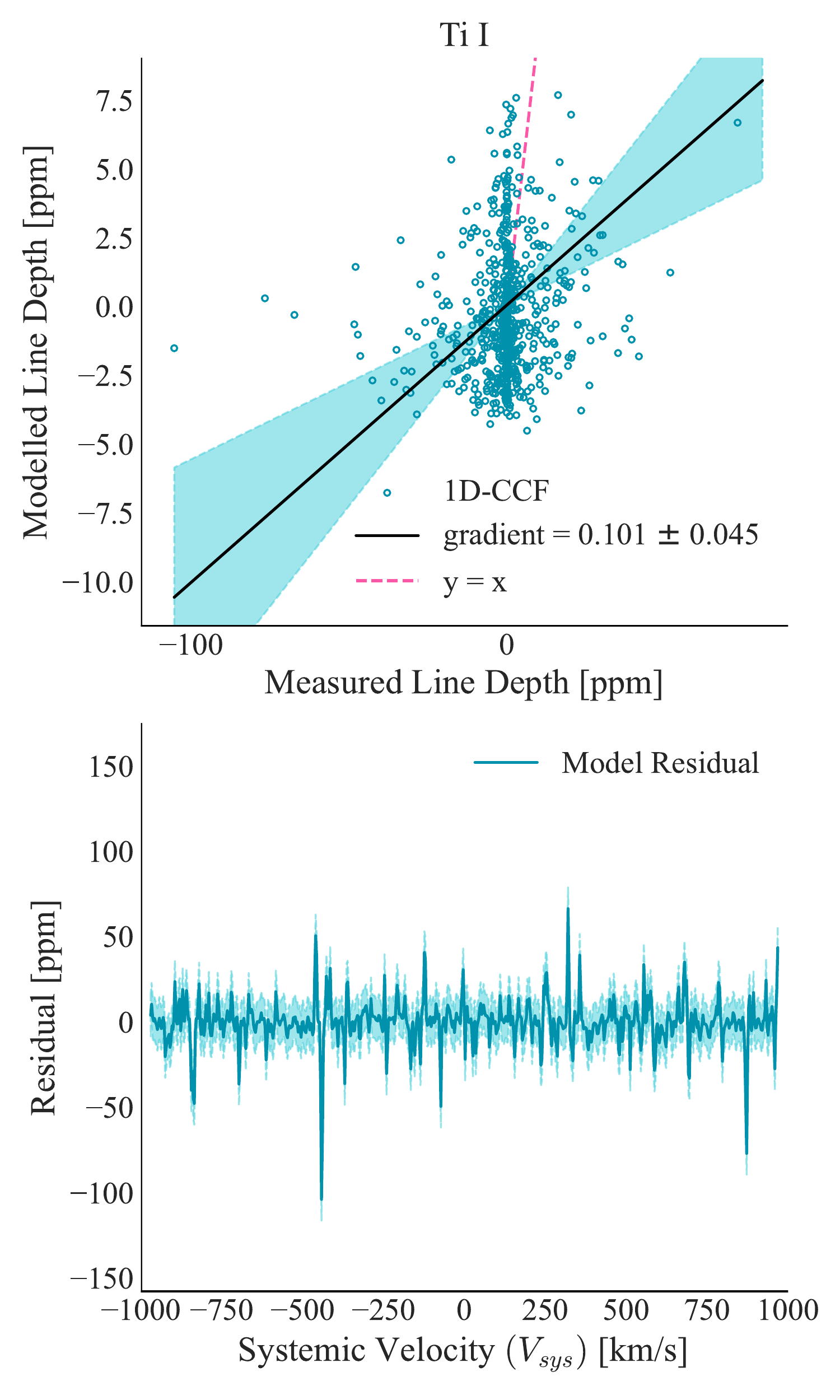}
\end{subfigure}
\hfill
\begin{subfigure}{0.35\textwidth}
    \includegraphics[width=\textwidth]{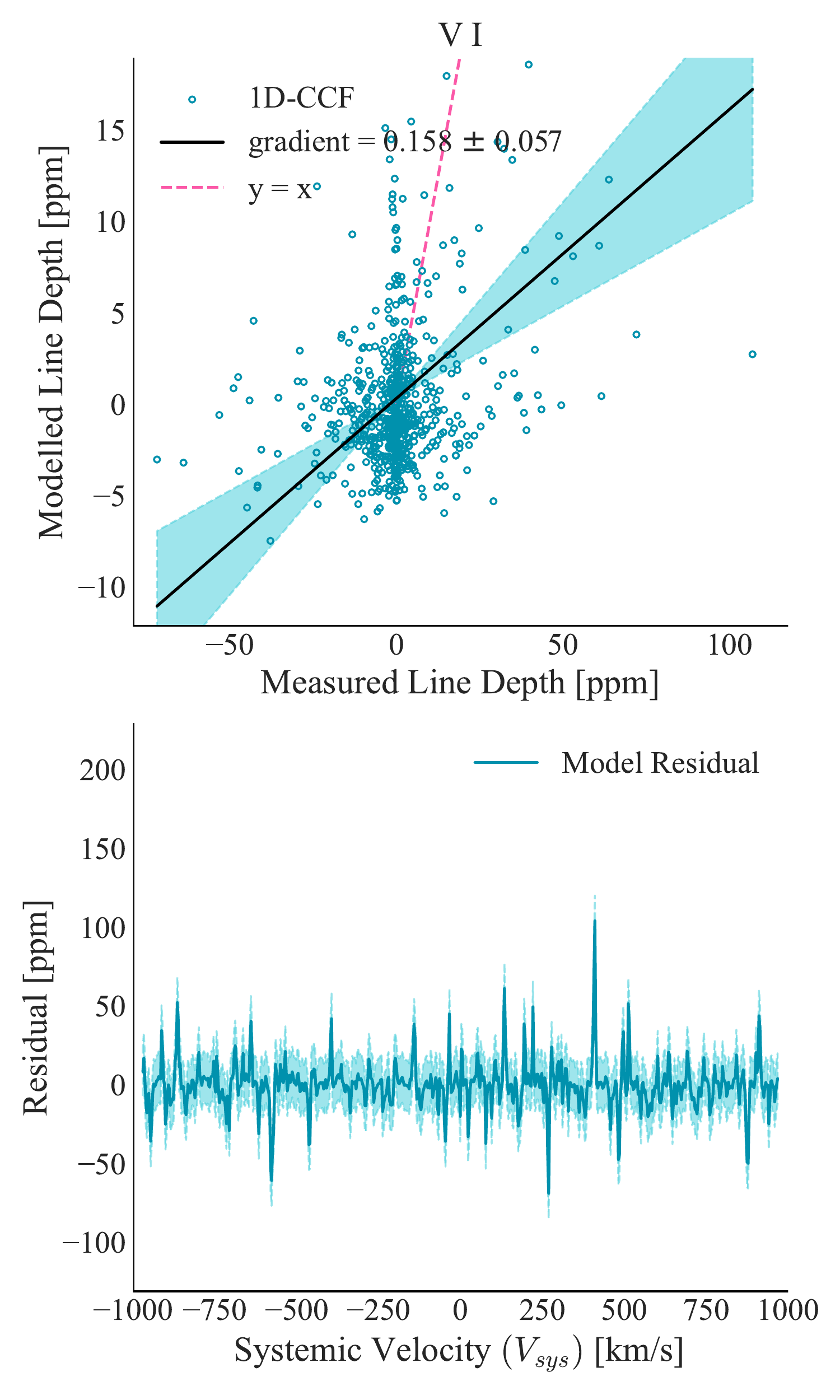}
\end{subfigure}
\begin{subfigure}{0.35\textwidth}
    \includegraphics[width=\textwidth]{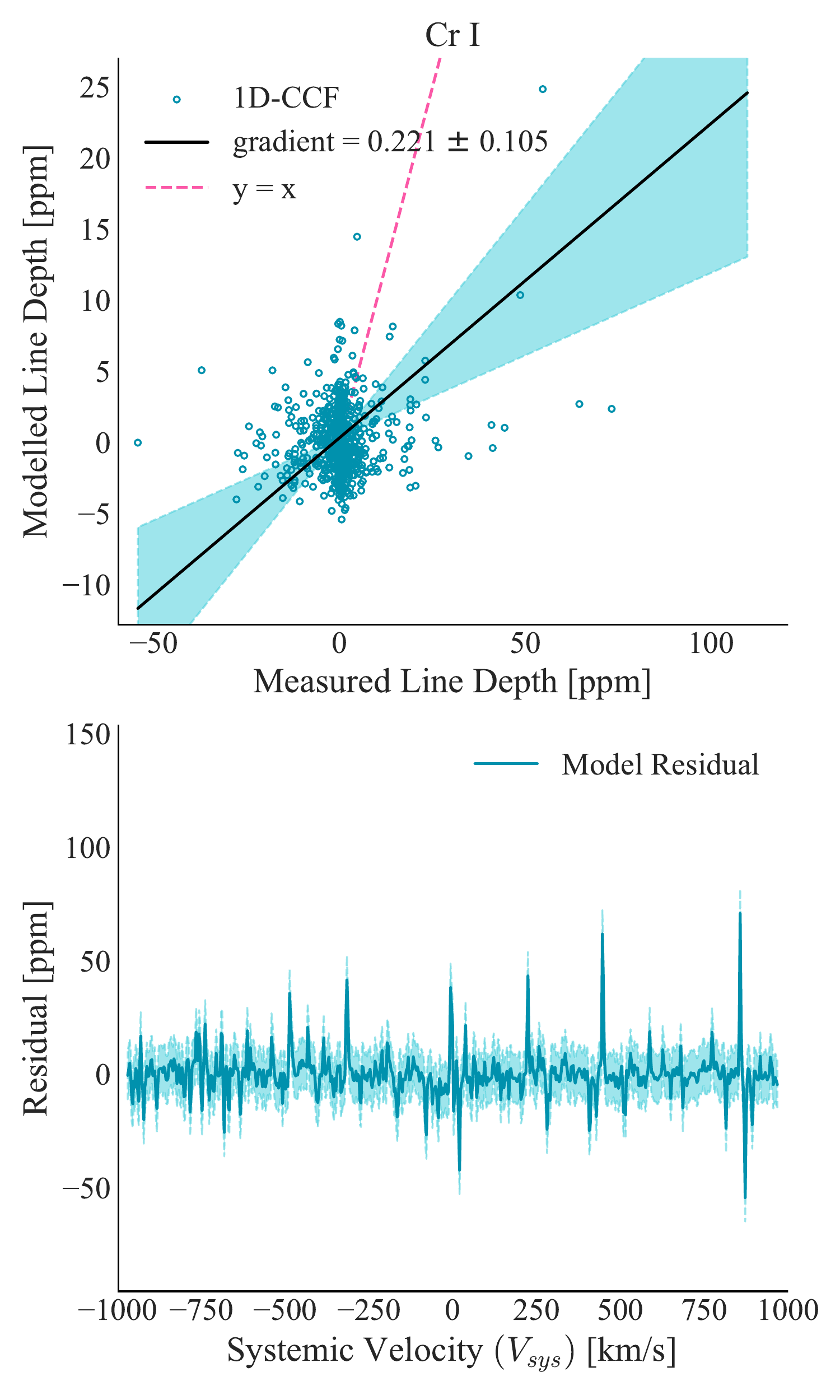}
\end{subfigure}
\caption{Response and residual plots for each newly detected species in KELT-9\,b's atmosphere. Each panel represents a different species. The order of the species is as follows (\textit{left to right}): Ca\,I, Ti\,I, V\,I, and Cr\,I. The format of the plot follows the same structure as Fig. \ref{fig:ConfirmedDetectionsModels}.}
\label{fig:NewDectionsModels}
\end{figure*}

\begin{figure*}
\centering
\begin{subfigure}{0.35\textwidth}
    \includegraphics[width=\textwidth]{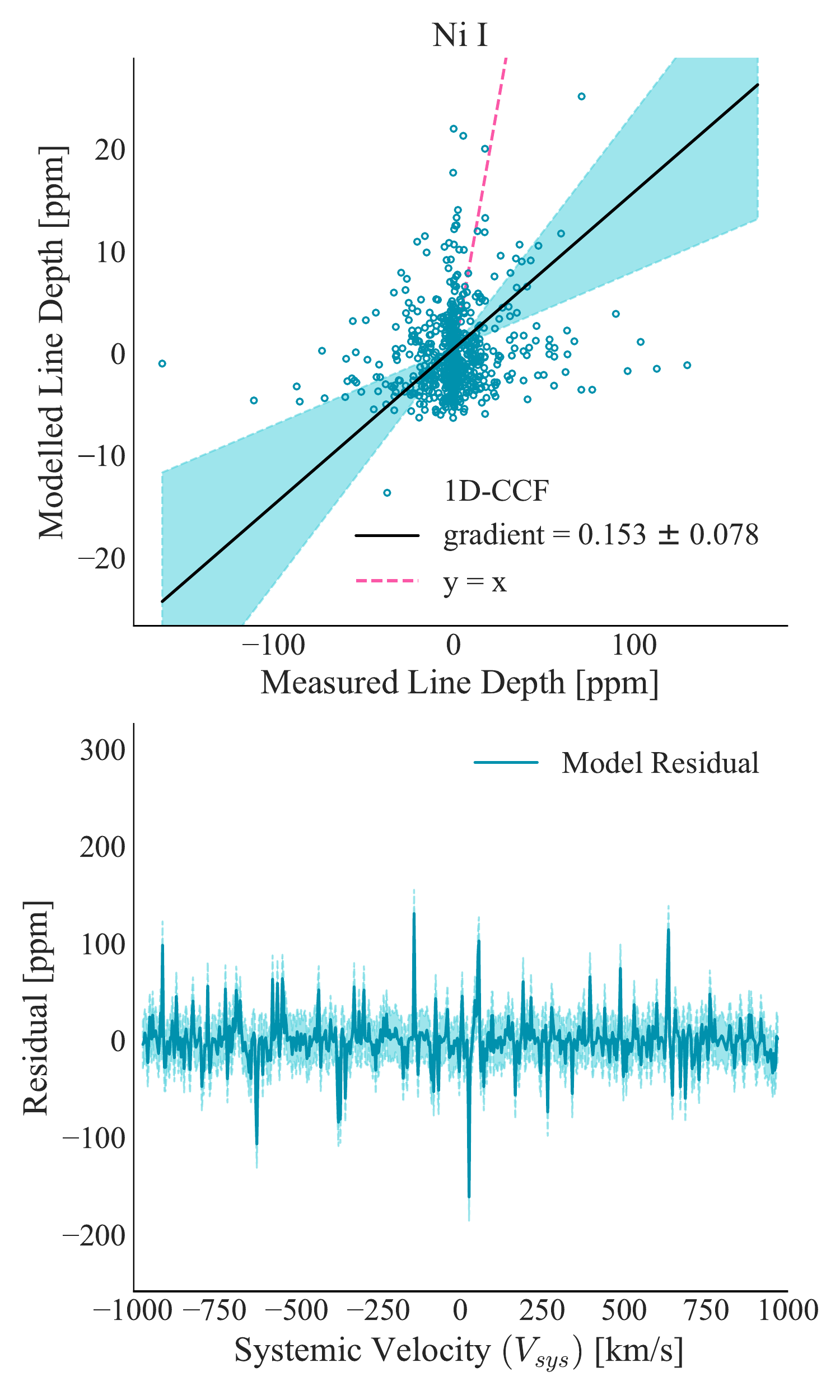}
\end{subfigure}
\begin{subfigure}{0.35\textwidth}
    \includegraphics[width=\textwidth]{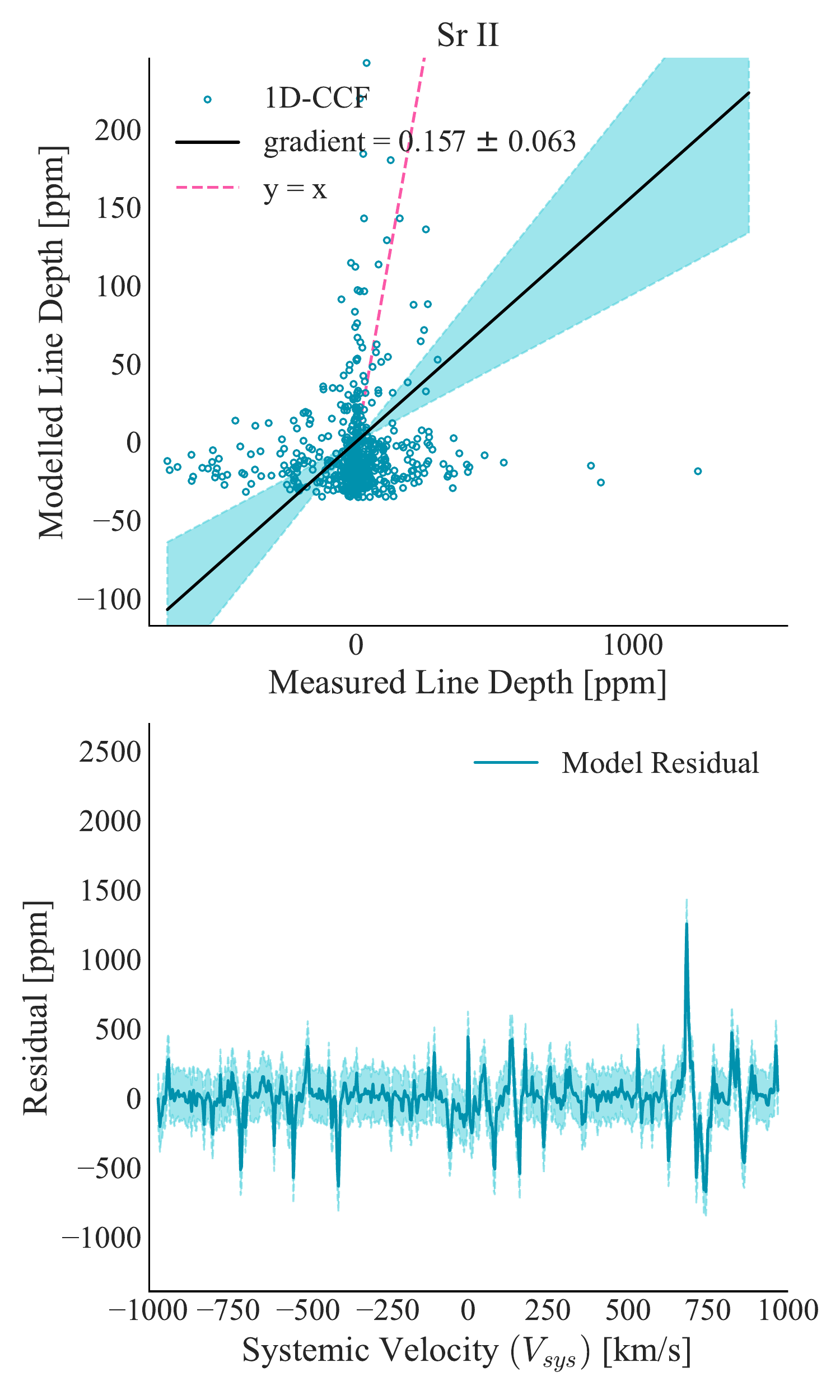}
\end{subfigure}
\hfill
\begin{subfigure}{0.35\textwidth}
    \includegraphics[width=\textwidth]{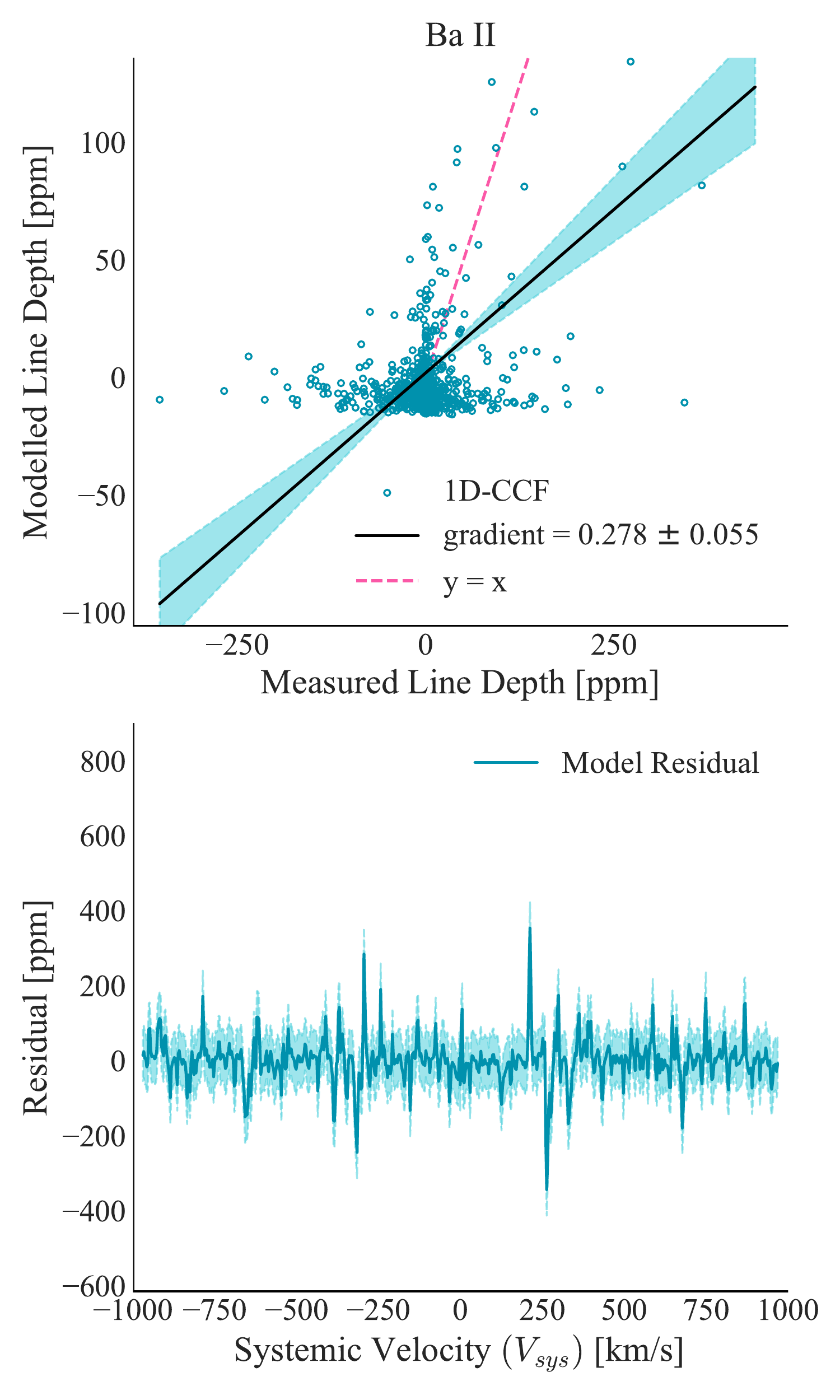}
\end{subfigure}
\begin{subfigure}{0.35\textwidth}
    \includegraphics[width=\textwidth]{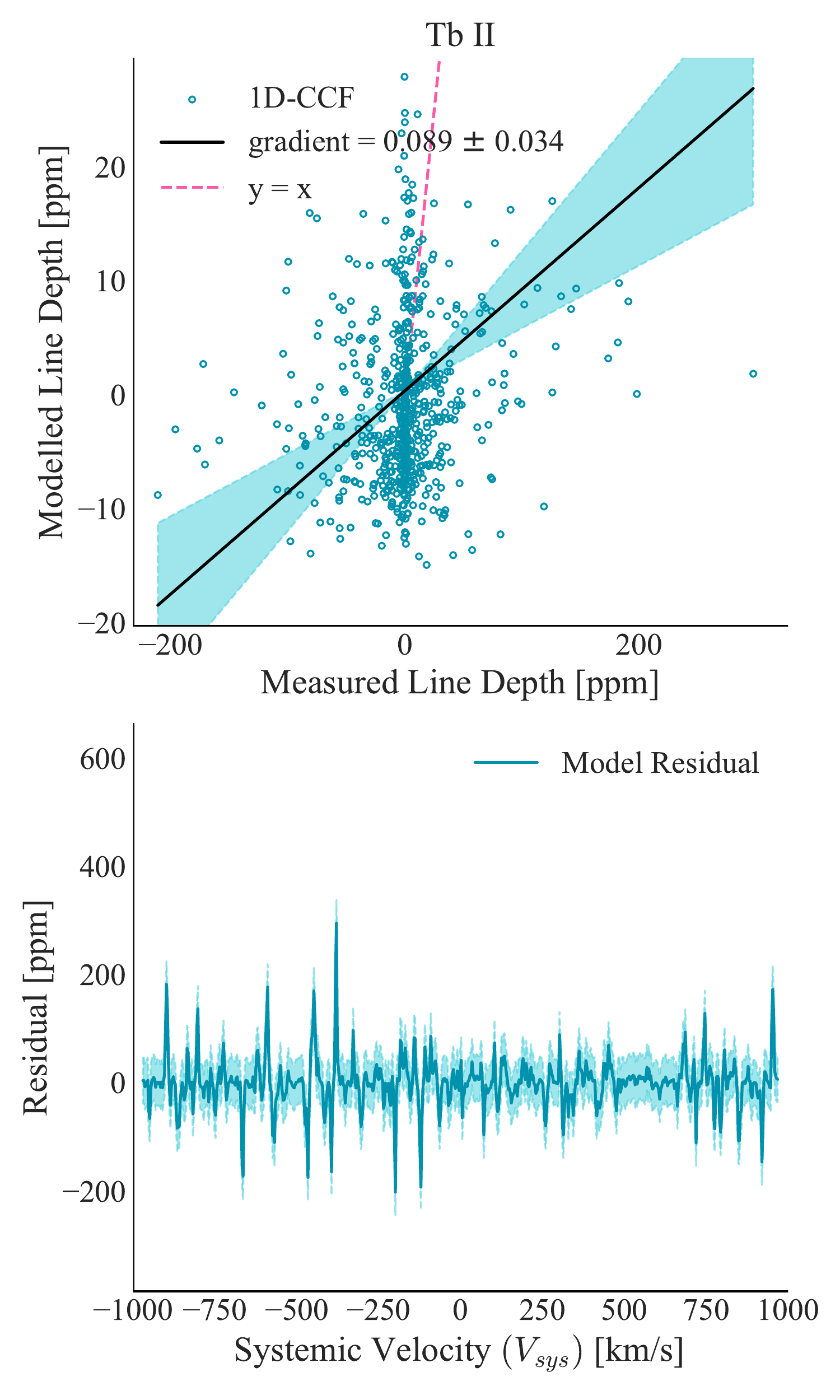}
\end{subfigure}
        
\caption{\textit{Continued: }Response and residual plots for each newly detected species in KELT-9\,b's atmosphere. The order of the species is as follows (\textit{left to right}): Ni\,I, Sr\,II, Ba\,II, and Tb\,II. The format of the plot follows the same structure as Fig. \ref{fig:ConfirmedDetectionsModels}.}
\label{fig:NewDectionsModels2}
\end{figure*}

\end{document}